\newif\ifeprint\eprinttrue
\def\figuredir{.}

\newlength\xx
\ifeprint

\documentclass
[rmp,twocolumn,amsfonts,amssymb,amsmath,showkeys,letterpaper,
raggedbottom,floatfix]{revtex4}
\makeatletter\def\raggedcolumn@skip{\vskip\z@\@plus.0001fil\relax}\makeatother
\def\eprint#1{E-print \urlalt{http://arxiv.org/abs/#1}{arXiv:#1}}

\else

\documentclass[final,twocolumn,natbib]{svjour3}
\journalname{J. Geod.}
\usepackage{amsmath}
\usepackage{amssymb}
\usepackage{mathptmx}
\def\eprint#1{preprint \url{http://arxiv.org/abs/#1}}

\fi

\usepackage{path}
\usepackage{array}
\date{\today}

\def\TITLE{Geodesics on an ellipsoid of revolution}
\usepackage{ifpdf}
\usepackage{times}
\usepackage[bookmarks=false]{hyperref}
\usepackage{breakurl}
\hypersetup{pdftitle={\TITLE},pdfauthor={C. F. F. Karney}}
\usepackage{graphicx}

\def\d{\mathrm d}
\newcount\minutes \newcount\hours
\hours=\time \divide\hours by 60 \multiply\hours by 60
\minutes=\time \advance\minutes by -\hours \divide\hours by 60

\def\twodigit#1{\ifnum#1<10 0\fi\the#1}

\ifpdf\def\urlalt#1#2{\burlalt{#2}{#1}}\else\let\urlalt=\burlalt\fi
\def\doi#1{\urlalt{http://dx.doi.org/#1}{doi:#1}}
\def\tfrac#1/#2 {{\textstyle\frac{#1}{#2}}}
\def\caesura{\bigm\Vert}
\def\brk{\,\caesura}
\def\v#1{\mathbf{#1}}
\newcommand{\ph}{\mathop{\mathrm{ph}}\nolimits}
\newcommand{\am}{\mathop{\mathrm{am}}\nolimits}
\newcommand{\dn}{\mathop{\mathrm{dn}}\nolimits}
\newcommand{\nd}{\mathop{\mathrm{nd}}\nolimits}
\newcommand{\cd}{\mathop{\mathrm{cd}}\nolimits}
\newcommand{\sn}{\mathop{\mathrm{sn}}\nolimits}
\def\abs#1{\left|#1\right|}
\def\sqrta#1{\sqrt{\vphantom{\sin^2k^2}\smash{#1}}}
\def\dlmf#1#2{\urlalt{http://dlmf.nist.gov/#2}{#1}}
\newcommand{\sign}{\mathop{\mathrm{sign}}\nolimits}
\newcommand{\atantwo}{\mathop{\mathrm{atan2}}\nolimits}
\newcommand{\exact}[1]{\mathbf{#1}}

\begin{document}

\ifeprint
\noindent\mbox{\begin{minipage}[b]{\textwidth}
\begin{flushright}
Link \url{http://geographiclib.sourceforge.net/geod.html}\par
\vspace{2ex}
\end{flushright}
\end{minipage}
\hspace{-\textwidth}}
\fi

\title{\TITLE}
\ifeprint
\author{\href{http://charles.karney.info}{Charles F. F. Karney}}
\email{charles.karney@sri.com}
\affiliation{\href{http://www.sri.com}{SRI International},
201 Washington Rd, Princeton, NJ 08543-5300, USA}
\else
\author{Charles F. F. Karney}
\institute{C. F. F. Karney \at
\href{http://www.sri.com}{SRI International},
201 Washington Rd\\Princeton, NJ 08543-5300, USA\\\email{charles.karney@sri.com}
}
\fi

\ifeprint\else\maketitle\fi

\begin{abstract}

Algorithms for the computation of the forward and inverse geodesic
problems for an ellipsoid of revolution are derived.  These are accurate
to better than $15\,\mathrm{nm}$ when applied to the terrestrial
ellipsoids.  The solutions of other problems involving geodesics
(triangulation, projections, maritime boundaries, and polygonal areas)
are investigated.

\ifeprint
\keywords{geometrical geodesy, geodesics, map projections,
triangulation, maritime boundaries, polygonal areas,
numerical methods}
\else
\keywords{geometrical geodesy \and geodesics \and map projections \and
triangulation \and maritime boundaries \and polygonal areas \and
numerical methods}
\fi
\end{abstract}

\ifeprint\maketitle\fi

\section{Introduction}\label{intro}

A geodesic is the natural ``straight line'', defined as the line of
minimum curvature, for the surface of the
earth \citep[pp.~220--222]{hilbert52}.  Geodesics are also of interest
because the shortest path between two points on the earth is always a
geodesic (although the converse is not necessarily true).  In most
terrestrial applications, the earth is taking to be an ellipsoid of
revolution and I adopt this model in this paper.

\begin{figure}[tb]
\begin{center}
\includegraphics[scale=0.75,angle=0]{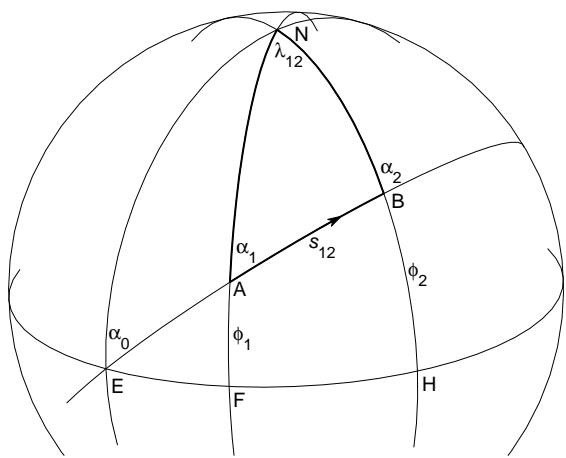}
\end{center}
\caption{\label{figtrig}
The ellipsoidal triangle $N\!AB$.  $N$ is the north pole, $N\!A$ and
$N\!B$ are meridians, and $AB$ is a geodesic of length $s_{12}$.  The
longitude of $B$ relative to $A$ is $\lambda_{12}$; the latitudes of $A$
and $B$ are $\phi_1$ and $\phi_2$.  $EFH$ is the equator with $E$ also
lying on the extension of the geodesic $AB$; and $\alpha_0$, $\alpha_1$,
and $\alpha_2$ are the azimuths of the geodesic at $E$, $A$, and $B$.}
\end{figure}%
Consider two points $A$, at latitude and longitude $(\phi_1,
\lambda_1)$, and $B$, at $(\phi_2, \lambda_2)$, on the surface of the
earth connected by a geodesic.  Denote the bearings of the geodesic
(measured clockwise from north) at $A$ and $B$ by $\alpha_1$ and
$\alpha_2$, respectively, and the length of the geodesic by $s_{12}$;
see Fig.~\ref{figtrig}.  There are two main ``geodesic problems'': the
{\it direct} problem, given $\phi_1$, $s_{12}$, and $\alpha_1$ determine
$\phi_2$, $\lambda_{12} = \lambda_2 - \lambda_1$, and $\alpha_2$; and
the {\it inverse} problem, given the $\phi_1$, $\phi_2$, and
$\lambda_{12}$, determine $s_{12}$, $\alpha_1$, and $\alpha_2$.
Considering the ellipsoidal triangle, $N\!AB$, in Fig.~\ref{figtrig},
where $N$ is the north pole, it is clear that both problems are
equivalent to solving the triangle given two sides and the included
angle.  In the first half of this paper
(Secs.~\ref{auxsphere}--\ref{errs}), I present solutions to these
problems: the accuracy of the solutions is limited only by the precision
of the number system of the computer; the direct solution is
non-iterative; the inverse solution is iterative but always converges in
a few iterations.  In the second half
(Secs.~\ref{spheroid-trig}--\ref{areasec}), I discuss several
applications of geodesics.

This paper has had a rather elephantine gestation.  My initial work in
this area grew out of a dissatisfaction with the widely used algorithms
for the main geodesic problems given by \citet{vincenty75a}.  These have
two flaws: firstly, the algorithms are given to a fixed order in the
flattening of the ellipsoid thereby limiting their accuracy; more
seriously, the algorithm for the inverse problem fails in the case of
nearly antipodal points.  Starting with the overview of the problem
given by \citet{williams02}, I cured the defects noted above and
included the algorithms in GeographicLib \citep{geographiclib17} in
March 2009.  At the same time, I started writing this paper and in the
course of this I came across references in, for
example, \citet{rainsford55} to work by Euler, Legendre, and Bessel.
Because no specific citations were given, I initially ignored these
references.  However, when I finally stumbled across Bessel's paper on
geodesics \citep{bessel25}, I was ``like some watcher of the skies when
a new planet swims into his ken''.  Overlooking minor quirks of notation
and the use of logarithms for numerical calculations, Bessel gives a
formulation and solution of the direct geodesic problem which is as
clear, as concise, and as modern as any I have read.  However, I was
surprised to discover that Bessel derived series expansions for the
geodesic integrals which are more economical than those used in the
English-language literature of the 20th century.  This prompted me to
undertake a systematic search for the other original papers on
geodesics, the fruits of which are available on-line \citep{geodbib}.
These contained other little known results---probably the most important
of which is the concept of the reduced length (Sect.~\ref{redlength})---
which have been incorporated into the geodesic classes in GeographicLib.

Some authors define ``spheroid'' as an ellipsoid of revolution.  In this
paper, I use the term in its more general sense, as an approximately
spherical figure.  Although this paper is principally concerned the
earth modeled as an ellipsoid of revolution, there are two sections
where the analysis is more general: (1) in the development of the
auxiliary sphere, Sect.~\ref{auxsphere} and Appendix~\ref{spheroid},
which applies to a spheroid of revolution; (2) in the generalization of
the gnomonic projection, Sect.~\ref{gnomproj}, which applies to a
general spheroid.

\section{Auxiliary Sphere}\label{auxsphere}

The study of geodesics on an ellipsoid of revolution was pursued by many
authors in the 18th and 19th centuries.  The important early papers are
by \citet{clairaut35}, \citet{euler55b}, \citet[Book~1,
Chaps.~1--3]{sejour89}, \citet{legendre89,legendre06},
and \citet{oriani06,oriani08,oriani10}.  \citet{clairaut35} found an
invariant for a geodesic (a consequence of the rotational symmetry of
the ellipsoid); this reduces the equations for the geodesic to
quadrature.  Subsequently, \citet{legendre06} and \citet{oriani06}
reduced the spheroidal triangle in Fig.~\ref{figtrig} into an equivalent
triangle on the ``auxiliary'' sphere.  \citet{bessel25} provided a
method (using tables that he supplied) to compute the necessary
integrals and allowed the direct problem to be solved with an accuracy
of a few centimeters.  In this section, I summarize this formulation of
geodesics; more details are given in Appendix \ref{spheroid} in which
the derivation of the auxiliary sphere is given.

I consider an ellipsoid of revolution with equatorial radius $a$, and
polar semi-axis $b$, flattening $f$, third flattening $n$, eccentricity
$e$, and second eccentricity $e'$ given by
\begin{align}
f &= (a - b) / a,\label{flat}\\
n &= (a - b)/(a + b) = f/(2-f),\label{flat3}\\
e^2 &= (a^2 - b^2)/a^2 = f(2-f),\label{ecc}\\
e'^2 &= (a^2 - b^2)/b^2 = e^2/(1-e^2).\label{ecc2}
\end{align}
In this paper, I am primarily concerned with oblate ellipsoids ($a >
b$); however, with a few exceptions, the formulas apply to prolate
ellipsoids merely by allowing $f < 0$ and $e^2 < 0$.
(Appendix \ref{prolate} addresses the modifications necessary to treat
prolate ellipsoids in more detail.) Most of the examples in this paper
use the WGS84 ellipsoid for which $a = \exact{6\,378.137}\,\mathrm{km}$
and $f = 1 / \exact{298.257\,223\,563}$.  (In the illustrative examples,
numbers given in boldface are exact.  The other numbers are obtained by
rounding the exact result to the given number of places.)  The surface
of the ellipsoid is characterized by its meridional and transverse radii
of curvature,
\begin{align}
\rho &= \frac a{1-f} w^3\label{rhoeq},\\
\nu &= \frac a{1-f} w\label{nueq},
\end{align}
respectively, where
\begin{equation}\label{weq0}
w = \frac1{\sqrt{1 + e'^2 \cos^2\phi}}.
\end{equation}

Consider a geodesic which intersects the equator, $\phi = 0$, in the
northwards direction with azimuth (measured clockwise from north)
$\alpha_0 \in [-\frac12\pi, \frac12\pi]$.  I denote this equatorial
crossing point $E$ and this is taken as the origin for the longitude
$\lambda$ and for measuring (signed) displacements $s$ along the
geodesic.  Because this definition of longitude depends on the geodesic,
longitude differences must be computed for points on the same geodesic.
Consider now a point $P$ with latitude $\phi$, longitude $\lambda$, a
displacement $s$ along the geodesic and form the ellipsoidal triangle
$N\!EP$ where $N$ represents the north pole.  The (forward) azimuth of
the geodesic at $P$ is $\alpha$ (i.e., the angle $N\!PE$ is
$\pi-\alpha$).

\begin{figure}[tb]
\begin{center}
\includegraphics[scale=0.75,angle=0]{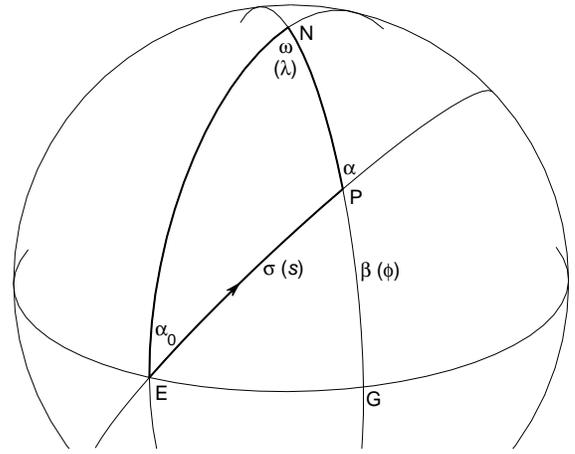}
\end{center}
\caption{\label{figtri}
The elementary ellipsoidal triangle $N\!EP$ mapped to the auxiliary
sphere.  $N\!E$ and $N\!PG$ are meridians; $EG$ is the equator; and $EP$
is the geodesic. The corresponding ellipsoidal variables are shown in
parentheses.}
\end{figure}%
Appendix \ref{spheroid} shows how this triangle may be transferred to the
auxiliary sphere where the latitude on the sphere is the {\it reduced
latitude} $\beta$, given by
\begin{equation}\label{redlat}
\tan\beta = (1-f)\tan\phi
\end{equation}
\citep[p.~136]{legendre06}, azimuths ($\alpha_0$ and $\alpha$) are
conserved, the longitude is denoted by $\omega$, and $EP$ is a portion
of a great circle with arc length $\sigma$; see Fig.~\ref{figtri}.
(\citet[p.~331]{cayley70} suggested the term {\it parametric latitude}
for $\beta$ because this is the angle most commonly used when the
meridian ellipse is written in parametric form.)  Applying Napier's
rules of circular parts \citep[\S66]{todhunter71} to the right triangle
$EPG$ in Fig.~\ref{figtri} gives a set of relations that apply
cyclically to $[\alpha_0, \frac12\pi - \sigma, \frac12\pi
- \alpha, \beta, \omega]$,
\begin{align}
\sin\alpha_0 &= \sin\alpha \cos\beta\label{napa0a}\\
&= \tan\omega \cot\sigma,\label{napa0b}\displaybreak[0]\\
\cos\sigma &= \cos\beta \cos\omega\label{napsa}\\
&= \tan\alpha_0 \cot\alpha,\label{napsb}\displaybreak[0]\\
\cos\alpha &= \cos\omega \cos\alpha_0\label{napaa}\\
&= \cot\sigma \tan\beta,\label{napab}\displaybreak[0]\\
\sin\beta &= \cos\alpha_0 \sin\sigma\label{napba}\\
&= \cot\alpha \tan\omega,\label{napbb}\displaybreak[0]\\
\sin\omega &= \sin\sigma \sin\alpha\label{napoa}\\
&= \tan\beta \tan\alpha_0.\label{napob}
\end{align}
In solving the main geodesic problems, I let $P$ stand for $A$ or $B$
with the quantities $\beta$, $\alpha$, $\sigma$, $\omega$, $s$, and
$\lambda$ acquiring a subscript $1$ or $2$.  I also define $\sigma_{12}
= \sigma_2 - \sigma_1$, the increase of $\sigma$ along $AB$, with
$\omega_{12}$, $s_{12}$, and $\lambda_{12}$ defined similarly.  Equation
(\ref{napa0a}) is Clairaut's characteristic equation for the geodesic,
Eq.~(\ref{clairaut}).  The ellipsoidal quantities $s$ and $\lambda$ are
then given by \citep[Eqs.~(5)]{bessel25}
\begin{equation}\label{geodeq}
\frac1a \frac{\d s}{\d\sigma} = \frac{\d\lambda}{\d\omega} = w.
\end{equation}
where $w$, Eq.~(\ref{weq0}), is now given in terms of $\beta$ by
\begin{equation}\label{weq}
w = \sqrt{1 - e^2 \cos^2\beta},
\end{equation}
and where the derivatives are taken holding $\alpha_0$ fixed (see
Appendix \ref{spheroid}).  Integrating the equation for $s$ and
substituting for $\beta$ from Eq.~(\ref{napba})
gives \citep[\S5]{bessel25}
\begin{equation}
\frac sb = \int_0^\sigma\sqrt{1 + k^2\sin^2\sigma'}\,\d\sigma', \label{disteq}
\end{equation}
where
\begin{equation}\label{keq}
k = e'\cos\alpha_0.
\end{equation}
As \citet[\S127]{legendre11} points out, the expression for $s$ given by
Eq.~(\ref{disteq}) is the same as that for the length along the
perimeter of an ellipse with semi-axes $b$ and $b\sqrt{1+k^2}$.  The
equation for $\lambda$ may also be expressed as an integral in $\sigma$
by using the second of Eqs.~(\ref{dsigeq}), $\d\omega/\d\sigma
= \sin\alpha/\cos\beta$; this gives \citep[\S9]{bessel25}
\begin{equation}
\lambda = \omega -
f\sin\alpha_0
\int_0^\sigma \frac{2-f}{1+(1-f)\sqrt{1 + k^2\sin^2\sigma'}}
\,\d\sigma'. \label{lameq}
\end{equation}
Consider a geodesic on the auxiliary sphere completely encircling the
sphere.  On the ellipsoid, the end point $B$ satisfies $\phi_2 = \phi_1$
and $\alpha_2 = \alpha_1$; however, from Eq.~(\ref{lameq}), the
longitude difference $\lambda_{12}$ falls short of $2\pi$ by
approximately $2\pi f\sin\alpha_0$.  As a consequence, geodesics on
ellipsoids (as distinct from spheres) are not, in general, closed.

In principle, the auxiliary sphere and Eqs.~(\ref{disteq}) and
(\ref{lameq}) enable the solution of all geodesic problems on an
ellipsoid.  However, the efficient solution of the inverse problem
requires knowledge of how neighboring geodesics behave.  This is
examined in the next section.

\section{Reduced Length}\label{redlength}

\begin{figure}[tb]
\begin{center}
\includegraphics[scale=0.75,angle=0]{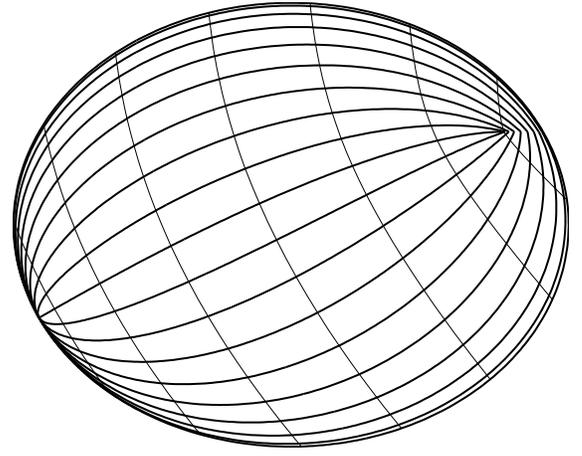}
\end{center}
\caption{\label{figall}
Geodesics from a point $\phi_1 = -30^\circ$.  The east-going geodesics
with azimuths $\alpha_1$ which are multiples of $10^\circ$ are shown as
heavy lines.  The spherical arc length of the geodesics is $\sigma_{12}
= 180^\circ$.  The geodesics are viewed from a distant point over the
equator at $\lambda - \lambda_1 = 90^\circ$.  The light lines show
equally spaced geodesic circles.  The flattening of the ellipsoid was
taken to be $f = \frac15$ for the purposes of this figure.}
\end{figure}%
Following Bessel's paper, \citet{gauss27} studied the properties of
geodesics on general surfaces.  Consider all the geodesics emanating
from a point $A$.  Define a {\it geodesic circle} centered at $A$ to be
the locus of points a fixed (geodesic) distance $s_{12}$ from $A$; this
is a straightforward extension of the definition of a circle on a
plane.  \citet[\S15]{gauss27} proved that geodesic circles intersect the
geodesics at right angles; see Fig.~\ref{figall}.

(Circles on a plane also have a second property: they are the curves
which enclose the maximum area for a given perimeter.  On an ellipsoid
such curves have constant geodesic
curvature \citep{minding30}.  \citet[\S652]{darboux94} adopts this as
his definition of the geodesic circle; however, in general, these curves
are different from the geodesic circles as defined in the previous
paragraph.)

\citet{gauss27} also introduced the concept of the
{\it reduced length} $m_{12}$ for the geodesic which is also the subject
of a detailed investigation by \citet{christoffel68} (the term is his
coinage).  Consider two geodesics of length $s_{12}$ departing from $A$
at azimuths $\alpha_1$ and $\alpha_1 + \d\alpha_1$.  On a flat surface
the end points are separated by $s_{12}\,\d\alpha_1$ (in the limit
$\d\alpha_1 \rightarrow 0$).  On a curved surface the separation is
$m_{12}\,\d\alpha_1$ where $m_{12}$ is the reduced
length.  \citet[\S19]{gauss27} showed that the reduced length satisfies
the differential equation
\begin{equation}\label{reducedeq}
\frac{\d^2m}{\d s^2} + K(s)m = 0,
\end{equation}
where $K(s)$ is the Gaussian curvature of the surface.  Let $m(s; s_1)$
be the solution to Eq.~(\ref{reducedeq}) subject to the initial
conditions
\[
m(s_1; s_1) = 0, \quad \left. \frac{\d m(s; s_1)}{\d s} \right|_{s = s_1} = 1;
\]
then $m_{12}$ is given by $m_{12} = m(s_2; s_1)$.  Equation
(\ref{reducedeq}) obeys a simple reciprocity relation $m(s_1; s_2) = -
m(s_2; s_1)$ which gives the result \citep[\S9]{christoffel68} that the
reduced length is invariant under interchange of the end points, i.e.,
$m_{21} = -m_{12}$.

For a geodesic on an ellipsoid of revolution, the Gaussian curvature is
given by
\begin{equation}
K = \frac1{\rho\nu} = \frac{ b^2 } { a^4 w^4 }
  = \frac1 {b^2 (1 + k^2 \sin^2\sigma)^2}. \label{curvature}
\end{equation}
\citet[\S6.5]{helmert80} solves Eq.~(\ref{reducedeq}) in this case
to give
\begin{align}
\label{meq}
m_{12}/b &=
  \sqrta{1 + k^2\sin^2\sigma_2} \cos\sigma_1 \sin\sigma_2 \notag\\
&\quad - \sqrta{1 + k^2\sin^2\sigma_1} \sin\sigma_1 \cos\sigma_2 \notag\\
&\quad - \cos\sigma_1 \cos\sigma_2 \bigl(J(\sigma_2) - J(\sigma_1)\bigr),
\end{align}
where
\begin{align}
\label{jeq}
J(\sigma) &=
\int_0^\sigma
\frac{k^2\sin^2\sigma'}{\sqrt{1 + k^2\sin^2\sigma'}}\,\d\sigma' \notag\\
&= \frac sb - \int_0^\sigma \frac 1{\sqrt{1 + k^2\sin^2\sigma'}}\,\d\sigma'.
\end{align}
In the spherical limit, Eq.~(\ref{meq}) reduces to
\[
m_{12} = a \sin\sigma_{12} = a \sin(s_{12}/a).
\]

\citet[\S23]{gauss27} also introduced what I call
the {\it geodesic scale} $M_{12}$, which gives the separation of close,
initially parallel, geodesics, and which is given by another solution of
Eq.~(\ref{reducedeq}), $M(s;s_1)$, with the initial conditions
\[
M(s_1; s_1) = 1, \quad \left. \frac{\d M(s; s_1)}{\d s} \right|_{s = s_1} = 0.
\]
\citet[\S633]{darboux94} shows how to construct the solution given two
independent solutions of Eq.~(\ref{reducedeq}) which are given, for
example, by Eq.~(\ref{meq}) with two different starting points.  This
gives
\begin{align}\label{Meq}
M_{12} &=
\cos\sigma_1 \cos\sigma_2 \notag\\
&\quad + \frac{\sqrt{1 + k^2\sin^2\sigma_2}}{\sqrt{1 + k^2\sin^2\sigma_1}}
\sin\sigma_1 \sin\sigma_2 \notag\\
&\quad - \frac{\sin\sigma_1 \cos\sigma_2 \bigl(J(\sigma_2) - J(\sigma_1)\bigr)}
{\sqrt{1 + k^2\sin^2\sigma_1}},
\end{align}
where $M_{12} = M(s_2; s_1)$.  Note that $M_{12}$ is {\it not} symmetric
under interchange of the end points.  In the spherical limit,
Eq.~(\ref{Meq}) reduces to
\[
M_{12} = \cos\sigma_{12} = \cos(s_{12}/a).
\]
By direct differentiation, it is easy to show that the Wronskian for the
two solutions $m_{12}$ and $M_{12}$ is a constant.  Substituting the
initial conditions then gives
\begin{equation}\label{wronski}
M_{12}\frac{\d m_{12}}{\d s_2}-m_{12}\frac{\d M_{12}}{\d s_2} = 1.
\end{equation}

There is little mention of the reduced length in the geodetic literature
in the English language of the last century.  An exception
is \citet[Prop.~IV]{tobey28} who derives an expression for $m_{12}$ as a
series valid for small $s_{12}/a$ \citep[\S4.22]{rapp91}.

\section{Properties of the integrals}\label{intprop}

The solution of the main geodesic problems requires the evaluation of
the three integrals appearing in Eqs.~(\ref{disteq}), (\ref{jeq}), and
(\ref{lameq}).  In order to approach the evaluation systematically, I
write these integrals as
\begin{align}
I_1(\sigma) &= \int_0^\sigma\sqrt{1 + k^2\sin^2\sigma'}\,\d\sigma',
\label{i1eq}\displaybreak[0]\\
I_2(\sigma) &= \int_0^\sigma\frac 1{\sqrt{1 + k^2\sin^2\sigma'}}\,\d\sigma',
\label{i2eq}\displaybreak[0]\\
I_3(\sigma) &= \int_0^\sigma
\frac{2-f}{1+(1-f)\sqrt{1 + k^2\sin^2\sigma'}}\,\d\sigma'.
\label{i3eq}
\end{align}
In terms of $I_j(\sigma)$, Eqs.~(\ref{disteq}), (\ref{jeq}), and
(\ref{lameq}) become
\begin{align}
s/b &= I_1(\sigma),\label{i1expr}\\
J(\sigma) &= I_1(\sigma) - I_2(\sigma),\label{i2expr}\\
\lambda &= \omega - f\sin\alpha_0 \,I_3(\sigma).\label{i3expr}
\end{align}

The integrals $I_j(\sigma)$ may be expressed in terms of elliptic
functions as \citep{jacobi55,luther56,forsyth96a}
\begin{align}
I_1(\sigma)&=k'_1 \int_0^{u_1} \nd^2(u',k_1) \,\d u',\label{ellfun1}
\displaybreak[0]\\
I_2(\sigma)&=k'_1 u_1,\label{ellfun2}\displaybreak[0]\\
I_3(\sigma)&=-\frac{(1-f)k'_1}{f\sin^2\alpha_0}
\int_0^{u_1} \frac{\nd^2(u', k_1)}{1 + \cot^2\alpha_0\cd^2(u', k_1)}\,\d u'
\notag\\
&\qquad+\frac{\tan^{-1}(\sin\alpha_0\tan\sigma)}{f\sin\alpha_0},
\label{ellfun3}
\end{align}
where $k_1 = k/\sqrt{1 + k^2}$, $k'_1 = \sqrta{1 - k_1^2} = 1/\sqrt{1 +
k^2}$,
\[
\am(u_1 - K(k_1), k_1) = \sigma - \tfrac1/2 \pi,
\]
$\cd(x,k)$ and $\nd(x,k)$ are Jacobian elliptic functions
\citep[\S\dlmf{22.2}{22.2}]{dlmf10}, $\am(x,k)$ is Jacobi's
amplitude function \citep[\S\dlmf{22.16(i)}{22.16.i}]{dlmf10}, and
$K(k)$ is the complete elliptic integral of the first
kind \citep[\S\dlmf{19.2(ii)}{19.2.ii}]{dlmf10}.  The integrals can also
be written in closed form as \citep[\S\S127--128]{legendre11}
\begin{align}
I_1(\sigma) &=\frac1{k'_1}E(\sigma-\tfrac1/2 \pi, k_1) - c_1,
\label{ellipt1}
\displaybreak[0]\\
I_2(\sigma) &= k'_1F(\sigma-\tfrac1/2 \pi, k_1) - c_2,
\label{ellipt2}
\displaybreak[0]\\
I_3(\sigma) &= -\frac{1-f}{fk'_1\sin^2\alpha_0} G(\sigma-\tfrac1/2 \pi,
-\cot^2\alpha_0, k_1)\notag\\
&\qquad+\frac{\tan^{-1}(\sin\alpha_0\tan\sigma)}{f\sin\alpha_0} -c_3,
\label{ellipt3}
\end{align}
where
\begin{align}
G(\phi,\alpha^2,k) &= \int_0^\phi
\frac{\sqrt{1 - k^2\sin^2\theta}}{1 - \alpha^2\sin^2\theta}\,\d\theta
\notag\\
&=\biggl(1-\frac{k^2}{\alpha^2}\biggr)\Pi(\phi, \alpha^2, k)
+\frac{k^2}{\alpha^2}F(\phi, k),
\label{G-eq}
\end{align}
the integration constants $c_j$ are given by the condition $I_j(0) = 0$,
and $F(\phi, k)$, $E(\phi, k)$, and $\Pi(\phi, \alpha^2, k)$ are
Legendre's incomplete elliptic integrals of the first, second, and third
kinds \citep[\S\dlmf{19.2(ii)}{19.2.ii}]{dlmf10}.

There are several ways that the integrals may be computed.  One
possibility is merely to utilize standard
algorithms \citep{bulirsch65,carlson95} for elliptic functions and
integrals for the evaluation of Eqs.~(\ref{ellipt1})--(\ref{ellipt3}).
Alternatively, some authors, for example \citet{saito70,saito79}, have
employed numerical quadrature on Eqs.~(\ref{i1eq}) and (\ref{i3eq}).
However, the presence of small parameters in the integrals also allow
the integrals to be expressed in terms of rapidly converging
series.  \citet{bessel25} used this approach and tabulated the
coefficients appearing in these series thereby allowing the direct
geodesic problem to be solved easily and with an accuracy of about 8
decimal digits.  I also use this technique because it allows the
integrals to be evaluated efficiently and accurately.

Before carrying out the series expansions, it is useful to established
general properties of the integrals.  As functions of $\sigma$, the
integrands are all even, periodic with period $\pi$, positive, and of
the form $1+O(f)$.  (Note that $k^2 = O(f)$.)  The integrals
$I_j(\sigma)$ can therefore be expressed as
\begin{equation}\label{ij}
I_j(\sigma) = A_j \bigl(\sigma + B_j(\sigma)\bigr),\quad
\text{for $j=1,2,3$},
\end{equation}
where the constant $A_j = 1 + O(f)$ and $B_j(\sigma) = O(f)$ is odd and
periodic with period $\pi$ and so may be written as
\begin{equation}\label{bj}
B_j(\sigma) = \sum_{l = 1}^\infty C_{jl}\sin 2l,\sigma\quad
\text{for $j=1,2,3$}.
\end{equation}
In addition, it is easy to show that $C_{jl} = O(f^l)$.  In order to
obtain results for $s$, $\lambda$, and $m_{12}$ accurate to order $f^L$,
truncate the sum in Eq.~(\ref{bj}) at $l = L$ for $j = 1$ and $2$ and at
$l = L - 1$ for $j = 3$.  (In the equation for $\lambda$,
Eq.~(\ref{i3expr}), $I_3(\sigma)$ is multiplied by $f$; so it is only
necessary to compute this integral to order $f^{L-1}$.)  Similarly, the
expansions for $A_j$ and $C_{jl}$ may be truncated at order $f^L$ for $j
= 1$ and $2$ and at order $f^{L-1}$ for $j = 3$.

The form of the trigonometric expansion, Eqs.~(\ref{ij}) and (\ref{bj}),
and the subsequent expansion of the coefficients as Taylor series in $f$
(or an equivalent small parameter), that I detail in the next section,
provide expansions for the integrals which, with a modest number of
terms, are valid for arbitrarily long geodesics.  This is to be
distinguished from a number of approximate methods for short
geodesics \citep[\S6]{rapp91}, which were derived as an aid to computing
by hand.

\section{Series expansions of the integrals}\label{intser}

Finding explicit expressions for $A_j$ and $C_{jl}$ is simply matter of
expanding the integrands for small $k$ and $f$ enabling the integrals to
be evaluated.  I used the algebra system \citet{maxima} to carry out the
necessary expansion, integration, and simplification.  Here, I present
the expansions to order $L = 8$.

The choice of expansion parameter affects the compactness of the
resulting expressions.  In the case of $I_1$, Bessel introduced a change
of variable,
\begin{align}
k^2 &= \frac{4\epsilon}{(1-\epsilon)^2},\label{epseq}\\
\epsilon &=\frac{\sqrt{1+k^2}-1}{\sqrt{1+k^2}+1}
    = \frac{k^2}{\bigl(\sqrt{1+k^2}+1\bigr)^2},\label{epseqa}
\end{align}
into Eq.~(\ref{i1eq}) to give
\begin{equation}\label{disteqa}
(1-\epsilon)I_1(\sigma) =
\int_0^\sigma\sqrt{1 - 2\epsilon \cos 2\sigma' + \epsilon^2}\,\d\sigma'.
\end{equation}
The integrand now exhibits the symmetry that it is invariant under the
transformation $\epsilon \rightarrow -\epsilon$ and
$\sigma \rightarrow \frac12 \pi - \sigma$.  This results in a series
with half the number of terms (compared to a simple expansion in $k^2$).
The relation between $k$ and $\epsilon$ is the same as that between the
second eccentricity and third flattening of an ellipsoid, $e'$ and $n$,
and frequently formulas for ellipsoids are simpler when expressed in
terms of $n$ because of the symmetry of its definition,
Eq.~(\ref{flat3}).  (Bessel undertook the task of tabulating the
coefficients in the series for $B_1(\sigma)$ for some 200 different
values of $k$.  This gave him with a strong incentive to find a way to
halve the amount of work required.)

The quantity $\epsilon$ is $O(f)$; thus expanding Eq.~(\ref{disteqa}) to
order $f^8$ is, asymptotically, equivalent to expanding to order
$\epsilon^8$.  Carrying out this expansion in $\epsilon$ then yields
\begin{align}
A_1 &= (1 - \epsilon)^{-1} \bigl(1 + \tfrac1/4 \epsilon^2
       + \tfrac1/64 \epsilon^4
       + \tfrac1/256 \epsilon^6 \notag\\&\qquad\qquad\qquad
       \brk+ \tfrac25/16384 \epsilon^8 + \cdots\bigr),
\label{A1}\displaybreak[0]\\
C_{11} &= - \tfrac1/2 \epsilon
       + \tfrac3/16 \epsilon^3
       - \tfrac1/32 \epsilon^5
       \brk+ \tfrac19/2048 \epsilon^7 + \cdots,\displaybreak[0]\notag\\
C_{12} &= - \tfrac1/16 \epsilon^2
       + \tfrac1/32 \epsilon^4
       - \tfrac9/2048 \epsilon^6
       \brk+ \tfrac7/4096 \epsilon^8 + \cdots,\displaybreak[0]\notag\\
C_{13} &= - \tfrac1/48 \epsilon^3
       + \tfrac3/256 \epsilon^5
       \brk- \tfrac3/2048 \epsilon^7 + \cdots,\displaybreak[0]\notag\\
C_{14} &= - \tfrac5/512 \epsilon^4
       + \tfrac3/512 \epsilon^6
       \brk- \tfrac11/16384 \epsilon^8 + \cdots,\displaybreak[0]\notag\\
C_{15} &= - \tfrac7/1280 \epsilon^5
       \brk+ \tfrac7/2048 \epsilon^7 + \cdots,\displaybreak[0]\notag\\
C_{16} &= - \tfrac7/2048 \epsilon^6
       \brk+ \tfrac9/4096 \epsilon^8 + \cdots,\displaybreak[0]\notag\\
C_{17} &= \brk- \tfrac33/14336 \epsilon^7 + \cdots,\displaybreak[0]\notag\\
C_{18} &= \brk- \tfrac429/262144 \epsilon^8 + \cdots.
\label{C1}
\end{align}
I use the caesura symbol, $\caesura$, to indicate where the series may
be truncated, at $O(f^6)$, while still giving full accuracy with
double-precision arithmetic for $\abs f \le 1/150$ (this is established
in Sect.~\ref{errs}).  Equation (\ref{C1}) is a simple extension of
series given by \citet[\S5]{bessel25}, except that I have divided out
the coefficient of the linear term $A_1$.  Bessel's formulation was used
throughout the 19th century and the series given here, truncated to
order $\epsilon^4$, coincides with \citet[Eq.~(5.5.7)]{helmert80}.
However, many later works, such as \citet[Eqs.~(18)--(19)]{rainsford55},
use less efficient expansions in $k^2$.

The expansion for $I_2$ proceeds analogously yielding
\begin{align}
A_2 &= (1 - \epsilon) \bigl(1 + \tfrac1/4 \epsilon^2
       + \tfrac9/64 \epsilon^4
       + \tfrac25/256 \epsilon^6 \notag\\&\qquad\qquad\quad
       \brk+ \tfrac1225/16384 \epsilon^8 + \cdots\bigr),
\label{A2}\displaybreak[0]\\
C_{21} &= \tfrac1/2 \epsilon
       + \tfrac1/16 \epsilon^3
       + \tfrac1/32 \epsilon^5
       \brk+ \tfrac41/2048 \epsilon^7 + \cdots,\displaybreak[0]\notag\\
C_{22} &= \tfrac3/16 \epsilon^2
       + \tfrac1/32 \epsilon^4
       + \tfrac35/2048 \epsilon^6
       \brk+ \tfrac47/4096 \epsilon^8 + \cdots,\displaybreak[0]\notag\\
C_{23} &= \tfrac5/48 \epsilon^3
       + \tfrac5/256 \epsilon^5
       \brk+ \tfrac23/2048 \epsilon^7 + \cdots,\displaybreak[0]\notag\\
C_{24} &= \tfrac35/512 \epsilon^4
       + \tfrac7/512 \epsilon^6
       \brk+ \tfrac133/16384 \epsilon^8 + \cdots,\displaybreak[0]\notag\\
C_{25} &= \tfrac63/1280 \epsilon^5
       \brk+ \tfrac21/2048 \epsilon^7 + \cdots,\displaybreak[0]\notag\\
C_{26} &= \tfrac77/2048 \epsilon^6
       \brk+ \tfrac33/4096 \epsilon^8 + \cdots,\displaybreak[0]\notag\\
C_{27} &= \brk  \tfrac429/14336 \epsilon^7 + \cdots,\displaybreak[0]\notag\\
C_{28} &= \brk  \tfrac6435/262144 \epsilon^8 + \cdots.
\label{C2}
\end{align}

The expansion of $I_3$ is more difficult because of the presence of two
parameters $f$ and $k$; this presented a problem for Bessel---it was a
practical impossibility for him to compile complete tables of
coefficients with two dependencies.  Later, when the flattening of the
earth was known with some precision, he might have contemplated
compiling tables for a few values of $f$.
Instead, \citet[\S8]{bessel25} employs a transformation to move the
dependence on the second parameter into a higher order term which he
then neglects.  The magnitude of the neglected term is about
$0.000\,003''$ for geodesics stretching half way around the WGS84
ellipsoid; this corresponds to an error in position of about
$0.1\,\mathrm{mm}$.  Although this is a very small error, there is no
need to resort to such trickery nowadays because computers can evaluate
the coefficients as needed.

Following \citet[Eq.~(5.8.14)]{helmert80}, I expand in $n$ and
$\epsilon$, both of which are $O(f)$, to give
\begin{align}
A_3 &= 1 - \bigl(\tfrac1/2 - \tfrac1/2 n\bigr) \epsilon
       - \bigl(\tfrac1/4 + \tfrac1/8 n - \tfrac3/8 n^2\bigr)
       \epsilon^2 \notag\\&\qquad{}
       - \bigl(\tfrac1/16 + \tfrac3/16 n + \tfrac1/16 n^2 \brk
       - \tfrac5/16 n^3\bigr) \epsilon^3 \notag\\&\qquad{}
       - \bigl(\tfrac3/64 + \tfrac1/32 n \brk
       + \tfrac5/32 n^2 + \tfrac5/128 n^3 + \cdots\bigr)
        \epsilon^4 \notag\\&\qquad{}
       - \bigl(\tfrac3/128 \brk+ \tfrac5/128 n + \tfrac5/256 n^2 + \cdots\bigr)
       \epsilon^5 \notag\\&\qquad{}
       \brk- \bigl(\tfrac5/256 + \tfrac15/1024 n + \cdots\bigr) \epsilon^6
       - \tfrac25/2048 \epsilon^7 + \cdots,
\label{A3}\displaybreak[0]\\
C_{31} &= \bigl(\tfrac1/4 - \tfrac1/4 n\bigr) \epsilon
        + \bigl(\tfrac1/8 - \tfrac1/8 n^2\bigr) \epsilon^2 \notag\\&\quad{}
        + \bigl(\tfrac3/64 + \tfrac3/64 n - \tfrac1/64 n^2 \brk
        - \tfrac5/64 n^3\bigr) \epsilon^3 \notag\\&\quad{}
        + \bigl(\tfrac5/128 + \tfrac1/64 n \brk
        + \tfrac1/64 n^2 - \tfrac1/64 n^3 + \cdots\bigr)
         \epsilon^4 \notag\\&\quad{}
        + \bigl(\tfrac3/128 \brk
        + \tfrac11/512 n + \tfrac3/512 n^2 + \cdots\bigr)
         \epsilon^5 \notag\\&\quad{}
        \brk+ \bigl(\tfrac21/1024 + \tfrac5/512 n + \cdots\bigr) \epsilon^6
        + \tfrac243/16384 \epsilon^7  + \cdots,\displaybreak[0]\notag\\
C_{32} &= \bigl(\tfrac1/16 - \tfrac3/32 n + \tfrac1/32 n^2\bigr)
        \epsilon^2 \notag\\&\quad{}
        + \bigl(\tfrac3/64 - \tfrac1/32 n - \tfrac3/64 n^2 \brk
        + \tfrac1/32 n^3\bigr) \epsilon^3 \notag\\&\quad{}
        + \bigl(\tfrac3/128 + \tfrac1/128 n \brk
        - \tfrac9/256 n^2 - \tfrac3/128 n^3 + \cdots\bigr)
        \epsilon^4 \notag\\&\quad{}
        + \bigl(\tfrac5/256 \brk+ \tfrac1/256 n - \tfrac1/128 n^2 + \cdots\bigr)
         \epsilon^5 \notag\\&\quad{}
        \brk+ \bigl(\tfrac27/2048 + \tfrac69/8192 n + \cdots\bigr) \epsilon^6
        + \tfrac187/16384 \epsilon^7 + \cdots, \displaybreak[0]\notag\\
C_{33} &= \bigl(\tfrac5/192 - \tfrac3/64 n + \tfrac5/192 n^2 \brk
        - \tfrac1/192 n^3\bigr) \epsilon^3 \notag\\&\quad{}
        + \bigl(\tfrac3/128 - \tfrac5/192 n \brk
        - \tfrac1/64 n^2 + \tfrac5/192 n^3 + \cdots\bigr)
        \epsilon^4 \notag\\&\quad{}
        + \bigl(\tfrac7/512 \brk
        - \tfrac1/384 n - \tfrac77/3072 n^2 + \cdots\bigr)
        \epsilon^5 \notag\\&\quad{}
        \brk+ \bigl(\tfrac3/256 - \tfrac1/1024 n + \cdots\bigr) \epsilon^6
        + \tfrac139/16384 \epsilon^7  + \cdots,\displaybreak[0]\notag\\
C_{34} &= \bigl(\tfrac7/512 - \tfrac7/256 n \brk
        + \tfrac5/256 n^2 - \tfrac7/1024 n^3 + \cdots\bigr)
        \epsilon^4 \notag\\&\quad{}
        + \bigl(\tfrac7/512 \brk
        - \tfrac5/256 n - \tfrac7/2048 n^2 + \cdots\bigr)
        \epsilon^5 \notag\\&\quad{}
        \brk+ \bigl(\tfrac9/1024 - \tfrac43/8192 n + \cdots\bigr) \epsilon^6
        + \tfrac127/16384 \epsilon^7  + \cdots,\displaybreak[0]\notag\\
C_{35} &= \bigl(\tfrac21/2560 \brk
        - \tfrac9/512 n + \tfrac15/1024 n^2 + \cdots\bigr)
        \epsilon^5 \notag\\&\quad{}
        \brk+ \bigl(\tfrac9/1024 - \tfrac15/1024 n + \cdots\bigr) \epsilon^6
        + \tfrac99/16384 \epsilon^7  + \cdots,\displaybreak[0]\notag\\
C_{36} &= \brk\bigl(\tfrac11/2048 - \tfrac99/8192 n + \cdots\bigr) \epsilon^6
        + \tfrac99/16384 \epsilon^7  + \cdots,\displaybreak[0]\notag\\
C_{37} &= \brk\tfrac429/114688 \epsilon^7 + \cdots.
\label{I3}
\end{align}
I continue these expansions out to order $f^7$, which is, as noted as
the end of Sect.~\ref{intprop}, consistent with expanding the other
integrals to order $f^8$.  All the parenthetical terms in
Eqs.~(\ref{A3}) and (\ref{I3}) are functions of $n$ only and so may be
evaluated once for a given ellipsoid.  Note that the coefficient of
$\epsilon^l$ is a {\it terminating} polynomial of order $l$ in $n$.
This is a curious degeneracy of this integral when expressed in terms of
$n$ and $\epsilon$.  \citet[Eqs.~(10)--(11)]{rainsford55} writes $k^2$
in Eq.~(\ref{lameq}) in terms of $f$ and $\cos^2\alpha_0$ and gives a
expansion for the integral in powers of $f$.  This results in an
expansion with more terms.

The direct geodesic problem requires solving Eq.~(\ref{disteq}) for
$\sigma$ in terms of $s$.  (There is a unique solution because $\d
s/\d\sigma > 0$.)  Equations (\ref{i1expr}) and (\ref{ij}), with $j =
1$, can be written as
\begin{equation}\label{tau1}
\tau = \sigma + B_1(\sigma),
\end{equation}
where
\begin{equation}\label{taudef}
\tau = s/(bA_1),
\end{equation}
which shows that finding $\sigma$ as a function of $s$ is equivalent to
inverting Eq.~(\ref{tau1}).  This may be accomplished
using \citet[\S16]{lagrange70} inversion, which gives
\begin{equation}\label{sigmaeq}
\sigma = \tau + B'_1(\tau),
\end{equation}
where
\[
B'_j(\tau) = \sum_{l = 1}^\infty
\frac{(-1)^l}{l!}
\left.
\frac{\d^{l-1}B_j(\sigma)^l}{\d\sigma^{l-1}}
\right|_{\sigma=\tau}.
\]
Carrying out these operations with \citet{maxima} gives
\begin{equation}\label{Bj1}
B'_j(\tau) = \sum_{l = 1}^\infty C'_{jl}\sin 2l\tau,
\end{equation}
where
\begin{align}
C'_{11} &= \tfrac1/2 \epsilon
        - \tfrac9/32 \epsilon^3
        + \tfrac205/1536 \epsilon^5
        \brk- \tfrac4879/73728 \epsilon^7 + \cdots,\displaybreak[0]\notag\\
C'_{12} &= \tfrac5/16 \epsilon^2
        - \tfrac37/96 \epsilon^4
        + \tfrac1335/4096 \epsilon^6
        \brk- \tfrac86171/368640 \epsilon^8 + \cdots,\displaybreak[0]\notag\\
C'_{13} &= \tfrac29/96 \epsilon^3
        - \tfrac75/128 \epsilon^5
        \brk+ \tfrac2901/4096 \epsilon^7 + \cdots,\displaybreak[0]\notag\\
C'_{14} &= \tfrac539/1536 \epsilon^4
        - \tfrac2391/2560 \epsilon^6
        \brk+ \tfrac1082857/737280 \epsilon^8 + \cdots,\displaybreak[0]\notag\\
C'_{15} &= \tfrac3467/7680 \epsilon^5
        \brk- \tfrac28223/18432 \epsilon^7 + \cdots,\displaybreak[0]\notag\\
C'_{16} &= \tfrac38081/61440 \epsilon^6
        \brk- \tfrac733437/286720 \epsilon^8 + \cdots,\displaybreak[0]\notag\\
C'_{17} &= \brk  \tfrac459485/516096 \epsilon^7
        + \cdots,\displaybreak[0]\notag\\
C'_{18} &= \brk  \tfrac109167851/82575360 \epsilon^8 + \cdots.
\label{C11}
\end{align}

\citet[\S13]{legendre06} makes a half-hearted attempt at inverting
Eq.~(\ref{disteq}) in terms of trigonometric functions of $s/b$ (instead
of $s/(bA_1)$).  Because the period is slightly different from $\pi$,
the result is a much more messy expansion than given
here.  \citet{oriani33} used Lagrange inversion to solve the distance
integral as a series in $k^2$.  Finally, \citet[Eq.~(5.6.8)]{helmert80}
carries out the inversion in terms of $\epsilon$ (as here) including
terms to order $\epsilon^3$.

Most other authors invert Eq.~(\ref{disteq}) iteratively.  Both
\citet{bessel25} and \citet{vincenty75a} use the scheme given by the
first line of
\begin{align*}
\sigma^{(i+1)} &= \tau - B_1(\sigma^{(i)})\\
&=\sigma^{(i)} + \frac{s/b - I_1(\sigma^{(i)})}{A_1},
\end{align*}
with $\sigma^{(0)} = \tau$.  This converges linearly and this is
adequate to achieve accuracies on the order of $0.1\,\mathrm{mm}$.  A
superior iterative solution is given by Newton's method which can be
written as
\[
\sigma^{(i+1)}
=\sigma^{(i)} + \frac{s/b - I_1(\sigma^{(i)})}{\sqrt{1 + k^2\sin^2\sigma^{(i)}}},
\]
which converges quadratically enabling the solution to full machine
precision to be found in a few iterations.  The simple iterative scheme
effectively replaces the denominator in the fraction above by its mean
value $A_1$.  By using the reverted series, I solve for $\sigma$
non-iteratively.  This does incur the cost of evaluating the
coefficients $C'_{1l}$; however, this cost can be amortized if several
points along the same geodesics are computed.

I give the expansions for $A_j$, $C_{jl}$, and $C'_{1l}$ to order $f^8$
above.  However, these expansions are generated by \citet{maxima} with
only a few dozen lines of code in about 4 seconds.  The expansions can
easily be extended to higher order just by changing one parameter in the
Maxima code; this has been tested by generating the series to order
$f^{30}$, which takes about 13 minutes.  Maxima is also used to generate
the C++ code for the expansions in GeographicLib.  Treating Maxima as a
preprocessor for the C++ compiler, the methods presented here can be
considered to be of arbitrary order.  In the current GeographicLib
implementation, the order of the expansion is a compile-time constant
which can be set to any $L \le 8$.  As a practical matter, the series
truncated at order $f^6$ suffice to give close to full accuracy with
double-precision arithmetic for terrestrial ellipsoids.

Although \citet{oriani06} and \citet{bessel25} give expressions for
general terms in their expansions, most subsequent authors are content
to work out just a few terms.  An exception is the work
of \citet{levallois52} who formulate the problem in terms of Wallis
integrals where the general term in the series is given by a recursion
relation, allowing the series to be extended to arbitrary order at
run-time \citep[Chap.~5]{levallois70}.  \citet{pittman86} independently
derived a similar method.  Unfortunately, Pittman uses $\beta$ as the
variable of integration (instead of $\sigma$); because the latitude does
not change monotonically along the geodesic, this choice leads to a loss
of numerical accuracy and to technical problems in following geodesics
through vertices (the positions of extrema of the latitude for the
geodesic).

Because of their widespread adoption, the expansions given by
\citet{vincenty75a} are of special interest.  He expresses
the distance integral in terms of $C_{11}$ and the longitude integral
in terms of $1-A_3$.  This leads to rather compact series when truncated
at order $f^3$.  However, much of the simplicity disappears at the next
higher order, at which point Vincenty's technique offers no particular
advantage.  Nevertheless, his procedure does expose the symmetry between
$k$ and $\sigma$ in $I_1(\sigma)$ even though his original expansion was
in $k^2$.  Belatedly, \citet[addendum]{vincenty75a} discovered the
economy of Bessel's change of variable, Eq.~(\ref{epseq}), to obtain the
same expansions for $A_1$ and $C_{11}$ as given here (truncated at
order $\epsilon^3$).  Incidentally, the key constraint that Vincenty
worked under was that his programs should fit onto calculators, such as
the Wang 720 \citep[p.~10]{vincenty75b}, which only had a few kilobytes
of memory; this precluded the use of higher-order expansions and allowed
for only a simple (and failure prone) iterative solution of the inverse
problem.  Vincenty cast the series in ``nested'', i.e., Horner, form, in
order to minimize program size and register use; I also use the Horner
scheme for evaluating the series because of its accuracy and speed.

\section{Direct problem}\label{direct}

The direct geodesic problem is to determine $\phi_2$, $\lambda_{12}$,
and $\alpha_2$, given $\phi_1$, $\alpha_1$ and $s_{12}$; see
Fig.~\ref{figtrig}.  The solution starts by finding $\beta_1$ using
Eq.~(\ref{redlat}); next solve the spherical triangle $N\!EA$, to give
$\alpha_0$, $\sigma_1$, and $\omega_1$, by means of Eqs.~(\ref{napa0a}),
(\ref{napab}), and (\ref{napa0b}).  With $\alpha_0$ known, the
coefficients $A_j$, $C_{jl}$, and $C'_{1l}$ may be computed from the
series in Sect.~\ref{intser}.  These polynomials are most easily computed
using the Horner scheme and the Maxima program accompanying
GeographicLib creates the necessary code.  The functions $B_j(\sigma)$
and $B'_1(\sigma)$, Eqs.~(\ref{bj}) and (\ref{Bj1}), may similarly be
evaluated for a given $\sigma$ using \citet{clenshaw55} summation,
wherein the truncated series,
\[
f(x) = \sum_{l=1}^L a_l \sin lx,
\]
is computed by determining
\begin{equation}\label{clenshaw}
b_l = \begin{cases}
0,& \text{for $l > L$},\\
a_l + 2 b_{l+1} \cos x - b_{l+2},&\text{otherwise},
\end{cases}
\end{equation}
and by evaluating the sum as
\[
f(x) = b_1 \sin x.
\]
Now $s_1$ and $\lambda_1$ can be determined using Eqs.~(\ref{i1expr}),
(\ref{i3expr}), and (\ref{ij}).  Compute $s_2 = s_1 + s_{12}$ and find
$\sigma_2$ using Eqs.~(\ref{taudef}) and (\ref{sigmaeq}).  Solve the
spherical triangle $N\!EB$ to give $\beta_2$, $\alpha_2$, and $\omega_2$
using Eqs.~(\ref{napba}), (\ref{napsb}), and (\ref{napa0b}).  Find
$\phi_2$ using Eq.~(\ref{redlat}) again.  With $\sigma_2$ and $\omega_2$
given, $\lambda_2$ can be found from Eq.~(\ref{i3expr}) which yields
$\lambda_{12} = \lambda_2 - \lambda_1$.

Although the reduced length $m_{12}$ is not needed to solve the direct
problem, it is nonetheless a useful quantity to compute.  It is found
using Eqs.~(\ref{meq}), (\ref{i2expr}), (\ref{ij}), (\ref{bj}),
(\ref{A2}), and (\ref{C2}).  In forming $I_1(\sigma) - I_2(\sigma)$ in
Eq.~(\ref{i2expr}), I avoid the loss of precision in the term
proportional to $\sigma$ by writing
\[
A_1\sigma - A_2\sigma = (A_1 - 1) \sigma - (A_2 - 1) \sigma,
\]
where $A_1 - 1$, for example, is given, from Eq.~(\ref{A1}), by
\[
A_1 - 1 = (1 - \epsilon)^{-1}\bigl(\epsilon + \tfrac1/4 \epsilon^2
       + \tfrac1/64 \epsilon^4 + \cdots\bigr).
\]
The geodesic scales $M_{12}$ and $M_{21}$ may be found similarly,
starting with Eq.~(\ref{Meq}).  This completes the solution of the
direct geodesic problem.

If several points are required along a single geodesic, many of the
intermediate expressions above may be evaluated just once; this includes
all the quantities with a subscript ``1'', and the coefficients $A_j$,
$C_{jl}$, $C'_{1l}$.  In this case the determination of the points
entails just two Clenshaw summations and a little spherical
trigonometry.  If it is only necessary to obtain points which are {\it
approximately} evenly spaced on the geodesic, replace $s_{12}$ in the
specification of the direct problem with the arc length on the auxiliary
sphere $\sigma_{12}$, in which case the conversion from $\tau$ to
$\sigma$ is avoided and only one Clenshaw summation is needed.

For speed and accuracy, I avoid unnecessarily invoking trigonometric and
inverse trigonometric functions.  Thus I usually represent an angle
$\theta$ by the pair the pair $(\sin\theta, \cos\theta)$; however, I
avoid the loss of accuracy that may ensue when computing $\cos\alpha_0$,
for example, using $\sqrt{1-\sin^2\alpha_0}$.  Instead, after finding
the sine of $\alpha_0$ using Eq.~(\ref{napa0a}), I compute its cosine
with
\[
\cos\alpha_0 = \sqrta{\cos^2\alpha + \sin^2\alpha\sin^2\beta};
\]
similarly, after finding $\sin\beta$ with Eq.~(\ref{napba}), I use
\[
\cos\beta = \sqrta{\sin^2\alpha_0 + \cos^2\alpha_0\cos^2\sigma}.
\]
In this way, the angles near the 4 cardinal directions can be
represented accurately.  In order to determine the quadrant of angles
correctly, I replace Eqs.~(\ref{napa0b}), (\ref{napsb}), and
(\ref{napab}) by
\begin{align}
\omega &= \ph(\cos\sigma + i \sin\alpha_0 \sin\sigma),\label{napa0c}\\
\alpha &= \ph(\cos\sigma \cos\alpha_0 + i \sin\alpha_0),\label{napsc}\\
\sigma &= \ph(\cos\alpha \cos\beta + i \sin\beta),\label{napac}
\end{align}
where $\theta = \ph(x + iy)$ is the phase of a complex
number \citep[\S\dlmf{1.9(i)}{1.9.i}]{dlmf10}, typically given by the
library function $\atantwo(y, x)$.  Equations~(\ref{napa0c}) and
(\ref{napsc}) become indeterminate at the poles, where $\sin\alpha_0
= \cos\sigma = 0$.  However, I ensure that $\omega$ (and hence
$\lambda$) and $\alpha$ are consistent with their interpretation for a
latitude very close to the pole (i.e., $\cos\beta$ is a small positive
quantity) and that the direction of the geodesic in three-dimensional
space is correct.  In some contexts, the solution requires explicit use
of the arc length $\sigma$ instead of its sine or cosine, for example,
in the term $A_j\sigma$ in Eq.~(\ref{ij}) and in determining $\sigma$
from Eq.~(\ref{sigmaeq}).

Geodesics which encircle the earth multiple times can be handled by
allowing $s_{12}$ and $\sigma_{12}$ to be arbitrarily large.
Furthermore, Eq.~(\ref{napa0c}) allows $\omega$ to be followed around
the circle in synchronism with $\sigma$; this permits the longitude to
be tracked continuously along the geodesic so that it increases by
$+360^\circ$ (resp.~$-360^\circ$) with each circumnavigation of the
earth in the easterly (resp.~westerly) direction.

The solution for the direct geodesic problem presented here is a
straightforward extension to higher order of Helmert's
method \citep[\S5.9]{helmert80}, which is largely based
on \citet{bessel25}.   These authors, in common with many more recent
ones, express the difference of the trigonometric terms which arise
when the sums, Eq.~(\ref{bj}), in $B_j(\sigma_2) - B_j(\sigma_1)$ are
expanded  as
\[
\sin 2l\sigma_2 - \sin 2l\sigma_1 =
2 \cos \bigl( l(\sigma_2+\sigma_1) \bigr) \sin l\sigma_{12}.
\]
This substitution is needed to prevent errors in the evaluation of the
terms on the left side of the equation causing large relative errors in
the difference when using low-precision arithmetic.  However, the use of
double-precision arithmetic renders this precaution unnecessary (see
Sect.~\ref{errs}); furthermore its use interferes with Clenshaw
summation and prevents the efficient evaluation of many points along a
geodesic.

\section{Behavior near the antipodal point}\label{antipodal}

Despite the seeming equivalence of the direct and inverse geodesic
problems when considered as exercises in ellipsoidal trigonometry, the
inverse problem is significantly more complex when transferred to the
auxiliary sphere.  The included angle for the inverse problem on the
ellipsoid is $\lambda_{12}$; however, the equivalent angle $\omega_{12}$
on the sphere cannot be immediately determined because the relation
between $\lambda$ and $\omega$, Eq.~(\ref{lameq}), depends on the
unknown angle $\alpha_0$.  The normal approach, epitomized by
\citet{rainsford55} and \citet{vincenty75a}, is to estimate $\alpha_1$
and $\alpha_2$, for example, by approximating the ellipsoid by a sphere
(i.e., $\omega_{12} = \lambda_{12}$), obtain a corrected $\omega_{12}$
from Eq.~(\ref{lameq}) and to iterate until convergence.  This procedure
breaks down if $\alpha_1$ and $\alpha_2$ depend very sensitively on
$\omega_{12}$, i.e., for nearly antipodal points.  Before tackling the
inverse problem, it is therefore useful to examine the behavior of
geodesics in this case.

\begin{figure}[tb]
\begin{center}
\includegraphics[scale=0.75,angle=0]{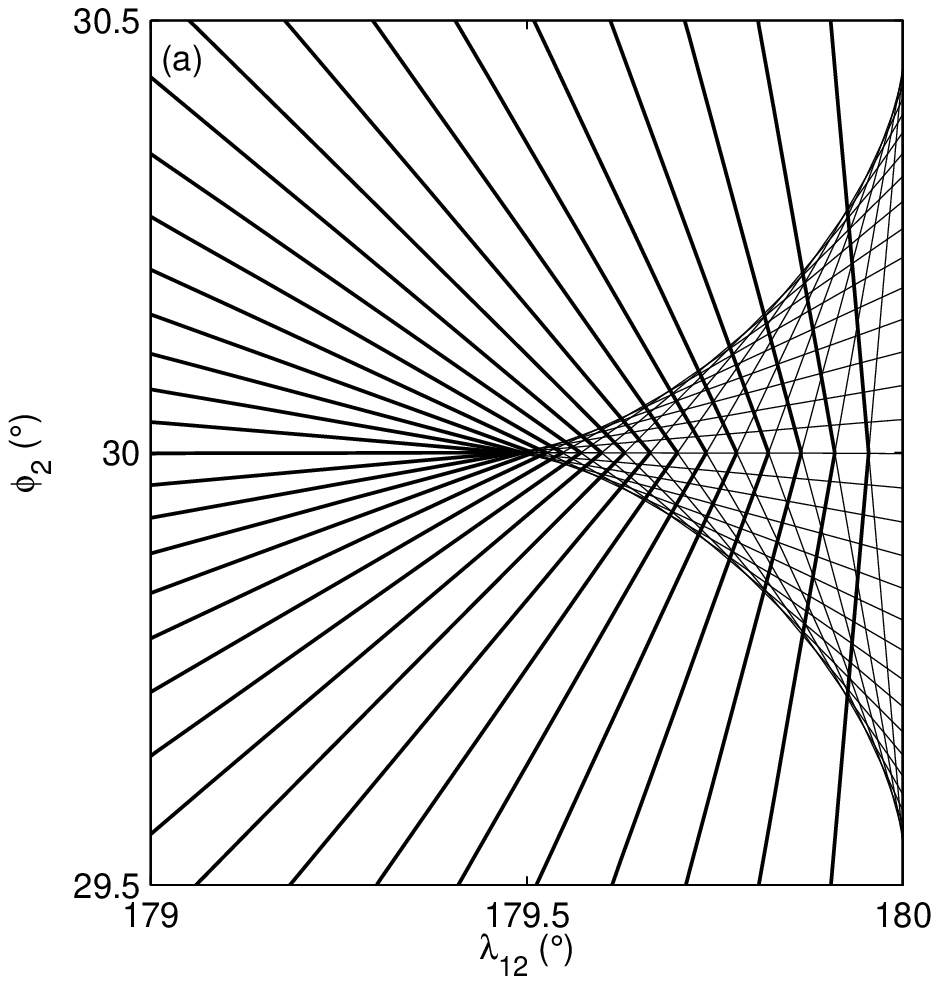}\\[10pt]
\includegraphics[scale=0.75,angle=0]{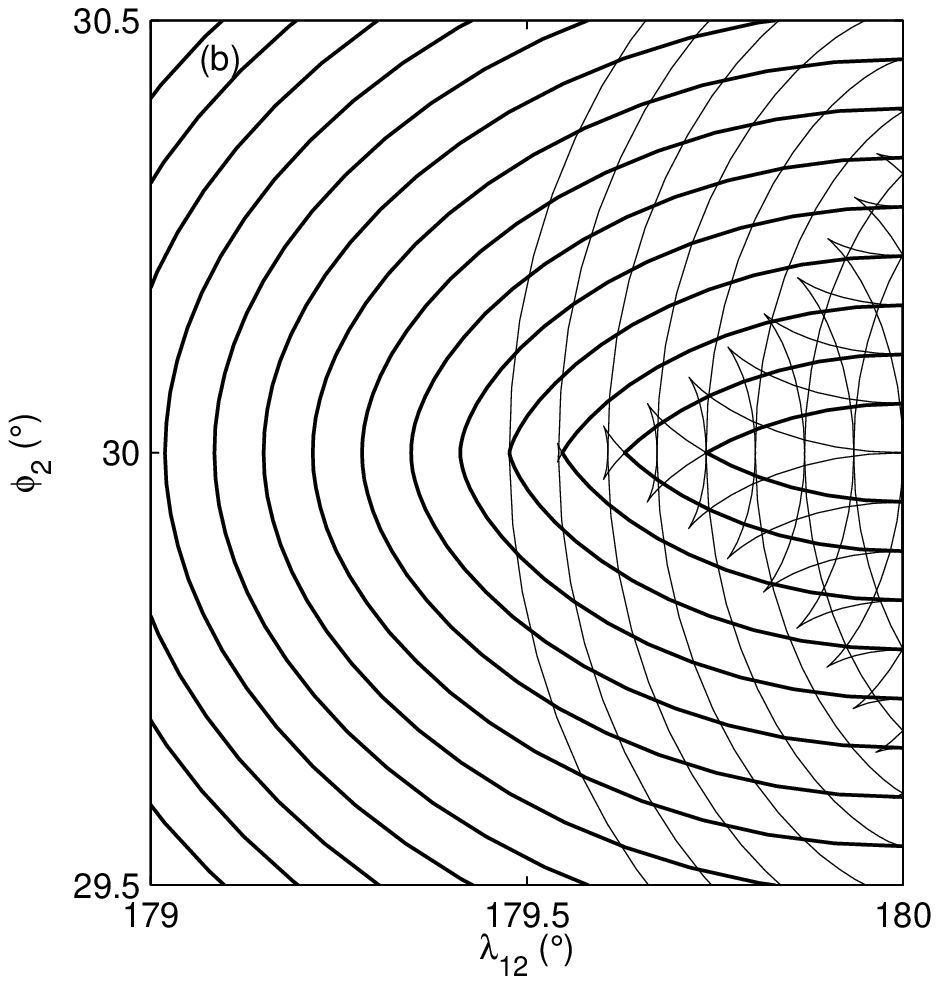}
\end{center}
\caption{\label{antipodalfig}
Geodesics in antipodal region.  (a)~The geodesics emanating from a point
$\phi_1 = -30^\circ$ are shown close in the neighborhood of the
antipodal point at $\phi_2 = 30^\circ$, $\lambda_{12} = 180^\circ$.  The
azimuths $\alpha_1$ are multiples of $5^\circ$ between $0^\circ$ and
$180^\circ$.  The geodesics are given in an equidistant cylindrical
projection with the scale set for $\phi_2 = 30^\circ$.  The heavy lines
are geodesics which satisfy the shortest distance property; the light
lines are their continuation.  The WGS84 ellipsoid is used. (b)~The
geodesic circles on the same scale.  The heavy (resp.~light) lines are
for geodesics which have (resp.~do not have) are property of being
shortest paths.  The cusps on the circles lie on the geodesic envelope
in (a).}
\end{figure}%
Consider two geodesics starting at $A$ with azimuths $\alpha_1$ and
$\alpha_1 + \d\alpha_1$.  On a closed surface, they will intersect at
some distance from $A$.  The first such intersection is the {\it
conjugate point} for the geodesic and it satisfies $m_{12} = 0$ (for
$s_{12} > 0$).  \citet{jacobi91} showed that the geodesics no longer
retain the property of being the shortest path beyond the conjugate
point \citep[\S623]{darboux94}.  For an ellipsoid with small flattening,
the conjugate point is given by
\begin{align*}
\phi_2 &= -\phi_1 - f\pi \cos^2\phi_1 \cos^3\alpha_1 + O(f^2),\\
\lambda_2 &= \lambda_1 + \pi - f\pi \cos\phi_1 \sin^3\alpha_1 + O(f^2).
\end{align*}
The envelope of the geodesics leaving $A$ is given by the locus of the
conjugate points and, in the case of an ellipsoid, this yields a
four-point star called an
{\it astroid} \citep[Eqs.~(16)--(17)]{jacobi91}, whose equation in
cartesian form is, after suitable scaling (see below),
\begin{equation}
x^{2/3} + y^{2/3} = 1,\label{astroidc}
\end{equation}
which is depicted in Fig.~\ref{antipodalfig}a; see also
\citet[Fig.~11]{jacobi91} and \citet[\S7.2]{helmert80}.  The angular
extent of the astroid is, to lowest order, $2f\pi\cos^2\phi_1$.
Figure~\ref{antipodalfig}b shows the behavior of the geodesic circles
in this region.

Although geodesics are no longer the shortest paths beyond the conjugate
point (where they intersect a nearby geodesic), in general, they loose
this property earlier when they first intersect any geodesic of the same
length emanating from the same starting point.  In the case of the
ellipsoid, it is easy to establish earlier intersection points.
Consider two geodesics leaving $A$ with azimuths $\alpha_1$ and $\pi
- \alpha_1$.  These intersect at $\abs{\omega_{12}} = \pi$ and $\phi_2 =
-\phi_1$ and, for oblate ellipsoids this intersection is earlier than
the conjugate points.  For prolate ellipsoids the corresponding pair of
azimuths are $\pm\alpha_1$ and these intersect at $\abs{\lambda_{12}}
= \pi$, also prior to the conjugate points.  Thus, an ellipsoidal
geodesic is the shortest path if, and only if,
\[
\max(\abs{\omega_{12}}, \abs{\lambda_{12}}) \le \pi.
\]
The only conjugate points lying on shortest paths are for the geodesics
with $\alpha_1 = \pm \frac12 \pi$ for oblate ellipsoids and $\alpha_1
= 0$ or $\pi$ for prolate ellipsoids.  Solving the inverse problem with
end points close to such conjugate pairs presents a challenge because
tiny changes in end points lead to large changes in the geodesic.

\begin{figure}[tb]
\begin{center}
\includegraphics[scale=0.75,angle=0]{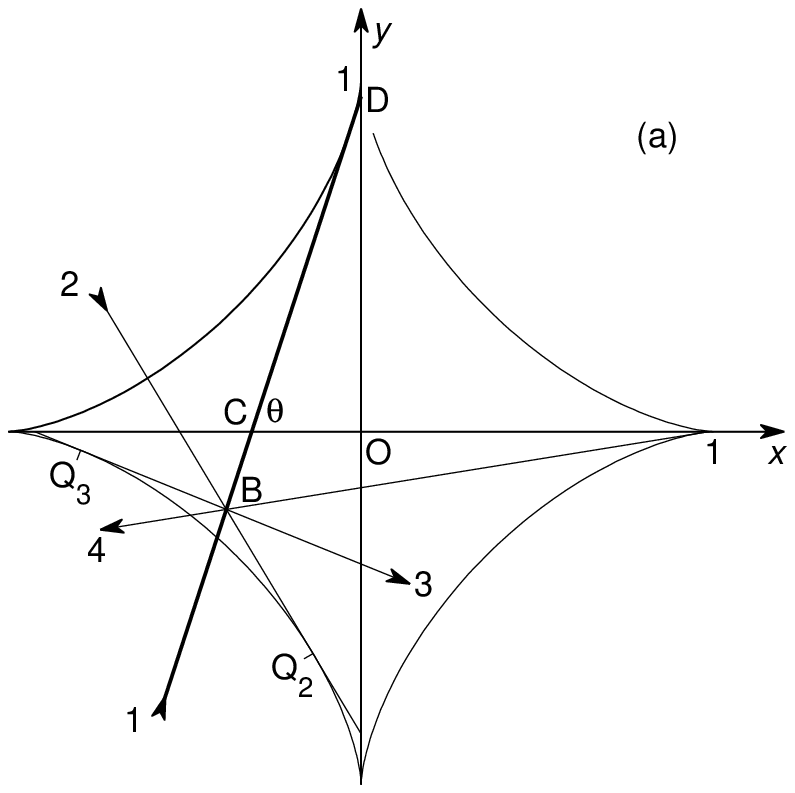}\\[0pt]
\hspace{10mm}\includegraphics[scale=0.75,angle=0]{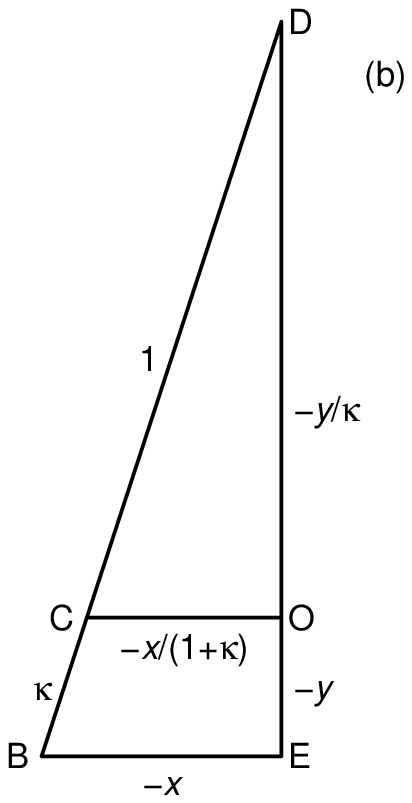}
\end{center}
\caption{\label{astroidfig}
The inverse geodesic problem for nearly antipodal points.  (a)~The heavy
line (labeled 1) shows the shortest geodesic from $A$ to $B$ continued
until it intersects the antipodal meridian at $D$.  The light lines
(2--4) show 3 other approximately hemispherical geodesics.  The
geodesics are all tangent (at their points of conjugacy) to the astroid
in this figure.  The points $Q_2$ and $Q_3$ are the points of conjugacy
for the geodesics 2 and 3.  (b)~The solution of the astroid equations by
similar triangles.}
\end{figure}

The inverse problem may be solved approximately in the case of nearly
antipodal points by considering the point $B$ together with envelope of
geodesics for $A$ (centered at the antipodes of $A$); see
Fig.~\ref{astroidfig}.  The coordinates near the antipodes can be
rescaled as
\[
x = \frac{\sin(\lambda - \lambda_{1} - \pi)}{\Delta\lambda},\quad
y = \frac{\sin(\beta + \beta_1)}{\Delta\beta},
\]
where
\[
\Delta\lambda = f\pi A_3 \cos\beta_1,\quad
\Delta\beta = \cos\beta_1 \Delta\lambda,
\]
and $A_3$ is evaluated with $\alpha_0 = \frac12 \pi -
\abs{\beta_1}$.  In the $(x,y)$ coordinate system, the conjugate point
for the geodesic leaving $A$ with $\alpha_1 = \pm \frac12 \pi$ is at
$(\mp 1, 0)$ and the scale in the $y$ direction compared to $x$ is $1 +
O(f)$.  The geodesic through $B$ is tangent to the astroid; plane
geometry can therefore be used to find its direction, using
\begin{equation}\label{anti-line}
\frac x{\cos\theta} - \frac y{\sin\theta} + 1 = 0,
\end{equation}
where $\theta = \frac12 \pi - \alpha_2 \approx \alpha_1 - \frac12 \pi$
is the angle of the geodesic measured anticlockwise from the $x$ axis.
The constant term on the left hand side of Eq.~(\ref{anti-line}), $1$,
reflects the property of tangents to the astroid, that the length $CD$
is constant.  (The astroid is the envelope generated as a ladder slides
down a wall.)  Note that geodesics are directed lines; thus a distinct
line is given by $\theta \rightarrow \theta + \pi$.  The point of
tangency of Eq.~(\ref{anti-line}) with the astroid is
\[
x_0 = - \cos^3\theta, \quad y_0 = \sin^3\theta ,
\]
which are the parametric equations for the astroid,
Eq.~(\ref{astroidc}); the line and the astroid are shown in
Fig.~\ref{astroidfig}a.  The goal now is, given $x$ and $y$ (the
position of $B$), to solve Eq.~(\ref{anti-line}) for $\theta$.  I follow
the method given by \citet{vermeille02} for converting from geocentric
to geodetic coordinates.  In Fig.~\ref{astroidfig}b, $COD$ and $BED$
are similar triangles; if the (signed) length $BC$ is $\kappa$, then an
equation for $\kappa$ can be found by applying Pythagoras' theorem to
$COD$,
\[
\frac{x^2}{(1+\kappa)^2} + \frac{y^2}{\kappa^2} = 1,
\]
which can be expanded to give a quartic equation in $\kappa$,
\begin{equation}\label{kapeq}
\kappa^4 + 2 \kappa^3 + (1 - x^2 - y^2) \kappa^2 - 2 y^2 \kappa - y^2 = 0.
\end{equation}
Once $\kappa$ is known, $\theta$ can be determined from the triangle
$COD$ in Fig.~\ref{astroidfig}b,
\begin{equation}\label{thetaeq}
\theta = \ph\bigl(-x/(1+\kappa) - i y/\kappa\bigr).
\end{equation}

The point $C$ in Fig.~\ref{astroidfig} corresponds to a spherical
longitude difference of $\pi$.  Thus the spherical longitude difference
for $B$, $\omega_{12}$, can be estimated as
\begin{equation}\label{omega-anti}
\omega_{12} \approx \pi + \frac{\kappa x}{1+\kappa}\Delta\lambda;
\end{equation}
compare with \citet[Eq.~(7.3.11)]{helmert80}.

In Appendix~\ref{geocent}, I summarize the closed form solution of
Eq.~(\ref{kapeq}) as given by \citet{vermeille02}.  I have modified this
solution so that it is applicable for all $x$ and $y$ and more stable
numerically.

Equation~(\ref{kapeq}) has 2 (resp.~4) real roots if $B$ lies outside
(resp.~inside) the astroid.  The methods given in Appendix~\ref{geocent}
(with $e$ set to unity) can be used to determine all these roots,
$\kappa$, and Eq.~(\ref{thetaeq}) then gives the corresponding angles of
the geodesics at $B$.  All the geodesics obtained in this way are
approximately hemispherical and that obtained using the largest value of
$\kappa$ is the shortest path.  If $B$ lies on the axes within the
astroid, then the limiting solutions Eqs.~(\ref{xsmall}) or
(\ref{ysmall}) should be used to avoid an indeterminate expression.  For
example, if $y=0$, substitute the largest $\kappa$ from
Eq.~(\ref{ysmall}) into Eq.~(\ref{thetaeq}) to give $\theta
= \ph\bigl(-x + i\sqrt{\max(0,1-x^2)}\bigr)$.

\begin{table}[tb]
\caption{\label{fourgeod}
The four approximately hemispherical solutions of the inverse geodesic
problem on the WGS84 ellipsoid for $\phi_1 = -\exact{30}^\circ$, $\phi_2
= \exact{29.9}^\circ$, $\lambda_{12} = \exact{179.8}^\circ$, ranked by
length $s_{12}$.}
\begin{center}
\begin{tabular}{@{\extracolsep{0.5\xx}}>{$}c<{$} >{$}r<{$} >{$}r<{$}
>{$}r<{$} >{$}r<{$} >{$}r<{$}}
\hline\hline\noalign{\smallskip}
\text{No.} &
\multicolumn{1}{c}{$\alpha_1\,(\mathrm{^\circ})$} &
\multicolumn{1}{c}{$\alpha_2\,(\mathrm{^\circ})$} &
\multicolumn{1}{c}{$s_{12}\,(\mathrm{m})$} &
\multicolumn{1}{c}{$\sigma_{12}\,(\mathrm{^\circ})$} &
\multicolumn{1}{c}{$m_{12}\,(\mathrm{m})$}
\\\noalign{\smallskip}\hline\noalign{\smallskip}
1& 161.891&  18.091& 19\,989\,833& 179.895&  57\,277\\
2&  30.945& 149.089& 20\,010\,185& 180.116&  24\,241\\
3&  68.152& 111.990& 20\,011\,887& 180.267& -22\,649\\
4& -81.076& -99.282& 20\,049\,364& 180.631& -68\,796\\
\noalign{\smallskip}\hline\hline
\end{tabular}
\end{center}
\end{table}%
Figure~\ref{astroidfig} shows a case where $B$ is within the astroid
resulting in 4 hemispherical geodesics which are listed in
Table~\ref{fourgeod} ranked by their length.  The values given here have
been accurately computed for the case of the WGS84 ellipsoid using the
method described in Sect.~\ref{inverse}.  The second and third geodesics
are eastward (the same sense as the shortest geodesic), while the last
is westward.  As $B$ crosses the boundary of the astroid the second and
third geodesics approach one another and disappear (leaving the first
and fourth geodesics).  Figure~\ref{astroidfig}a also illustrates that
the envelope is an evolute of the geodesics; in particular,
\citet[\S94]{eisenhart09} shows that the length of geodesic 2 from $A$ to
its conjugate point $Q_2$ exceeds the length of geodesic 3 from $A$ to
$Q_3$ by the distance along the envelope from $Q_3$ to $Q_2$; see
also \citet[\S9.2]{helmert80}.

\citet{schmidt00} uses Helmert's method for estimating the azimuth.
\citet{bowring96} proposed a solution of the astroid problem where he
approximates the 4 arcs of the astroid by quarter circles.
\citet[Table~1.6, p.~54]{rapp93} gives a similar set of hemispherical
geodesics to those given in Table~\ref{fourgeod}.  However, this table
contains two misprints: $\phi_2$ should be $-40^\circ01'05.759\,32''$ and
not $-40^\circ00'05.759\,32''$; for method 3, $\alpha_{12}$ should be
$86^\circ20'38.153\,06''$ and not $87^\circ20'38.153\,06''$.

\section{Inverse problem}\label{inverse}

Recall that the inverse geodesic problem is to determine $s_{12}$,
$\alpha_1$ and $\alpha_2$ given $\phi_1$, $\phi_2$, and $\lambda_{12}$.
I begin by reviewing the solution of the inverse problem assuming that
$\omega_{12}$ is given (which is equivalent to seeking the solution of
the inverse problem for a sphere).  Write the cartesian coordinates for
the two end points on the auxiliary sphere (with unit radius) as $\v A =
[\cos\beta_1, 0, \sin\beta_1]$ and $\v B = [\cos\beta_2\cos\omega_{12},
\allowbreak \cos\beta_2\sin\omega_{12}, \allowbreak \sin\beta_2]$.
A point on the geodesic a small spherical arc length $\d\sigma$ from $A$
is at position $\v A + \d\v A$ where $\d\v A = [-\sin\beta_1\cos\alpha_1,
\allowbreak \sin\alpha_1, \allowbreak \cos\beta_1\cos\alpha_1]\,\d\sigma$.
The azimuth $\alpha_1$ can be found by demanding that $\v A$, $\v B$,
and $\d\v A$ be coplanar or that $\v A \times \v B$ and $\v A \times \d\v
A$ be parallel (where $\times$ here denotes the vector cross product),
and similarly for $\alpha_2$.  Likewise, the spherical arc length
$\sigma_{12}$ is given by $\ph(\v A \cdot \v B + i \abs{\v A \times \v
B})$, where $\cdot$ denotes the vector dot product.  Evaluating these
expressions gives
\begin{align}
z_1 &=\cos\beta_1\sin\beta_2 - \sin\beta_1\cos\beta_2 \cos\omega_{12}
\notag\\&\qquad
+i\cos\beta_2\sin\omega_{12},\displaybreak[0]\notag\\
z_2 &= -\sin\beta_1\cos\beta_2 + \cos\beta_1\sin\beta_2 \cos\omega_{12}
\notag\\&\qquad
+i\cos\beta_1\sin\omega_{12},\displaybreak[0]\notag\\
\alpha_1 &= \ph z_1, \label{alpha1-spher}\displaybreak[0]\\
\alpha_2 &= \ph z_2, \label{alpha2-spher}\displaybreak[0]\\
\sigma_{12} &=
\ph(\sin\beta_1\sin\beta_2 + \cos\beta_1\cos\beta_2 \cos\omega_{12}
+ i \abs{z_1}). \label{sigma-spher}
\end{align}
In order to maintain accuracy when $A$ and $B$ are nearly coincident or
nearly antipodal, I evaluate the real part of $z_1$ as
\[
\sin(\beta_2\mp\beta_1)
\pm \frac{\sin^2\omega_{12}\sin\beta_1\cos\beta_2}
{1\pm\cos\omega_{12}},
\]
where the upper and lower signs are for $\cos\omega_{12}\gtrless 0$.
The evaluation of $z_2$ is handled in the same way.  This completes the
solution of the inverse problem for a sphere.

In the ellipsoidal case, the inverse problem is just a two-dimensional
root finding problem.  Solve the direct geodesic starting at $A$ and
adjust $\alpha_1$ and $s_{12}$ (subject to the shortest distance
constraint), so that the terminal point of the geodesic matches $B$.  In
order to convert this process into an algorithm, a rule needs to be
given for adjusting $\alpha_1$ and $\sigma_{12}$ so that the process
converges to the true solution.

The first step in finding such a rule is to convert the two-dimensional
problem into a one-dimensional root-finding one.  I begin by putting the
points in a canonical configuration,
\begin{equation}\label{canon}
\phi_1 \le 0,\quad \phi_1 \le \phi_2 \le -\phi_1,\quad
 0\le \lambda_{12} \le \pi.
\end{equation}
This may be accomplished swapping the end points and the signs of the
coordinates if necessary, and the solution may similarly be transformed
to apply to the original points.  Referring to Fig.~\ref{figall}, note
that, with these orderings of the coordinates, all geodesics with
$\alpha_1 \in [0,\pi]$ intersect latitude $\phi_2$ with
$\lambda_{12} \in [0,\pi]$.  Furthermore, the search for solutions can
be restricted to $\alpha_2 \in [0, \frac12 \pi]$, i.e., when the
geodesic {\it first} intersects latitude $\phi_2$.  (For $\phi_2 =
-\phi_1$, there is a second shortest path with $\alpha_2 \in
[\frac12 \pi, \pi]$ if $\lambda_{12}$ is nearly equal to $\pi$.  But
this geodesic is easily derived from the first.)

Meridional geodesics are treated as a special case.  These include the
cases $\lambda_{12} = 0$ or $\pi$ and $\beta_1 = -\frac12 \pi$ with any
$\lambda_{12}$.  This also includes the case where $A$ and $B$ are
coincident.  In these cases, set $\alpha_1 = \lambda_{12}$ and $\alpha_2
= 0$.  (This value of $\alpha_1$ is consistent with the prescription for
azimuths near a pole given at the end of Sect.~\ref{direct}.)

\begin{figure}[tb]
\begin{center}
\includegraphics[scale=0.75,angle=0]{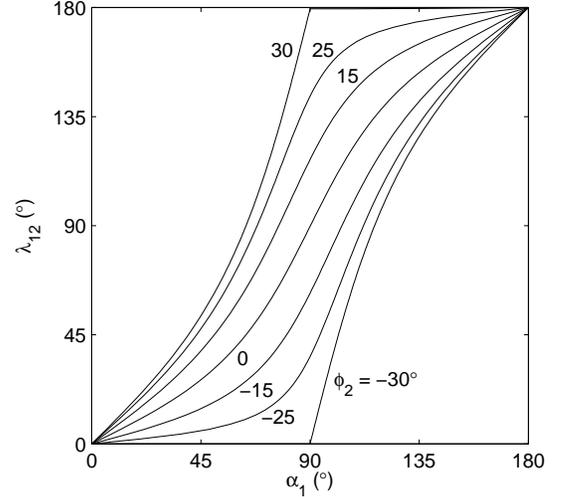}
\end{center}
\caption{\label{figalp1}
The variation of $\lambda_{12}$ as a function of $\alpha_1$.  The
latitudes are $\phi_1 = -30^\circ$ and $\phi_2 = -30^\circ$,
$-25^\circ$, $-15^\circ$, $0^\circ$, $15^\circ$, $25^\circ$, $30^\circ$.
For $\phi_1 < 0$ and $\phi_1 < \phi_2 < -\phi_1$, the curves are
strictly increasing, while for $\phi_1 < 0$ and $\phi_2 = \pm\phi_1$,
the curves are non-decreasing with discontinuities in the slopes at
$\alpha_1 = 90^\circ$ (see Fig.~\ref{figalp2}).  The WGS84 ellipsoid is
used.}
\end{figure}%
\begin{figure}[tb]
\begin{center}
\includegraphics[scale=0.75,angle=0]{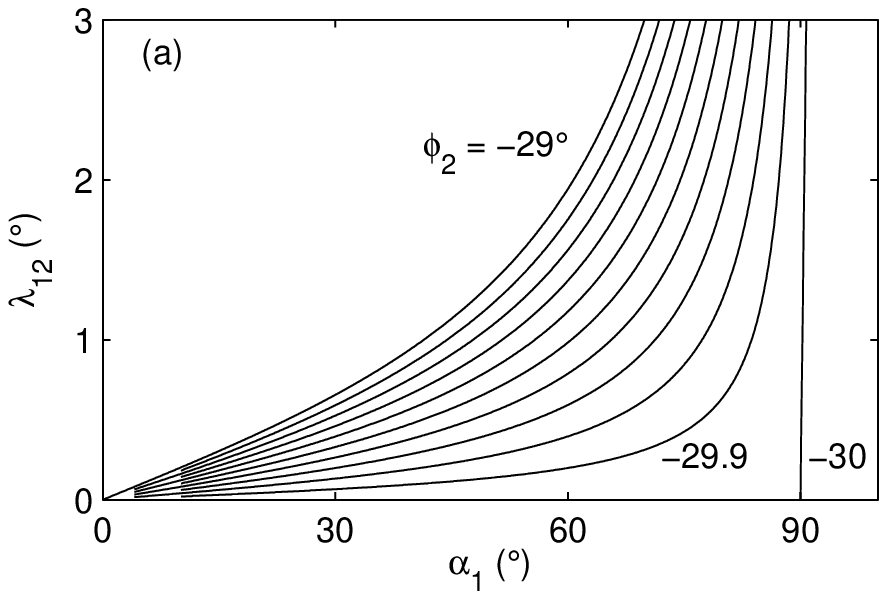}\\[10pt]
\includegraphics[scale=0.75,angle=0]{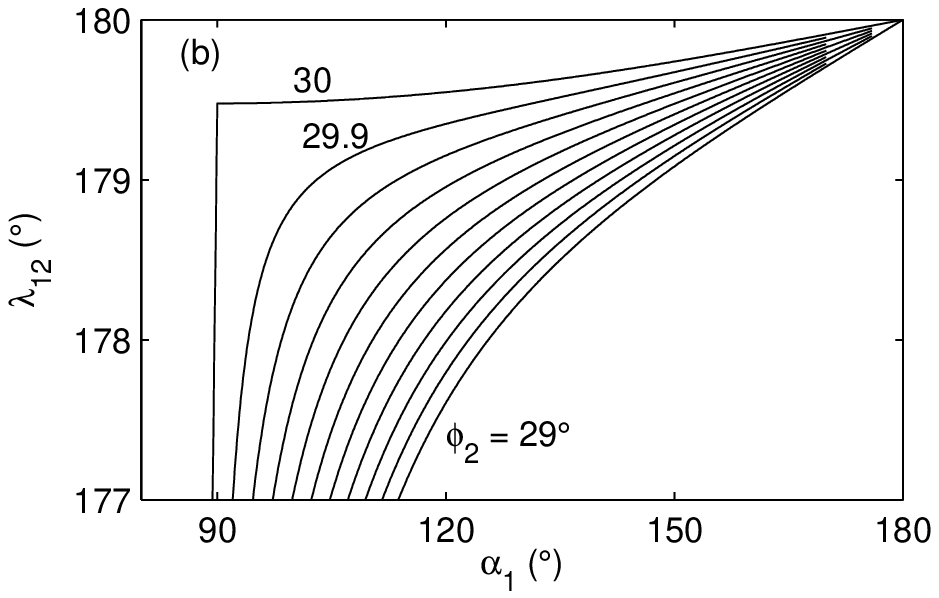}
\end{center}
\caption{\label{figalp2}
Details of $\lambda_{12}$ as a function of $\alpha_1$ near the
discontinuities in the slopes.  This shows blow-ups of the two
``corner'' regions in Fig.~\ref{figalp1} (for $\phi_1 = -30^\circ$).
(a)~The short line limit, where $\phi_2 \approx \phi_1$ and
$\lambda_{12} \approx 0$; here, $\phi_2$ is in $[-30^\circ, -29^\circ]$
at intervals of $0.1^\circ$.  At $\phi_2 = \phi_1 = -30^\circ$ and
$\alpha_1 = 90^\circ$, the slope changes from zero to a finite value.
(b)~The nearly antipodal limit, where $\phi_2 \approx -\phi_1$ and
$\lambda_{12} \approx 180^\circ$; here, $\phi_2$ is in $[29^\circ,
30^\circ]$ at intervals of $0.1^\circ$.  At $\phi_2 = -\phi_1 =
30^\circ$ and $\alpha_1 = 90^\circ$, the slope changes from a finite
value to zero.}
\end{figure}%
Define now a variant of the direct geodesic problem: given $\phi_1$,
$\phi_2$, subject to Eq.~(\ref{canon}), and $\alpha_1$, find
$\lambda_{12}$.  Proceed as in the direct problem up to the solution of
the triangle $N\!EA$.  Find $\beta_2$ from Eq.~(\ref{redlat}) and solve
the triangle $N\!EB$ for $\alpha_2 \in [0, \frac12 \pi]$, $\sigma_2$,
$\omega_2$ from Eqs.~(\ref{napa0a}), (\ref{napab}), and (\ref{napa0b}).
Finally, determine $\lambda_{12}$ as in the solution to the direct
problem.  In determining $\alpha_2$ from Eq.~(\ref{napa0a}), I use in
addition
\begin{equation}\label{alp2-eq}
\cos\alpha_2 = \frac{+\sqrt{\cos^2\alpha_1 \cos^2\beta_1 +
 (\cos^2\beta_2 - \cos^2\beta_1)}}{\cos\beta_2}, \end{equation} where
 the parenthetical term under the radical is computed by $(\cos\beta_2
- \cos\beta_1) (\cos\beta_2 + \cos\beta_1)$ if $\beta_1 <
-\frac14 \pi$ and by $(\sin\beta_1 - \sin\beta_2) (\sin\beta_1
+ \sin\beta_2)$ otherwise.  It remains to determine the value of
$\alpha_1$ that results in the given value of $\lambda_{12}$.

I show the behavior of $\lambda_{12}$ as a function of $\alpha_1$ in
Figs.~\ref{figalp1}--\ref{figalp2}.  For an oblate ellipsoid and
$\abs{\beta_2} < -\beta_1$, $\lambda_{12}$ is a strictly increasing
function of $\alpha_1$.  For $\beta_2 = \beta_1$, $\lambda_{12}$
vanishes for $0 \le \alpha_1 < \frac12 \pi$; for $\beta_2 =
- \beta_1$, $\d\lambda_{12}/\d\alpha_1$ vanishes for $\alpha_1
= \frac12 \pi +$.  Therefore if $\beta_1 = \beta_2 = 0$,
$\lambda_{12}$ is discontinuous at $\alpha_1 = \frac12 \pi$ jumping
from $0$ to $(1-f)\pi$.  This is the case of equatorial end points---if
$\lambda_{12} \le (1-f)\pi$, the geodesic lies along the equator, with
$\alpha_1 = \alpha_2 = \frac12 \pi$.

Thus with the ordering given by Eq.~(\ref{canon}), simple root finding
methods, such as binary search or {\it regula falsi}, will allow
$\alpha_1$ to be determined for a given $\lambda_{12}$.  Because such
methods converge slowly, I instead solve for $\alpha_1$ using Newton's
method.

\begin{figure}[tb]
\begin{center}
\includegraphics[scale=0.75,angle=0]{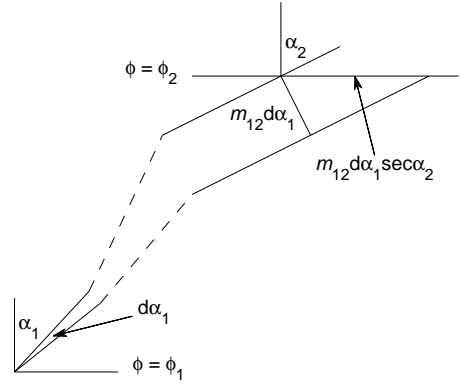}
\end{center}
\caption{\label{derivfig}
Finding $\d\lambda_{12}/\d\alpha_1$ with $\phi_1$ and $\phi_2$ held
fixed.}
\end{figure}%
First, I compute the necessary derivative.  Consider a trial geodesic
with initial azimuth $\alpha_1$.  If the azimuth is increased to
$\alpha_1 + \d\alpha_1$ with the length held fixed, then the other end of
the geodesic moves by $m_{12}\,\d\alpha_1$ in a direction $\frac12 \pi
+ \alpha_2$.  If the geodesic is extended to intersect the parallel
$\phi_2$ once more, the point of intersection moves by
$m_{12}\,\d\alpha_1/\cos\alpha_2$; see Fig.~\ref{derivfig}.  The radius
of this parallel is $a\cos\beta_2$, thus the rate of change of the
longitude difference is
\begin{equation}\label{lambda-deriv}
\left.
\frac{\d\lambda_{12}}{\d\alpha_1}
\right|_{\phi_1, \phi_2}
= \frac{m_{12}}a
\frac1{\cos\alpha_2 \cos\beta_2},
\end{equation}
where the subscripts on the derivative indicate which quantities are
held fixed in taking the derivative.  The denominator can vanish if
$\beta_2 = \abs{\beta_1}$ and $\alpha_2 = \frac12 \pi$; in this case,
use
\begin{equation}\label{lambda-deriv-0}
\left.
\frac{\d\lambda_{12}}{\d\alpha_1}
\right|_{\phi_1, \phi_2}
= -\frac{\sqrt{1-e^2 \cos^2\beta_1}}{\sin\beta_1}
\bigl(1 \mp \sign(\cos\alpha_1) \bigr),
\end{equation}
for $\beta_2 = \pm \beta_1$.  For Newton's method, pick the positive
derivative, i.e., take $1 \mp \sign(\cos\alpha_1) = 2$, which
corresponds to $\alpha_1 = 90^\circ\pm$ for $\phi_2 = \pm\phi_1$ in
Fig.~\ref{figalp1}.

\begin{figure}[tb]
\begin{center}
\includegraphics[scale=0.75,angle=0]{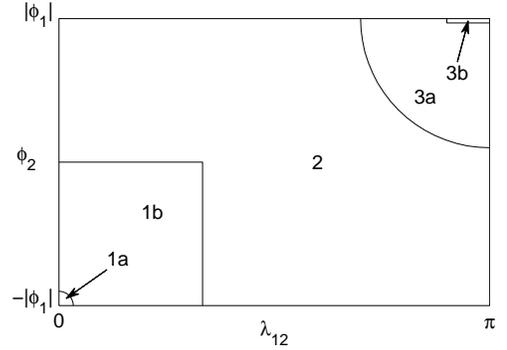}
\end{center}
\caption{\label{inversefig}
Schematic showing the 5 regions for the inverse problem.  Region 1 is
given by $\lambda_{12} < \frac16 \pi$ and $\beta_2 - \beta_1
< \frac16 \pi$.  Region 1a is given by $\sigma_{12} < \sqrt\delta /
\max\bigl(0.1, \abs{e^2}\bigr)$.  Region 3 is given by $\sigma_{12}
> \pi \bigl(1 - 3 \abs f A_3 \cos^2\beta_1\bigr)$.  Region 3b is given
by $y > -100 \delta$ and $x > -1 - 1000 \sqrt\delta$.}
\end{figure}%
Newton's method requires a sufficiently accurate starting guess for
$\alpha_1$ to converge.  To determine this, I first estimate
$\omega_{12}$, the longitude difference on the auxiliary sphere, and
then find $\alpha_1$ using Eq.~(\ref{alpha1-spher}).  To obtain this
estimate for $\omega_{12}$, I distinguish three regions; see
Fig.~\ref{inversefig} (the caption gives the precise boundaries of the
regions).  (1)~Short lines: If $\lambda_{12}$ and $\beta_2 - \beta_1$
are reasonably small, then use $\omega_{12} \approx \lambda_{12}/w_1$,
where $w_1$ is given by Eq.~(\ref{weq}).  (2)~Intermediate lines: Assume
$\omega_{12} = \lambda_{12}$, provided that the resulting $\sigma_{12}$
is sufficiently less than $\pi$.  (3)~Long lines: Analyze the problem
using the methods of Sect.~\ref{antipodal}, evaluate $x < 0$ and $y\le
0$, and use Eq.~(\ref{omega-anti}) as an estimate of $\omega_{12}$.
However, if $y$ is very small and $-1\le x \le 0$, then $\omega_{12}$ is
nearly equal to $\pi$ and Eq.~(\ref{alpha1-spher}) becomes
indeterminate; in this case, estimate $\alpha_1$ directly using
$\alpha_1 \approx \theta +\frac12 \pi$, with $\theta$ given by
Eq.~(\ref{thetaeq}) (region 3b in Fig.~\ref{inversefig}).  This rule is
also applied for $x$ slightly less than $-1$, to ensure that Newton's
method doesn't get tripped by the discontinuity in the slope of
$\lambda_{12}(\alpha_1)$ in Fig.~\ref{figalp2}b.

This provides suitable starting values for $\alpha_1$ for use in
Newton's method.  Carrying out Newton's method can be avoided in case 1
above if $\sigma_{12}$, Eq.~(\ref{sigma-spher}), is sufficiently small
(case 1a in Fig.~\ref{inversefig}), in which case the full solution is
given by Eqs.~(\ref{alpha1-spher})--(\ref{sigma-spher}) with $s_{12} = a
w_1 \sigma_{12}$.  This also avoids the problem of maintaining accuracy
when solving for $\lambda_{12}$ given $\phi_2\approx\phi_1$ and
$\alpha_1\approx \frac12 \pi$.  Details of the convergence of Newton's
method are given in Sect.~\ref{errs}.  The boundaries of regions 1a and
3b in Fig.~\ref{inversefig} depend on the precision of the
floating-point number system.  This is characterized by $\delta =
1/2^{p-1}$ where $p$ is the number of bits of precision in the number
system and $1 + \delta$ is the smallest representable number greater
than $1$.  Typically $p=53$ and $\delta = 2.2\times 10^{-16}$ for double
precision.

Once a converged value of $\alpha_1$ has been found, converged values of
$\alpha_2$, $\sigma_1$, and $\sigma_2$ are also known (during the course
of the final Newton iteration), and $s_{12}$ can be found using
Eq.~(\ref{i1expr}).  The quantities $m_{12}$, $M_{12}$, and $M_{21}$ can
also be computed as in the solution of the direct problem.  This
completes the solution of the inverse problem for an ellipsoid.

In the following cases, there are multiple solutions to the inverse
problem and I indicate how to find them all given one solution.  (1)~If
$\phi_1 + \phi_1 = 0$ (and neither point is at the pole) and if
$\alpha_1 \ne \alpha_2$, a second geodesic is obtained by setting
$\alpha_1 = \alpha_2$ and $\alpha_2 = \alpha_1$.  (This occurs when
$\lambda_{12} \approx \pm 180^\circ$ for oblate ellipsoids.)  (2)~If
$\lambda_{12} = \pm 180^\circ$ (and neither point is at a pole) and if
the geodesic in not meridional, a second geodesic is obtained by setting
$\alpha_1 = -\alpha_1$ and $\alpha_2 = -\alpha_2$. (This occurs when the
$\phi_1 + \phi_2 \approx 0$ for prolate ellipsoids.)  (3)~If $A$ and $B$
are at opposite poles, there are infinitely many geodesics which can be
generated by setting $\alpha_1 = \alpha_1 + \gamma$ and $\alpha_2
= \alpha_2 - \gamma$ for arbitrary $\gamma$.  (For spheres, this
prescription applies when $A$ and $B$ are antipodal.)  (4)~If $s_{12} =
0$ (coincident points), there are infinitely many geodesics which can be
generated by setting $\alpha_1 = \alpha_1 + \gamma$ and $\alpha_2
= \alpha_2 + \gamma$ for arbitrary $\gamma$.

The methods given here can be adapted to return geodesics which are not
shortest paths provided a suitable starting point for Newton's method is
given.  For example, geodesics 2--4 in Table~\ref{fourgeod} can be found
by using the negative square root in the equation for $\cos\alpha_2$,
Eq.~(\ref{alp2-eq}); and for geodesic 4, solve the problem with $2\pi
- \lambda_{12} = \exact{180.2}^\circ$ as the longitude difference.  In
these cases, the starting points are given by the multiple solutions of
the astroid equation, Eq.~(\ref{kapeq}).  Geodesics that wrap around the
globe multiple times can be handled similarly.

Although the published inverse method of \citet{vincenty75a} fails to
converge for nearly antipodal points, he did give a modification of his
method that deals with this case \citep{vincenty75b}.  The method
requires one minor modification: following his Eq.~(10), insert
$\cos\sigma = - \sqrt{1-\sin^2\sigma}$.  The principal drawback of his
method (apart from the limited accuracy of his series) is its very slow
convergence for nearly conjugate points---in some cases, many thousands
of iterations are required.  In contrast, by using Newton's method, the
method described here converges in only a few
iterations.  \citet{sodano58}, starting with the same formulation as
given here \citep{bessel25,helmert80} derives an approximate
non-iterative solution for the inverse problem; however, this may fail
for antipodal points \citep[\S1.3]{rapp93}.  Sodano's justification for
his method is illuminating: a non-iterative method is better suited to
the mechanical and electronic computers of his day.  (See also my
comment about Vincenty's use of programmable calculators at the end of
Sect.~\ref{intser}.)  The situation now is, of course, completely
different: the series of Bessel and Helmert are readily implemented on
modern computers and iterative methods are frequently key to efficient
and accurate computational algorithms.

\section{Errors} \label{errs}

Floating-point implementations of the algorithms described in
Secs.~\ref{direct} and \ref{inverse} are included in
GeographicLib \citep{geographiclib17}.  These suffer from two sources of
error: truncation errors because the series in Sect.~\ref{intser}, when
truncated at order $L$, differ from the exact integrals; and round-off
errors due to evaluating the series and solving the resulting problem in
spherical trigonometry using inexact (floating-point) arithmetic.  In
order to assess both types of error, it is useful to be able to compute
geodesics with arbitrary accuracy.  For this purpose, I used Maxima's
Taylor package to expand the series to 30th order and its ``bigfloat''
package to solve the direct problem with 100 decimal digits.  The
results obtained in this way are accurate to at least 50 decimal digits
and may be regarded as ``exact''.

\begin{table}[tb]
\caption{\label{truncerr}
Truncation errors for the main geodesic problems.  $\Delta_{\mathrm d}$
and $\Delta_{\mathrm i}$ are approximate upper bounds on the truncation
errors for the direct and inverse problems.  The parameters of the WGS84
and the SRMmax ellipsoids are used.  The SRMmax ellipsoid, $a
= \exact{6400}\,\mathrm{km}$, $f = 1/\exact{150}$, is an ellipsoid with
an exaggerated flattening introduced by the National
Geospatial-Intelligence Agency for the purposes of algorithm testing.}
\begin{center}
\def\num[#1,#2]{#1\times10^{#2}}%
\begin{tabular}{@{\extracolsep{0.75\xx}}>{$}r<{$}
>{$}l<{$} >{$}l<{$} >{$}l<{$} >{$}l<{$}}
\hline\hline\noalign{\smallskip}
&\multicolumn{2}{c}{WGS84}&\multicolumn{2}{c}{SRMmax}\\
\multicolumn{1}{c}{$L$} &
\multicolumn{1}{c}{$\Delta_{\mathrm d}\,(\mathrm m)$} &
\multicolumn{1}{c}{$\Delta_{\mathrm i}\,(\mathrm m)$} &
\multicolumn{1}{c}{$\Delta_{\mathrm d}\,(\mathrm m)$} &
\multicolumn{1}{c}{$\Delta_{\mathrm i}\,(\mathrm m)$} \\
\noalign{\smallskip}\hline\noalign{\smallskip}
  2 & \num[2.6, -2] & \num[2.6, -2] & \num[2.1, -1] & \num[2.1, -1]\\
  3 & \num[3.7, -5] & \num[1.6, -5] & \num[5.8, -4] & \num[2.5, -4]\\
  4 & \num[1.1, -7] & \num[3.2, -8] & \num[3.3, -6] & \num[1.0, -6]\\
  5 & \num[2.5,-10] & \num[2.3,-11] & \num[1.6, -8] & \num[1.5, -9]\\
  6 & \num[7.7,-13] & \num[5.3,-14] & \num[9.6,-11] & \num[6.6,-12]\\
  7 & \num[2.1,-15] & \num[4.1,-17] & \num[5.2,-13] & \num[1.0,-14]\\
  8 & \num[6.8,-18] & \num[1.1,-19] & \num[3.4,-15] & \num[5.0,-17]\\
  9 & \num[2.0,-20] & \num[8.0,-23] & \num[2.0,-17] & \num[7.9,-20]\\
 10 & \num[6.6,-23] & \num[2.1,-25] & \num[1.3,-19] & \num[4.1,-22]\\
 12 & \num[6.7,-28] & \num[4.6,-31] & \num[5.2,-24] & \num[3.6,-27]\\
 14 & \num[7.0,-33] & \num[1.1,-36] & \num[2.2,-28] & \num[3.3,-32]\\
 16 & \num[7.7,-38] & \num[2.5,-42] & \num[9.4,-33] & \num[3.0,-37]\\
 18 & \num[8.5,-43] & \num[5.8,-48] & \num[4.2,-37] & \num[2.8,-42]\\
 20 & \num[9.7,-48] & \num[1.4,-53] & \num[1.9,-41] & \num[2.7,-47]\\
\noalign{\smallskip}\hline\hline
\end{tabular}
\end{center}
\end{table}%
I first present the truncation errors.  I carry out a sequence of direct
geodesic computations with random $\phi_1$, $\alpha_1$, and $s_{12}$
(subject to the shortest path constraint) comparing the position of the
end point ($\phi_2$ and $\lambda_{12}$) computed using the series
truncated to order $L$ with the exact result (i.e., with $L=30$), in
both cases using arithmetic with 100 decimal digits.  The results are
shown in Table \ref{truncerr} which shows the approximate maximum
truncation error as a function of $L \le 20$.  The quantity
$\Delta_{\mathrm d}$ gives the truncation error for the method as given
in Sect.~\ref{direct}; this scales as $f^L$.  On the other hand,
$\Delta_{\mathrm i}$ is the truncation error where, instead of solving
$\sigma$ in terms of $s$ using the truncated reverted series,
Eq.~(\ref{sigmaeq}), I invert the truncated series for $s$,
Eqs.~(\ref{i1expr}) and (\ref{ij}), to give $\sigma$ in terms of $s$.
(This is done ``exactly'', i.e., using Newton's method and demanding
convergence to 100 decimal places.)  This is representative of the
truncation error in the solution of inverse geodesic problem, because
the inverse problem does not involve determining $\sigma$ in terms of
$s$.  $\Delta_{\mathrm i}$ scales approximately as $(\frac12 f)^L$.  For
comparison, the truncation errors for Vincenty's algorithm are
$9.1\times10^{-5}\,\mathrm m$ and $1.5\times10^{-3}\,\mathrm m$ for the
WGS84 and SRMmax ellipsoids; these errors are about $2.5$ times larger
than $\Delta_{\mathrm d}$ for $L = 3$ (the order of Vincenty's series).

I turn now to the measurement of the round-off errors.  The limiting
accuracy, assuming that the fraction of the floating-point
representation contains $p = 53$ bits, is about
$20\,000\,\mathrm{km}/2^{53} \approx 2\,\mathrm{nm}$ (where
$20\,000\,\mathrm{km}$ is approximately half the circumference of the
earth).  From Table \ref{truncerr}, the choice $L=6$ ensures that the
truncation error is negligible compared to the round-off error even for
$f = 1/150$.  I assembled a large set of exact geodesics for the WGS84
ellipsoid to serve as test data.  These were obtained by solving the
direct problem using Maxima using the protocol described at the
beginning of this section.  All the test data satisfies $\sigma_{12} \le
180^\circ$, so that they are all shortest paths.  Each test geodesic
gives accurate values for $\phi_1$, $\alpha_1$, $\phi_2$,
$\lambda_{12}$, $s_{12}$, $\sigma_{12}$, and $m_{12}$.  In this list
$\phi_1$, $\alpha_1$, and $s_{12}$ are ``input'' values for the direct
problem.  The other values are computed and then rounded to the nearest
$0.1\,\mathrm{pm}$ in the case of $m_{12}$ and $(10^{-18})^\circ$ in the
case of the angles.  The test data includes randomly chosen geodesics
together with a large number of geodesics chosen to uncover potential
numerical problems.  These include nearly meridional geodesics, nearly
equatorial ones, geodesics with one or both end points close to a pole,
and nearly antipodal geodesics.

For each test geodesic, I use the floating-point implementations of the
algorithms to solve the direct problem from each end point and to solve
the inverse problem.  Denoting the results of the computations with an
asterisk, I define the error in a computed quantity $x$ by $\delta x =
x^* - x$.  For the direct problem computed starting at the first end
point, I compute the error in the computed position of the second end
point as
\[
\abs{\rho_2 \delta\phi_2 + i \cos\phi_2 \nu_2 \delta\lambda_2}.
\]
I convert the error in the azimuth into a distance via
\[
a\abs{\delta\alpha_2-\delta\lambda_{12}\sin\phi_2},
\]
which is proportional to the error in the direction of the geodesic at
$B$ in three dimensions and accounts for the coupling of $\alpha_2$ and
$\lambda_{12}$ near the poles.  I compute the corresponding errors when
solving the direct problem starting at the second end point.

\begin{table}[tb]
\caption{\label{closegeod}
Two close geodesics.  The parameters of the WGS84 ellipsoid are used.}
\begin{center}
\begin{tabular}{@{\extracolsep{1\xx}}>{$}c<{$} >{$}c<{$} >{$}c<{$}}
\hline\hline\noalign{\smallskip}
 & \text{Case 1} & \text{Case 2} \\
\noalign{\smallskip}\hline\noalign{\smallskip}
\phi_1 & \multicolumn{2}{c}{$-\exact{30}^\circ$\qquad\qquad} \\
\phi_2 & \exact{30}^\circ & (\exact{30}-\mbox{}\qquad\qquad\\
& & \qquad\exact{4\times 10^{-15}})^\circ \\
\lambda_{12} & \multicolumn{2}{c}{$\exact{179.477\,019\,999\,756\,66}^\circ\qquad$}\\
\alpha_1 & 90.000\,008^\circ & 90.001\,489^\circ \\
\alpha_2 & 89.999\,992^\circ & 89.998\,511^\circ \\
s_{12} & \multicolumn{2}{c}{$19\,978\,693.309\,037\,086\,\mathrm m\qquad$}\\
\sigma_{12} & \exact{180}^\circ & 180.000\,000^\circ \\
m_{12} & 1.1\,\mathrm{nm} & 51\,\mathrm{\mu m}\\
\noalign{\smallskip}\hline\hline
\end{tabular}
\end{center}
\end{table}%
For the inverse problem, I record the error in the length
\[
\abs{\delta s_{12}}.
\]
I convert the errors in the azimuths into a length using
\[
\max(\abs{\delta\alpha_1}, \abs{\delta\alpha_2})\abs{m_{12}};
\]
the multiplication by the reduced length accounts for the sensitivity of
the azimuths to the positions of the end points.  An obvious example of
such sensitivity is when the two points are close to opposite poles or
when they are very close to each other.  A less obvious case is
illustrated in Table \ref{closegeod}.  The second end points in cases 1
and 2 are only $0.4\,\mathrm{nm}$ apart; and yet the azimuths in the two
cases differ by $-5.3''$ and the two geodesics are separated by about
$160\,\mathrm m$ at their midpoints.  (This ``unstable'' case finding
the conjugate point for a geodesic by solving the direct problem with
$\alpha_1 = 90^\circ$ and spherical arc length of $\sigma_{12} =
180^\circ$.)

This provides a suitable measure of the {\it accuracy} of the computed
azimuths for the inverse problem.  I also check their {\it consistency}.
If $s_{12} \ge a$, I determine the midpoint of the computed geodesic by
separate direct geodesic calculations starting at either end point with
the respective computed azimuths and geodesic lengths $\pm \frac12
s_{12}^*$ and I measure the distance between the computed midpoints.
Similarly for shorter geodesics, $s_{12} < a$, I compare the computed
positions of a point on the geodesic a distance $a$ beyond the second
point with separate direct calculations from the two endpoints and
repeat such a comparison for a point a distance $a$ before the first end
point.

In this way, all the various measures of the accuracy of the direct and
inverse geodesic are converted into comparable ground distances, and the
maximum of these measures over a large number of test geodesics is a
good estimate of the combined truncation and round-off errors for the
geodesic calculations.  The maximum error using double precision ($p =
53$) is about $15\,\mathrm{nm}$.  With extended precision ($p = 64$),
the error is about $7\,\mathrm{pm}$, consistent with 11 bits of
additional precision.  The test data consists of geodesics which are
shortest paths, the longest of which is about $20\,000\,\mathrm{km}$.
If the direct problem is solved for longer geodesics (which are not
therefore shortest paths), the error grows linearly with length.  For
example, the error in a geodesic of length $200\,000\,\mathrm{km}$ that
completely encircles the earth 5 times is about $150\,\mathrm{nm}$ (for
double precision).

Another important goal for the test set was to check the convergence of
the inverse solution.  Usually a practical convergence criterion for
Newton's method is that the relative change in the solution is less than
$O(\sqrt\delta)$; because of the quadratic convergence of the method,
this ensures that the error in the solution is less that $O(\delta)$.
However, this reasoning breaks down for the inverse geodesic problem
because the derivative of $\lambda_{12}$ with respect to $\alpha_1$ can
become arbitrarily small; therefore a more conservative convergence
criterion is used.  Typically 2--4 iterations of Newton's method are
required.  A small fraction of geodesics, those with nearly conjugate
end points, require up to 16 iterations.  No convergence failures are
observed.

\section{Ellipsoidal trigonometry}\label{spheroid-trig}

\begin{table}[tb]
\caption{\label{spheroidprob}
Ellipsoidal trigonometry problems.  Here, $\xi$, $\zeta$, and $\theta$ are
the three given quantities.  The ``notes'' column gives the number
assigned by \citet[p.~48]{oriani10} in his ``index of spheroidal
problems'' and by \citet[p.~521]{puissant31} in his enumeration of
solutions.  See the text for an explanation of the other columns.}
\begin{center}
\begin{tabular}{@{\extracolsep{0.7\xx}}>{}r<{} >{$}c<{$} >{$}c<{$}
>{}c<{} >{}c<{} >{}c<{}}
\hline\hline\noalign{\smallskip}
No. & \xi, \zeta, \theta & \psi & Ref. &
$\displaystyle\left.\frac{\d\theta}{\d\psi}\right|_{\xi, \zeta}$
& \multicolumn{1}{c}{Notes}\\\noalign{\smallskip}\hline\noalign{\smallskip}
 1& \phi_1,\alpha_1,\phi_2 & & & & 1, 1 \\
 2& \phi_1,\alpha_1,\alpha_2 & & & & 3, 12 \\
 3& \phi_1,\alpha_1,s_{12} & \sigma_{12} & & Eq.~(\ref{prob-2}) & 5, 2 \\
 4& \phi_1,\alpha_1,\lambda_{12} & \sigma_{12} & & Eq.~(\ref{prob-6}) & 11, 6 \\
 5& \phi_1,\phi_2,s_{12} & \alpha_1 & 1 &Eq.~(\ref{prob-3}) & 7, 3 \\
 6& \phi_1,\phi_2,\lambda_{12} & \alpha_1 & 1 &Eq.~(\ref{prob-7}) & 13, 7 \\
 7& \phi_1,\alpha_2,s_{12} & \alpha_1 & 2 &Eq.~(\ref{prob-4}) & 8, 4 \\
 8& \phi_1,\alpha_2,\lambda_{12} & \alpha_1 & 2 &Eq.~(\ref{prob-8}) & 14, 8 \\
 9& \phi_1,s_{12},\lambda_{12} & \alpha_1 & 3 &Eq.~(\ref{prob-11}) & 19, 11 \\
10& \alpha_1,s_{12},\lambda_{12} & \phi_1 & 3 &Eq.~(\ref{prob-10}) & 17, 10 \\
11& \alpha_1,\alpha_2,s_{12} & \phi_1 & 2 &Eq.~(\ref{prob-5}) & 10, 5 \\
12& \alpha_1,\alpha_2,\lambda_{12} & \phi_1 & 2 &Eq.~(\ref{prob-9}) & 16, 9 \\
\noalign{\smallskip}\hline\hline
\end{tabular}
\end{center}
\end{table}%
The direct and inverse geodesic problems are two examples of solving the
ellipsoidal triangle $N\!AB$ in Fig.~\ref{figtrig} given two sides and
the included angle.  The sides of this triangle are given by $N\!A =
aE(\frac12 \pi - \beta_1, e)$, $N\!B = aE(\frac12 \pi - \beta_2, e)$,
and $AB = s_{12}$ and its angles are $N\!AB = \alpha_1$, $N\!BA = \pi
- \alpha_2$, and $AN\!B = \lambda_{12}$.  The triangle is fully solved
if $\phi_1$, $\alpha_1$, $\phi_2$, $\alpha_2$, $s_{12}$, and
$\lambda_{12}$ are all known.  The typical problem in ellipsoidal
trigonometry is to solve the triangle if just three of these quantities
are specified.  Considering that $A$ and $B$ are interchangeable, there
are 12 distinct such problems which are laid out in
Table~\ref{spheroidprob}.  (In plane geometry, there are four distinct
triangle problems.  On a sphere, the constraint on the sum of the angles
of a triangle is relaxed, leading to six triangle problems.)

\citet[p.~48]{oriani10} and \citet[p.~521]{puissant31} both gave
similar catalogs of ellipsoidal problems as Table~\ref{spheroidprob}.
Here (and in Sect.~\ref{triangulation}), I do not give the full solution
of the ellipsoidal problems nor do I consider how to distinguish the
cases where there may be 0, 1, or 2 solutions.  Instead, I indicate how,
in each case, an accurate solution may be obtained using Newton's method
assuming that a sufficiently accurate starting guess has been found.
This might be obtained by approximating the ellipsoid by a sphere and
using spherical trigonometry \citep[Chap.~6]{todhunter71} or by using
approximate ellipsoidal methods \citep[\S6]{rapp91}.  In
Sect.~\ref{gnomproj}, I also show how the gnomonic projection may be
used to solve several ellipsoidal problems using plane geometry.

In treating these problems, recall that the relation between $\phi$ and
$\beta$ is given by Eq.~(\ref{redlat}) and depends only on the
eccentricity of the ellipsoid.  On the other hand, the relations between
$s_{12}$ and $\lambda_{12}$ and the corresponding variables on the
auxiliary sphere, $\sigma_{12}$ and $\omega_{12}$, depend on the
geodesic (specifically on $\alpha_0$).

In Table~\ref{spheroidprob}, $\xi, \zeta, \theta$ are the given
quantities.  In problems 1 and 2, the given quantities are all directly
related to corresponding quantities for the triangle on the auxiliary
sphere.  This allows the auxiliary triangle to be solved and the
ellipsoidal quantities can then be obtained.  (Problem 1 was used in
solving the inverse problem in Sect.~\ref{inverse}.)

Problem 3 is the direct problem whose solution is given in
Sect.~\ref{direct}.  Here, however, I give the solution by Newton's
method to put this problem on the same footing as the other problems.
The solution consists of treating $\sigma_{12}$ (the column labeled
$\psi$) as a ``control variable''.  Assume a value for this quantity,
solve the problem with given $\phi_1, \alpha_1, \sigma_{12}$ (i.e.,
$\xi, \zeta, \psi$) for $s_{12}$ (i.e., $\theta$) using
Eq.~(\ref{i1expr}).  The value thus found for $s_{12}$ will, of course,
differ from the given value and a better approximation for $\sigma_{12}$
is found using Newton's method with
\begin{equation}\label{prob-2}
\left. \frac{\d s_{12}}{\d\sigma_{12}}\right|_{\phi_1, \alpha_1}
= a w_2.
\end{equation}
The equation for the derivative needed for Newton's method is given in
the column labeled $\left.{\d\theta}/{\d\psi}\right|_{\xi, \zeta}$ in the
table.  Problem 4 is handled similarly except that Eq.~(\ref{i3expr}) is
used to give $\lambda_{12}$ and the necessary derivative for Newton's
method is
\begin{equation}\label{prob-6}
\left. \frac{\d\lambda_{12}}{\d\sigma_{12}}\right|_{\phi_1, \alpha_1}
= \frac{ w_2 \sin\alpha_2 }{\cos\beta_2}.
\end{equation}

Problems 1--4 are the simplest ellipsoidal trigonometry problems with
$\phi$ and $\alpha$ specified at the same point, so that it is possible
to determine $\alpha_0$ which fixes the relation between $s_{12}$ and
$\sigma_{12}$ and between $\lambda_{12}$ and $\omega_{12}$.  In the
remaining problems it is necessary to assume a value for $\phi_1$ or
$\alpha_1$ thereby reducing the problem to one of the reference problems
1--3 (the column labeled ``Ref.'').  Thus, given $\xi,\zeta,\theta$,
assume a value $\psi^{(0)}$ for $\psi$; solve the reference problem
$\xi,\zeta,\psi^{(i)}$ to determine $\theta^{(i)}$; find a more accurate
approximation to $\psi$ using
\[
\psi^{(i+1)} = \psi^{(i)} - (\theta^{(i)} - \theta)
\biggl(
\displaystyle\left.\frac{\d\theta^{(i)}}{\d\psi^{(i)}}\right|_{\xi, \zeta}
\biggr)^{-1},
\]
and iterate until convergence.  The remaining derivatives are
\begin{align}
\label{prob-3}
\left. \frac{\d s_{12}}{\d\alpha_1}\right|_{\phi_1, \phi_2} &=
m_{12}\tan\alpha_2,
\displaybreak[0]\\
\label{prob-7}
\left.
\frac{\d\lambda_{12}}{\d\alpha_1}\right|_{\phi_1, \phi_2} &=
\frac{m_{12}}a \frac1{\cos\alpha_2 \cos\beta_2},
\displaybreak[0]\\
\label{prob-4}
\left.\frac{\d s_{12}}{\d\alpha_1}\right|_{\phi_1, \alpha_2} &=
m_{12}\tan\alpha_2 - \frac{aw_2} {\tan\alpha_1\tan\beta_2\cos\alpha_2},
\displaybreak[0]\\
\label{prob-8}
\left. \frac{\d\lambda_{12}}{\d\alpha_1}\right|_{\phi_1, \alpha_2} &=
\frac{m_{12}/a}{\cos\alpha_2\cos\beta_2}
- \frac{w_2\tan\alpha_2}{\tan\alpha_1\sin\beta_2},
\displaybreak[0]\\
\label{prob-11}
\left. \frac{\d\lambda_{12}}{\d\alpha_1}\right|_{\phi_1, s_{12}} &= \frac{m_{12}}a
\frac{\cos\alpha_2}{\cos\beta_2},
\displaybreak[0]\\
\left. \frac{\d\lambda_{12}}{\d\phi_1}\right|_{\alpha_1, s_{12}} &=
\frac{ w_1^3 }{(1-f)\sin\alpha_1}\times\notag\\
&\qquad\qquad
\biggl(\frac{\cos\alpha_1}{\cos\beta_1}
-\frac{\cos\alpha_2}{\cos\beta_2} N_{12}
\biggr),
\label{prob-10}
\displaybreak[0]\\
\left. \frac{\d s_{12}}{\d\phi_1}\right|_{\alpha_1, \alpha_2} &=
a\frac{w_1^3}{1-f}
\biggl(
-\frac{N_{12}\tan\alpha_2}{\sin\alpha_1}\notag\\ &\qquad\qquad\qquad
+\frac{\tan\beta_1}{\cos\alpha_2\tan\beta_2}
\frac{w_2}{w_1}
\biggr),
\label{prob-5}
\displaybreak[0]\\
\left. \frac{\d\lambda_{12}}{\d\phi_1}\right|_{\alpha_1, \alpha_2} &=
\frac{w_1^3}{1-f}
\biggl(\frac{\cos\alpha_1}{\sin\alpha_1\cos\beta_1}
-\frac{N_{12}\sec\alpha_2}{\sin\alpha_1\cos\beta_2}\notag\\
&\qquad\qquad\qquad +\frac{\tan\beta_1\tan\alpha_2}{\sin\beta_2}
\frac{w_2}{w_1}
\biggr),
\label{prob-9}
\end{align}
where
\[
N_{12} = M_{12} - \frac{(m_{12}/a) \cos\alpha_1\tan\beta_1}{w_1}.
\]

The choice of $\psi$ in these solutions is somewhat arbitrary; other
choices may be preferable in some cases.  These formulas for the
derivatives are obtained with constructions similar to
Fig.~\ref{derivfig}.  Equations (\ref{prob-10})--(\ref{prob-9}) involve
partial derivatives taken with $\alpha_1$ held constant; the role of
$N_{12}$ in these equation can be contrasted with that of $M_{12}$ as
follows.  Consider a geodesic from $A$ to $B$ with length $s_{12}$ and
initial azimuth $\alpha_1$.  Construct a second geodesic of the same
length from $A'$ to $B'$ where $A'$ is given by moving a small distance
$\d t$ from $A$ in a direction $\alpha_1 + \frac12 \pi$.  If the initial
direction of the second geodesic is $\alpha'_1 = \alpha_1$
(resp.~parallel to the first geodesic), then the distance from $B$ to
$B'$ is $N_{12}\,\d t$ (resp.~$M_{12}\,\d t$).  (Because meridians converge,
two neighboring geodesics with the same azimuth are not, in general,
parallel.)

Problem 6 is the geodesic inverse problem solved in Sect.~\ref{inverse}
and I have repeated Eq.~(\ref{lambda-deriv}) as Eq.~(\ref{prob-7}).
Problem 7 is the ``retro-azimuthal'' problem for which \citet{hinks29}
gives an interesting application.  For many years a radio at Rugby
transmitted a long wavelength time signal.  Hinks' retro-azimuthal
problem is to determine the position of an unknown point with knowledge
of the distance and bearing to Rugby.

\section{Triangulation from a baseline}\label{triangulation}

\begin{figure}[tb]
\begin{center}
\includegraphics[scale=0.75,angle=0]{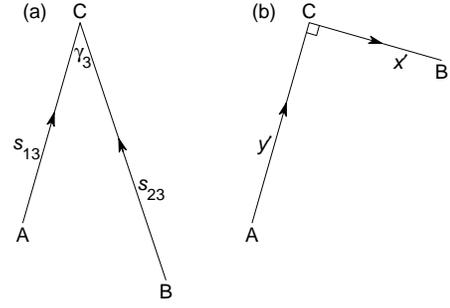}
\end{center}
\caption{\label{figbase}
Two ellipsoidal triangle problems: (a)~triangulating from a baseline and
(b)~rectangular geodesic (or oblique Cassini--Soldner) coordinates.}
\end{figure}%
The ellipsoidal triangle considered in Sect.~\ref{spheroid-trig} is
special in that one of its vertices is a pole so that two of its sides
are meridians.  The next class of ellipsoidal problems treated is
solving a triangle $ABC$ with a known baseline, see Fig.~\ref{figbase}a.
Here, the goal is to determine the position of $C$ if the positions if
$A$ and $B$ are known, i.e., if $\phi_1$, $\phi_2$, and $\lambda_{12}$
are given (and hence, from the solution of the inverse problem for $AB$,
the quantities $\alpha_1$, $\alpha_2$, and $s_{12}$ are known).  This
corresponds to a class of triangulation problems encountered in field
surveying: a line $AB$ is measured and is used as the base of a
triangulation network.  However, in the present context, $A$ need not be
visible from $B$ and I consider both the problems of triangulation and
trilateration.  (In addition, remember that the angles measured by a
theodolite are not the angles between geodesics but between normal
sections.)

Because the problems entail consideration of more than a single general
geodesic, it is necessary to generalize the notation for azimuthal angles
to make clear which geodesic line is being measured.  I define
$\alpha_{i(j)}$ as the azimuth of the geodesic line passing through
point $i$ where $j$ is some other point on the same line.  Each geodesic
line is assigned a unique direction, indicated by arrows in the figures,
and all azimuths are forward azimuths (as before).

\begin{table}[tb]
\caption{\label{triangtab}
Ellipsoidal triangulation problems.  In these problems, the positions of
$A$ and $B$ (i.e., $\phi_1$, $\phi_2$, and $\lambda_{12}$) are given and
the position of $C$ is sought.  The quantities $\xi$ and $\zeta$ are the
additional given quantities.}
\begin{center}
\begin{tabular}{@{\extracolsep{1\xx}}>{}r<{} >{$}c<{$} >{$}c<{$}
>{}c<{}}
\hline\hline\noalign{\smallskip}
No. & \xi, \zeta & \psi &
$\displaystyle\left.\frac{\d\zeta}{\d\psi}\right|_{\xi}$
\\\noalign{\smallskip}\hline\noalign{\smallskip}
 0& \alpha_{1(3)},s_{13} \\
 1& s_{13},s_{23} & \alpha_{1(3)} & Eq.~(\ref{two-side}) \\
 2& \alpha_{1(3)},\alpha_{2(3)} & s_{13} & Eq.~(\ref{two-angle}) \\
 3& \alpha_{1(3)},s_{23} & s_{13} & Eq.~(\ref{side-angle}) \\
 4& s_{13},\gamma_3 & \alpha_{1(3)} & Eq.~(\ref{oppangle-side}) \\
 5& \alpha_{1(3)},\gamma_3 & s_{13} & Eq.~(\ref{oppangle-angle}) \\
\noalign{\smallskip}\hline\hline
\end{tabular}
\end{center}
\end{table}%
With one side specified and two additional quantities needed to solve
the triangle, there are 10 possible problems to solve.  Of these, six
are distinct (considering the interchangeability of $A$ and $B$) and are
listed in Table~\ref{triangtab}.  For the angle at $C$, I assume that
$\gamma_3 = \alpha_{3(1)}-\alpha_{3(2)}$, the difference in the bearings
of $A$ and $B$, is given; this is included angle at $C$ in
Fig.~\ref{figbase}a.  In other words, I do not presume that the
direction of due north is known, {\it a priori}, at $C$.  This is the
common situation with theodolite readings and it also includes the
important case where $\gamma$ is required to be $\pm \frac12 \pi$ which
allows the shortest distance from a point to a geodesic to be
determined.

Problem 0 is just the direct geodesic problem (problem 3 of
Sect.~\ref{spheroid-trig}).  The remaining 5 problems may be solved by
Newton's method, in a similar fashion as in Sect.~\ref{spheroid-trig}, as
follows: replace the second given quantity $\zeta$ by one of the
unknowns $\psi$; estimate a value of $\psi$; solve the problem with
$\xi$ and $\psi$ (which, in each case, is a direct geodesic problem
from $A$) to determine a trial position for $C$; solve the inverse
geodesic problem between $B$ and the trial position for $C$ to obtain a
trial value for $\zeta$; update the value of $\psi$ using Newton's
method so that the resulting value of $\zeta$ matches the given value.
The derivatives necessary for Newton's method are given in
Eqs.~(\ref{two-side})--(\ref{oppangle-angle}):
\begin{align}
\left. \frac{\d s_{23}}{\d\alpha_{1(3)}}\right|_{\phi_1, \phi_2, \lambda_{12}, s_{13}} &=
-m_{13}\sin\gamma_3,\label{two-side}
\displaybreak[0]\\
\left. \frac{\d\alpha_{2(3)}}{\d s_{13}}\right|_{\phi_1, \phi_2, \lambda_{12}, \alpha_{1(3)}} &=
\frac1{m_{23}}\sin\gamma_3,\label{two-angle}
\displaybreak[0]\\
\left. \frac{\d s_{23}}{\d s_{13}}\right|_{\phi_1, \phi_2, \lambda_{12}, \alpha_{1(3)}} &=
\cos\gamma_3,\label{side-angle}
\displaybreak[0]\\
\left. \frac{\d\gamma_3}{\d\alpha_{1(3)}}
\right|_{\phi_1, \phi_2, \lambda_{12}, s_{13}} &=
M_{31}-M_{32}\frac{m_{13}}{m_{23}}\cos\gamma_3,\label{oppangle-side}
\displaybreak[0]\\
\left. \frac{\d\gamma_3}{\d s_{13}}
\right|_{\phi_1, \phi_2, \lambda_{12}, \alpha_{1(3)}} &=
\frac{M_{32}}{m_{23}}\sin\gamma_3.\label{oppangle-angle}
\end{align}

The triangulation problems 1, 2, and 3, entail specification of either
the distance or the bearing to $C$ from each of $A$ and $B$.  These can
also be solved by generalizing the method given
by \citet[\S5]{sjoeberg02} for the solution of problem 2.  The technique
is to treat the latitude of $C$, $\phi_3$, as the control variable.
Thus, start with an estimate for $\phi_3$; if the bearing of
(resp.~distance to) $C$ from a base point is given, then solve the
intermediate problem $\phi_i$, $\phi_3$, $\alpha_{i(3)}$
(resp.~$s_{i3}$) where $i = 1$ or $2$ for base points $A$ or $B$, i.e.,
problem 1 (resp.~5) in Sect.~\ref{spheroid-trig} to give $\lambda_{i3}$;
and evaluate $\lambda_{12} = \lambda_{13} - \lambda_{23}$.  Now adjust
$\phi_3$ using Newton's method so that the $\lambda_{12}$ matches the
known value using
\begin{align}
\left. \frac{\d\lambda_{i3}}{\d\phi_3}\right|_{\phi_i, \alpha_{i(3)}} &=
\frac{w_3^3}{1-f} \frac{\tan\alpha_{3(i)}}{\cos\beta_3},
\displaybreak[0]\\
\left. \frac{\d\lambda_{i3}}{\d\phi_3}\right|_{\phi_i, s_{i3}} &=
-\frac{w_3^3}{1-f} \frac{\cot\alpha_{3(i)}}{\cos\beta_3}.
\end{align}
This method of solution essentially factors the problem into two simpler
problems of the type investigated in Sect.~\ref{spheroid-trig}.

A similar approach can also be applied to problem 5.  Guess a value of
$s_{13}$, solve problems 3 and 1 of Sect.~\ref{spheroid-trig} to
determine successively the positions of $C$ and $B$.  Adjust $s_{13}$
using Newton's method so that the correct value of $\lambda_{12}$ is
obtained, which requires the use of the derivative
\begin{equation}
\left. \frac{\d\lambda_{12}}{\d s_{13}}\right|_{\phi_1, \phi_2, \alpha_{1(3)}, \gamma_3} =
-\frac{M_{32}}a \frac{\sin\gamma_3\sec\alpha_{2(3)}}{\cos\beta_2}.
\end{equation}

Knowledge of reduced length and the geodesic scale allow errors to be
propagated through a calculation.  For example, if the measurements of
$\alpha_{1(3)}$ and $\alpha_{2(3)}$ are subject to an instrumental error
$\delta\alpha$, then the error ellipse in the position of $C$ when
solving problem 2 will have a covariance which depends on
$m_{13}\,\delta\alpha$, $m_{23}\,\delta\alpha$, and $\gamma_3$.  The
effects of errors in the positions of $A$ and $B$ and in the
measurements of $s_{13}$ and $s_{23}$ can be similarly estimated.

\citet[\S\S A2.4.1--3]{rnav07} also presents solution for
problems 1--3.  However, these use the secant method and so converge more
slowly that the methods given here.

\section{Geodesic projections}\label{geodproj}

Several map projections are defined in terms of geodesics.  In the
azimuthal equidistant projection \citep[\S25]{snyder87} the distance and
bearing from a central point $A$ to an arbitrary point $B$ is
preserved.  \citet[\S19]{gauss27} lays out the problem for a general
surface: the point $B$ is projected to plane cartesian coordinates,
\[
x = s_{12}\sin\alpha_1, \quad y = s_{12}\cos\alpha_1.
\]
\citet[\S16]{bagratuni67} calls these ``geodetic polar coordinates''.
\citet[\S15]{gauss27} proves that the geodesics (lines of constant
$\alpha_1$) and the geodesic circles (lines of constant $s_{12}$),
which, by construction, intersect at right angles in the projection,
also intersect at right angles on the ellipsoid (see
Sect.~\ref{redlength}).  The scale in the radial direction is unity,
while the scale in the azimuthal direction is $s_{12}/m_{12}$; the
projection is conformal only at the origin.  The forward and reverse
projections are given by solving the inverse and direct geodesic
problems.  The entire ellipsoid maps to an approximately elliptical
area, with the azimuthal scale becoming infinite at the two boundary
points on the $x$ axis.  The projection can be continued beyond the
boundary giving geodesics which are no longer shortest lines and
negative azimuthal scales.  \citet[p.~197]{snyder87} gives the formulas
for this projection for the ellipsoid only for the case where the center
point is a pole.  For example, if $A$ is at the north pole then the
projection becomes
\[
s_{12} = aE(\tfrac1/2 \pi - \beta_2, e), \quad m_{12} = a\cos\beta_2.
\]
However, the method given here is applicable for any center point.  The
projection is useful for showing distances and directions from a central
transportation hub.

\citet[\S23]{gauss27} also describes another basic geodesic projection,
called ``right-angle spheroidal coordinates''
by \citet[\S17]{bagratuni67}.  Consider a reference geodesic passing
through the point $A$ at azimuth $\alpha_{1(3)}$.  The reverse
projection, $B$, of the point $x, y$ is given by the following
operations which are illustrated in Fig.~\ref{figbase}b: resolve the
coordinates into the directions normal and parallel to the initial
heading of the reference geodesic,
\begin{align*}
x' &= \cos\alpha_{1(3)} x - \sin\alpha_{1(3)} y,\\
y' &= \sin\alpha_{1(3)} x + \cos\alpha_{1(3)} y;
\end{align*}
starting at $A$ proceed along the reference geodesic a distance $y'$ to
$C$; then proceed along the geodesic with azimuth $\alpha_{3(2)}
= \alpha_{3(1)} + \frac12 \pi$ a distance $x'$ to point $B$.  (Here, the
``forward'' direction on the geodesic $CB$ is to the right of the
reference geodesic $AC$ which is the opposite of the convention in
Sect.~\ref{triangulation}.)  \citet[\S16]{gauss27} proves that geodesics
(lines of constant $y'$) and the geodesic ``parallels'' (lines of
constant $x'$) intersect at right angles on the ellipsoid.  At $B$, the
scale in the $x'$ direction is unity while the scale in the $y'$
direction is $1/M_{32}$ which is unity on $x' = 0$; thus the projection
is conformal on $x' = 0$.  The definition of the mapping given here
provides the prescription for carrying out the reverse projections.  The
forward projection is solved as follows: determine the point $C$ on the
reference geodesic which is closest to $B$ (problem 5 of
Sect.~\ref{triangulation}); set $x'$ to the distance $CB$ signed
positive or negative according to whether $B$ is to the right or left of
the reference geodesic; set $y'$ to the distance $AC$ signed positive or
negative according to whether $C$ is ahead or behind $A$ on the
reference geodesic; finally transform the coordinate frame
\begin{align*}
x &= \cos\alpha_{1(3)} x' + \sin\alpha_{1(3)} y',\\
y &=-\sin\alpha_{1(3)} x' + \cos\alpha_{1(3)} y'.
\end{align*}
In the case where the reference geodesic is the equator, the projection
is the ellipsoidal generalization of the so-called ``equidistant
cylindrical'' projection \citep[\S12]{snyder87}.  Solving for the point
$C$ is trivial; the coordinates are given by
\[
x = a\lambda_{12},\quad
y = a\bigl(E(e) - E(\tfrac1/2 \pi - \beta_2, e)\bigr), \quad
M_{32} = \cos\beta_2,
\]
where $E(k)$ is the complete elliptic integral of the second
kind \citep[\S\dlmf{19.2(ii)}{19.2.ii}]{dlmf10}; see
also \citet[\S2.1.4]{bugayevskiy95}.  The whole ellipsoid is mapped to a
rectangular region with the poles mapped to lines (where the scale in
the $x$ direction is infinite).  If the reference geodesic is a
``central'' meridian, the projection is called
``Cassini--Soldner'' \citep[\S13]{snyder87} and $C$ may most simply be
found by finding the midpoint of the geodesic $BD$ where $D$ is the
reflection of $B$ in the plane of the central meridian.  This allows the
Cassini--Soldner mapping to be be solved accurately for the whole
ellipsoid, in contrast to the series method presented
in \citet[p.~95]{snyder87} which is only valid near the central
meridian.  The ellipsoid maps to an approximately rectangular region
with the scale in the $y$ direction divergent where the equator
intersects the boundary.  The Cassini--Soldner was widely used for
large-scale maps until the middle of the 20th century when it was almost
entirely replaced by the transverse Mercator
projection \citep[\S8]{snyder87}.  Tasks such as navigation and
artillery aiming were much more easily accomplished with a conformal
projection, such as transverse Mercator, compared to Cassini--Soldner
with its unequal scales.  The general case of this mapping may be termed
the ``oblique Cassini--Soldner'' projection.  Because the reference
geodesic is not closed in this case, it is not convenient to use this
projection for mapping the entire ellipsoid because there may be
multiple candidates for $C$, the position on the reference geodesic
closest to $B$.

\begin{figure}[tb]
\begin{center}
\includegraphics[scale=0.75,angle=0]{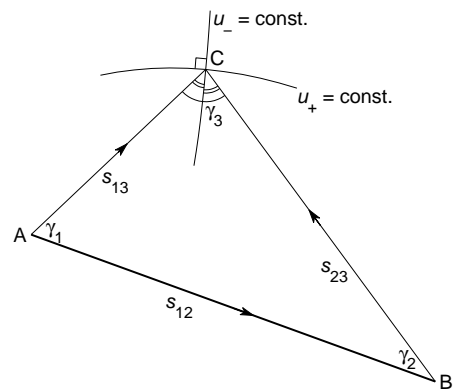}
\end{center}
\caption{\label{figequi}
The doubly equidistant projection.  Also shown are portions of the
geodesic ellipse and hyperbola through $C$.}
\end{figure}%
The doubly equidistant projection has been used in small-scale maps to
minimize the distortions of large land
masses \citep[\S7.9]{bugayevskiy95}.  In this projection, the distances
to an arbitrary point $C$ from two judiciously chosen reference points
$A$ and $B$ are preserved; see Fig.~\ref{figequi}.  The formulas are
usually given for a sphere; however, the generalization to an ellipsoid
is straightforward.  For the forward projection, fix the baseline $AB$
with $A$ and $B$ separated by $s_{12}$; solve the inverse geodesic
problems for $AC$ and $BC$ and use the distances $s_{13}$ and $s_{23}$
together with elementary trigonometry to determine the position of $C$
in the projected space; there are two solutions for the position of $C$
either side of the baseline; the desired solution is the one that lies
on the same side of the baseline as $C$ on the ellipsoid.  The reverse
projection is similar, except that the position of $C$ on the ellipsoid
is determined by solving problem 1 in Sect.~\ref{triangulation}.  The
projection is only well defined if $\gamma_1$ and $\gamma_2$ are the
same sign (consistent with a planar triangle).  This is always the case
for a sphere; the entire sphere projects onto an ellipse.  However, on
an ellipsoid, the geodesic connecting $A$ and $B$ is not closed in
general and thus does not divide the ellipsoid into two halves.  As a
consequence, there may be a portion of the ellipsoid which is on one
side of the baseline geodesic as seen from $A$ but on the other side of
it as seen from $B$; such points cannot be projected.  For example if $A
= (35^\circ\mathrm N,40^\circ\mathrm E)$ and $B = (35^\circ\mathrm
N,140^\circ\mathrm E)$, then the point $(43.5^\circ\mathrm
S,60.5^\circ\mathrm W)$ cannot be projected because it is north of the
baseline as seen by $A$ but south of the baseline relative to $B$

The scales of the doubly equidistant projection can be determined as
follows.  By construction, geodesic ellipses and hyperbolae, defined by
$u_\pm = \frac12 (s_{13}\pm s_{23}) = \text{const.}$, map to ellipses
and hyperbolae under this projection; see
Fig.~\ref{figequi}.  \citet{weingarten63} establishes these results for
geodesic ellipses and hyperbolae \citep[\S90]{eisenhart09}: they are
orthogonal; the geodesic hyperbola through $C$ bisects the angle
$\gamma_3$ made by the two geodesics from $A$ and $B$; and the scales in
the $u_\pm$ directions are $\cos\frac12\gamma_3$ and
$\sin\frac12\gamma_3$, respectively.  The same relations hold, of
course, for the projected ellipses and hyperbolae, except that the angle
$ACB$ takes on a different (smaller) value $\gamma_3'$.  Thus, the
elliptic and hyperbolic scale factors for the double equidistant
projection may be written as
\[
\frac{\cos\frac12\gamma_3}{\cos\frac12\gamma_3'},
\quad
\frac{\sin\frac12\gamma_3}{\sin\frac12\gamma_3'},
\]
respectively.  Evaluating $\gamma_3'$ using the cosine rule for the
plane triangle $ABC$ gives
\[
\frac{2\sqrt{s_{13}s_{23}}\cos\frac12\gamma_3}
{\sqrt{(s_{13}+s_{23})^2 - s_{12}^2}},
\quad
\frac{2\sqrt{s_{13}s_{23}}\sin\frac12\gamma_3}
{\sqrt{s_{12}^2 - (s_{13}-s_{23})^2}}.
\]
The results generalize those of \citet{cox46,cox51} for the projection
of a sphere.  In the limit $s_{12}\rightarrow 0$, this projection
reduces to the azimuthal equidistant projection and these scales reduce
to the radial scale, $1$, and the azimuthal scale, $s_{13}/m_{13}$.

\section{Spheroidal gnomonic projection}\label{gnomproj}

The gnomonic projection of the sphere, which is obtained by a central
projection of the surface of the sphere onto a tangent plane, has the
property that all geodesics on the sphere map to straight
lines \citep[\S22]{snyder87}.  Such a projection is impossible for an
ellipsoid because it does not have constant
curvature \citep{beltrami65}.  However, a spheroidal generalization of
the gnomonic projections can be constructed for which geodesics are very
nearly straight.  First recall that the doubly azimuthal
projection \citep[\S7.8]{bugayevskiy95} of the sphere, where the
bearings from two points $A$ and $B$ to $C$ are preserved, gives the
gnomonic projection which is compressed in the direction parallel to
$AB$.  In the limit as $B$ approaches $A$, the pure gnomonic projection
is recovered.

\begin{figure}[tb]
\begin{center}
\includegraphics[scale=0.75,angle=0]{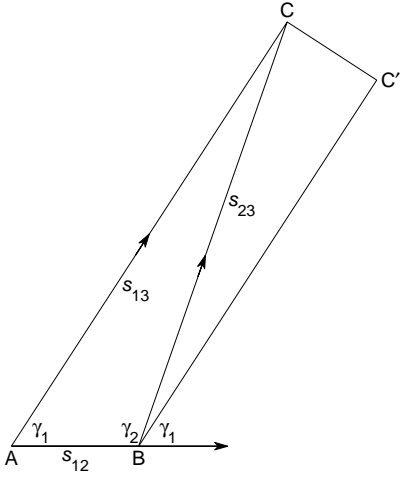}
\end{center}
\caption{\label{gnomconstr}
The construction of the spheroidal gnomonic projection as the limit of a
doubly azimuthal projection.}
\end{figure}%
The construction of the spheroidal gnomonic projection proceeds in the
same way; see Fig.~\ref{gnomconstr}.  Draw a geodesic $BC'$ such that it
is parallel to the geodesic $AC$ at $B$.  Its initial separation from
$AC$ is $s_{12}\sin\gamma_1$; at $C'$, the point closest to $C$, the
separation becomes $M_{13}s_{12}\sin\gamma_1$ (in the limit
$s_{12}\rightarrow 0$).  Thus the difference in the azimuths of the
geodesics $BC$ and $BC'$ at $B$ is $(M_{13}/m_{13})s_{12}\sin\gamma_1$,
which gives $\gamma_1 + \gamma_2 = \pi - (M_{13}/m_{13})
s_{12}\sin\gamma_1$.  Now, solving the planar triangle problem with
$\gamma_1$ and $\gamma_2$ as the two base angles gives the distance $AC$
in the projected space as $m_{13}/M_{13}$.

Thus leads to the following specification for the spheroidal gnomonic
projection.  Let the center point be $A$; for an arbitrary point $B$,
solve the inverse geodesic problem between $A$ and $B$; then point $B$
projects to the point
\begin{equation}\label{gnom-eq}
x = \rho\sin\alpha_1, \quad
y = \rho\cos\alpha_1, \quad
\rho = m_{12}/M_{12};
\end{equation}
the projection is undefined if $M_{12} \le 0$.  In the spherical limit,
this becomes the standard gnomonic projection, $\rho =
a \tan\sigma_{12}$ \citep[p.~165]{snyder87}.  The azimuthal scale is
$1/M_{12}$ and the radial scale, found by computing $\d\rho/\d s_{12}$ and
using Eq.~(\ref{wronski}), is $1/M_{12}^2$; the projection is therefore
conformal at the origin.  The reverse projection is found by $\alpha_1
= \ph(y + ix)$ and by solving for $s_{12}$ using Newton's method with
$\d\rho/\d s_{12} = 1/M_{12}^2$ (i.e., the radial scale).  Clearly the
projection preserves the bearings from the center point and all lines
through the center point are geodesics.  Consider now a straight line
$BC$ in the projection and project this line on the spheroid.  The
distance that this deviates from a geodesic is, to lowest order,
\begin{equation}
h = \frac{l^2}{32} (\nabla K \cdot \v t) \v t,
\end{equation}
where $l$ is the length of the geodesic, $K$ is the Gaussian curvature,
and $\v t$ is the perpendicular vector from the center of projection to
the geodesic.  I obtained this result semi-empirically: numerically, I
determined that the maximum deviation was for east-west geodesics; I
then found, by Taylor expansion, the deviation for the simple case in
which the end points are equally distant from the center point at
bearings $\pm\alpha$; finally, I generalized the resulting expression
and confirmed this numerically.  The deviation in the azimuths at the
end points is about $4h/l$ and the length is greater than the
geodesic distance by about $\frac83 h^2/l$.  For an ellipsoid,
the curvature is given by Eq.~(\ref{curvature}), which gives
\begin{equation}
\nabla K = -\frac{4a}{b^4}
e^2(1-e^2\sin^2\phi)^{5/2}\cos\phi\sin\phi;
\end{equation}
the direction of $\nabla K$ is along the meridian towards the equator.
Bounding $h$ over all the geodesics whose end-points lie within a
distance $r$ of the center of projection, gives (in the limit that $e$
and $r$ are small)
\begin{equation}\label{gnomonic-err}
\abs{h} \le \frac f8 \frac{r^3}{a^3} r.
\end{equation}
The limiting value is attained when the center of projection is at $\phi
 = \pm 45^\circ$ and the geodesic is running in an east-west direction
 with the end points at bearings $\pm 45^\circ$ or $\pm 135^\circ$ from
 the center.

\begin{figure}[tb]
\begin{center}
\includegraphics[scale=0.75,angle=0]{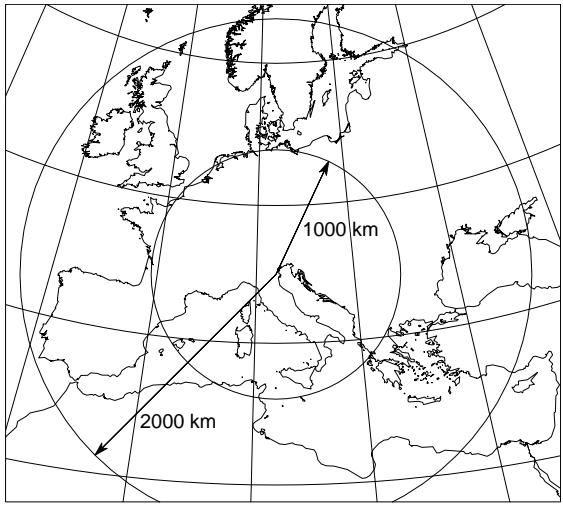}
\end{center}
\caption{\label{gnomfig}
The coast line of Europe and North Africa in the ellipsoidal gnomonic
projection with center at $(45^\circ\mathrm N, 12^\circ\mathrm E)$ near
Venice.  The graticule lines are shown at multiples of $10^\circ$.  The
two circles are centered on the projection center with (geodesic) radii
of $1000\,\mathrm{km}$ and $2000\,\mathrm{km}$.  The data for the coast
lines is taken from GMT \citep{gmt455} at ``low'' resolution.}
\end{figure}%
\citet{bowring97} and \citet{williams97} have proposed an alternate
ellipsoidal generalization of the gnomonic projection as a central
projection of the ellipsoid onto a tangent plane.  In such a mapping,
great ellipses project to straight lines.  Empirically, I find that the
deviation between straight lines in this mapping and geodesics is
\[
\abs{h} \le \frac f2 \frac ra r.
\]
\citet{letovaltsev63} suggested another gnomonic projection in which
normal sections through the center point map to straight lines.  The
corresponding deviation for geodesics is
\[
\abs{h} \le \frac {3f}8 \frac{r^2}{a^2} r,
\]
which gives a more accurate approximation to geodesics than great
ellipses.  However, the new definition of the spheroidal gnomonic
projection, Eq.~(\ref{gnom-eq}), results in an even smaller error,
Eq.~(\ref{gnomonic-err}), in estimating geodesics.  As an illustration,
consider Fig.~\ref{gnomfig} in which a gnomonic projection of Europe is
shown.  The two circles are geodesic circles of radii
$1000\,\mathrm{km}$ and $2000\,\mathrm{km}$.  If the geodesic between
two points within one of these circles is estimated by using a straight
line on this figure, the maximum deviation from the true geodesic will
be about $1.7\,\mathrm m$ and $28\,\mathrm m$, respectively.  The
maximum changes in the end azimuths are $1.1''$ and $8.6''$ and the
maximum errors in the lengths are only $5.4\,\mu\mathrm m$ and
$730\,\mu\mathrm m$.

At one time, the gnomonic projection was useful for determining
geodesics graphically.  However, the ability to determine geodesics
paths computationally renders such use of the projection an anachronism.
Nevertheless, the projection can be a used within an algorithm to solve
some triangulation problems.  For example, consider a variant of the
triangulation problem 2 of Sect.~\ref{triangulation}: determine the point
of intersection of two geodesics between $A$ and $B$ and between $C$ and
$D$.  This can be solved using the ellipsoidal gnomonic projection as
follows.  Guess an intersection point $O^{(0)}$ and use this as the
center of the gnomonic projection; define $\v a$, $\v b$, $\v c$, $\v d$
as the positions of $A$, $B$, $C$, $D$ in the gnomonic projection; find
the intersection of of $AB$ and $CD$ in this projection, i.e.,
\[
\v o = \frac
{ (\v{\hat z} \cdot \v c \times \v d) (\v b - \v a) -
  (\v{\hat z} \cdot \v a \times \v b) (\v d - \v c)}
{\v{\hat z} \cdot (\v b - \v a) \times (\v d - \v c) },
\]
where $\v{\hat{\text{\ }}}$ indicates a unit vector ($\v{\hat a} = \v
a/a$) and $\v{\hat z} = \v{\hat x} \times \v{\hat y}$ is in the
direction perpendicular to the projection plane.  Project $\v o$ back to
geographic coordinates $O^{(1)}$ and use this as a new center of
projection; iterate this process until $O^{(i)} = O^{(i-1)}$ which is
then the desired intersection point.  This algorithm converges to the
exact intersection point because the mapping projects all geodesics
through the center point into straight lines.  The convergence is rapid
because projected geodesics which pass near the center point are very
nearly straight.  Problem 5 of Sect.~\ref{triangulation} can be solving
using the gnomonic projection in a similar manner.  If the point $O$ on
$AB$ which closest to $C$ is to be found, the problem in the gnomonic
space becomes
\[
\v o = \frac
{\v c \cdot (\v b - \v a) (\v b - \v a)
- (\v{\hat z} \cdot \v a \times \v b) \v{\hat z} \times (\v b - \v a)}
{\abs{\v b - \v a}^2};
\]
in this case, the method relies on the preservation of azimuths about
the center point.

Another application of the gnomonic projection is in solving for region
intersections, unions, etc.  For example the intersection of two
polygons can be determined by projecting the polygons to planar polygons
with the gnomonic projection about some suitable center.  Any place were
the edges of the polygons intersect {\it or nearly intersect} in the
projected space is a candidate for an intersection on the ellipsoid
which can be found exactly using the techniques given above.  The
inequality (\ref{gnomonic-err}) can be used to define how close to
intersection the edges must be in projection space be candidates for
intersection on the ellipsoid.

The methods described here suffer from the drawback that the gnomonic
projection can be used to project only about one half of the ellipsoid
about a given center.  This is unlikely to be a serious limitation in
practice and can, of course, be eliminated by partitioning a problem
covering a large area into a few smaller sub-problems.

\section{Maritime boundaries}\label{median}

Maritime boundaries are defined to be a fixed distance from the coast of
a state or, in the case of adjacent states or opposite states, as the
``median line'' between the states \citep[Chaps.~5--6]{talos06}.  In the
application of these rule, distances are defined as the geodesic
distance on a reference ellipsoid to the nearest point of a state and
the extent of a state is defined either by points on the low water mark
or straight lines closing off bays or joining islands to the mainland.

For median lines, several cases can then be
enumerated \citep[Chap.~6]{talos06}: the median is determined by two
points, a point and a line, or two lines.  In the first case, the
boundary is analogous to the perpendicular bisector in plane geometry.
It may be constructed by determining the midpoint of the geodesic
joining the two points and then marking off successive points either
side of the geodesic by solving the 2-distance triangulation problem
(problem 1 in Sect.~\ref{triangulation}) using increasing distances.
This continues until some other coastal point becomes closer, at which
point the median line changes direction and continues as the
perpendicular bisector of a new pair of points.  Such turning points are
called ``tri-points'' and are equidistant from two points of one state
and two point of the other; such a point is the center of the geodesic
circle circumscribing the triangle formed by the three points.

\begin{figure}[tb]
\begin{center}
\includegraphics[scale=0.75,angle=0]{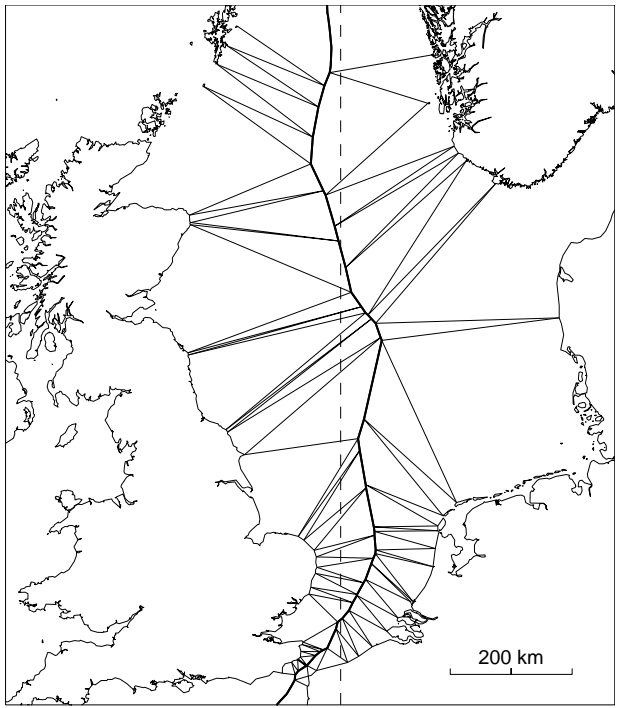}
\end{center}
\caption{\label{medianfig}
The median line (shown as a heavy line) between Britain and her North
Sea neighbors.  Light lines connect the median line to the controlling
boundary points.  The Cassini--Soldner projection is used with the
central meridian (shown as a dashed line) equal to
$2^\circ20'15''\mathrm E$ (the longitude of Paris).  The data for the
coast lines is taken from GMT \citep{gmt455} at ``intermediate''
resolution.}
\end{figure}%
I consider first the problem of determining a tri-point $O$ given the
three coastal points $A$, $B$, and $C$.  The solution is an iterative
one which is conveniently described in terms of the azimuthal
equidistant projection.  Make an initial guess $O^{(0)}$ for the
position of the tri-point.  Map $A$, $B$, and $C$ to the azimuthal
equidistant projection with $O^{(0)}$ as the center and denote their
positions in this projection as $\v a$, $\v b$, and $\v c$.  Compute the
center $\v o$ of the circle circumscribing the triangle formed by these
three points,
\begin{equation}\label{tri-a}
\v o = \frac
{ \bigl(a^2 (\v b - \v c) + b^2 (\v c - \v a) + c^2 (\v a - \v b)\bigr)
\times \v{\hat z}}
{ 2 (\v a - \v b) \times (\v b - \v c) \cdot \v{\hat z}}.
\end{equation}
Project $\v o$ to geographic coordinates $O^{(1)}$ and use this as the
new center of projection.  Repeat these steps until convergence.  This
process converges to the required tri-point because of the equidistant
property of the projection and it converges rapidly because of the
projection is azimuthal.  If the points are sufficiently distant (in
other words if the center of the projection is sufficiently close to the
tri-point), Eq.~(\ref{tri-a}) can be replaced by
\begin{equation}\label{tri-b}
\v o = -\frac
{ \bigl((a - b) \v{\hat c} + (b - c) \v{\hat a} + (c - a) \v{\hat b}\bigr)
\times \v{\hat z}}
{(\v{\hat a} - \v{\hat b}) \times (\v{\hat b} - \v{\hat c}) \cdot \v{\hat z}}.
\end{equation}
Figure~\ref{medianfig} shows the result of using this method to
determine the median line separating Britain and her North Sea
neighbors.

With slight modifications this procedure can be applied if any of the
coast points are replaced by lines.  For example, assume that $A$ is
replaced by a line and let $A$ now denote the point on the line closest
to $O$.  At the same time as picking $O^{(0)}$, provide an estimate
$A^{(0)}$ for the position of $A$.  (These can be the result of solving
problem 5 of Sect.~\ref{triangulation}; however, it's more efficient to
interleave this solution with the iteration for the tri-point.)
Transform $A^{(0)}$, $B$, and $C$ to the azimuthal projection and obtain
$O^{(1)}$ using either Eq.~(\ref{tri-a}) or (\ref{tri-b}).  Also update
the estimates for $A$ to $A^{(1)}$ using one step of Newton's method
with Eq.~(\ref{oppangle-angle}).  In this update, use the computed angle
between the line and the geodesic from $O^{(0)}$ to $A^{(0)}$ corrected
for the anticipated change due to $\v o$; this is $\v o \times \v{\hat
a} \cdot \v{\hat z}$ divided by the reduced length of the geodesic from
$O^{(0)}$ to $A^{(0)}$.  Repeat these steps until convergence.

The median line between two points $A$ and $B$ (or lines) can be found
similarly.  In this case, introduce a third point $C$ (or line), use
the distance to $C$ as a control variable, and determine the point which
is equidistant from $A$ and $B$ and a distance $c_0$ from $C$.  By
adjusting $c_0$ a set of regularly spaced points along the median line
can be found.  The algorithm is similar to the solving for the tri-point
using Eq.~(\ref{tri-b}); however, the updated position of the median
point in the projected space is
\begin{equation}\label{median-a}
\v o = - \frac
{ \bigl((c-c_0)(\v{\hat a} - \v{\hat b}) - (a - b) \v{\hat c}\bigr)
\times \v{\hat z} }
{ (\v{\hat a} - \v{\hat b}) \times \v{\hat c} \cdot \v{\hat z} }.
\end{equation}

Boundaries which are a fixed distance from a state, typically
$12\,\mathrm{NM}$ for territorial seas or $200\,\mathrm{NM}$ for
exclusive economic zones, can be found using the same
machinery \citep[Chap.~5]{talos06}.  If the coast is defined by a set of
points, the boundary is a set of circular arcs which meet at points
which are equidistant from two coastal points.  The general problem is
to determine the point with is a distance $a_0$ from $A$ and $b_0$ from
$B$; $A$ is a coast point or a point on a coastal line and $a_0 =
12\,\mathrm{NM}$; $B$ is another such point when determining where two
circular arcs meet in which case $b_0 = a_0$, or is a control point in
which case $b_0$ measures off the distance along a circular arc.  This
is just problem 1 of Sect.~\ref{triangulation}; however, it can also be
solving using the azimuthal equidistant projection in a similar manner
to finding the median line.  The formula for updating the boundary point
in this case is
\begin{equation}\label{boundary-a}
\v o = \frac
{ \bigl((a-a_0)\v{\hat b} - (b-b_0)\v{\hat a}\bigr)
\times \v{\hat z} }
{ \v{\hat a} \times \v{\hat b} \cdot \v{\hat z} }.
\end{equation}

Figure \ref{medianfig} was obtained by exhaustively computing the
distances to all the coastal points at each point along the median line.
This is reasonably fast for small data sets, $N \lesssim 1000$ points.
For larger data sets, the points should be organized in such a way that
the distance to the closest point can be computed quickly.  For example,
the points can be organized into quad trees \citep{finkel74} where every
node in the tree can be characterized by a center and a radius $r$.  The
geodesic triangle inequality can be used to give bounds on the distance
$s$ from a point $P$ to any point within the node in terms of the
distance $s_0$ to its center, namely $\max(0, s_0 - r) \le s \le s_0 +
r$.  Once the quad tree has been constructed, which takes $O(N\log N)$
operations, computing the closest distance takes only $O(\log N)$
operations.

\section{Areas of a geodesic polygon}\label{areasec}

The last geodesic problem I consider is the computation of the area of a
geodesic polygon.  Here, I extend the method of \citet{danielsen89} to
higher order so that the result is accurate to round-off, and I recast
his series into a simple trigonometric sum which is amenable to Clenshaw
summation.  In formulating the problem, I follow \citet{sjoeberg06}.

The area of an arbitrary region on the ellipsoid is given by
\begin{equation}\label{area}
T = \int \d T,
\end{equation}
where $\d T = \cos\phi\,\d\phi\,\d\lambda/K$ is an element of area and $K$
is the Gaussian curvature.  Compare this with the Gauss--Bonnet
theorem \citep[\S34]{eisenhart40}
\citep[\S105]{bonnet48}
\begin{equation}\label{gauss-bonnet}
\Gamma = \int K \,\d T,
\end{equation}
where $\Gamma = 2\pi - \sum_j \theta_j$ is the geodesic excess.  This
form of the theorem applies only for a polygon whose sides are geodesics
and the sum is over its vertices and $\theta_j$ is the exterior angle at
vertex $j$.  Sj\"oberg combines Eqs.~(\ref{area}) and
(\ref{gauss-bonnet}) to give
\begin{align}
T &= c^2 \Gamma + \int \biggl(\frac1K - c^2\biggr)\cos\phi\,\d\phi\,\d\lambda
\notag\\
&=c^2 \Gamma + \int \biggl(
\frac{b^2}{(1 - e^2\sin^2\phi)^2} - c^2
\biggr)\cos\phi\,\d\phi\,\d\lambda,\label{area2}
\end{align}
where $c$ is a constant, and $K$ has been evaluated using
Eqs.~(\ref{weq0}) and (\ref{curvature}).  Now apply Eq.~(\ref{area2}) to
the geodesic quadrilateral $AFHB$ in Fig.~\ref{figtrig} for which
$\Gamma = \alpha_2 - \alpha_1$ and the integration over $\phi$ may be
performed to give
\begin{align}
S_{12}&=c^2 (\alpha_2 - \alpha_1)
+ b^2 \int_{\lambda_1}^{\lambda_2} \biggl(
\frac1{2(1 - e^2\sin^2\phi)}
\notag\\&\qquad\qquad{}+
\frac{\tanh^{-1}(e \sin\phi)}{2e \sin\phi}
- \frac{c^2}{b^2}\biggr)\sin\phi
\,\d\lambda,\label{area3}
\end{align}
where $S_{12}$ is the area of the geodesic quadrilateral and integral is
over the geodesic line (so that $\phi$ is implicitly a function of
$\lambda$).  Convert this to an integral over the spherical arc length
$\sigma$ using a similar technique to that used in deriving
Eq.~(\ref{lameq}).  Sj\"oberg chooses $c=b$; however, this leads to a
singular integrand when the geodesics pass over a pole.  (In addition,
he expresses the integral in terms of the latitude which leads to
greater errors in its numerical evaluation.)  In contrast, I define
\begin{equation}\label{authalic}
c^2 = R_q^2 = \frac{a^2}2 + \frac{b^2}2 \frac{\tanh^{-1}e}e,
\end{equation}
which is, in effect, the choice that Danielsen makes and which leads to
a non-singular integrand.  The quantity $R_q$ is the authalic radius,
the radius of the sphere with the same area as the ellipsoid.
Expressing $S_{12}$ in terms of $\sigma$ gives
\begin{align}
S_{12} &= S(\sigma_2) - S(\sigma_1),\label{S12}\\
S(\sigma) &= R_q^2\alpha + e^2a^2\cos\alpha_0 \sin\alpha_0 \,I_4(\sigma),
\label{Sdef}
\end{align}
where
\begin{align}
I_4(\sigma) &= -\int_{\pi/2}^\sigma
\frac{t(e'^2) - t(k^2\sin^2\sigma')}{e'^2-k^2\sin^2\sigma'}
\frac{\sin\sigma'}2 \,\d\sigma',
\label{I4eq}
\end{align}
and
\begin{equation}\label{t-eq}
t(x) = x + \sqrt{x^{-1} + 1}\sinh^{-1}\sqrt x,
\end{equation}
and I have chosen the limits of integration in Eq.~(\ref{I4eq}) so that
the mean value of the integral vanishes.  Expanding the integrand in
powers of $k^2$ and $e'^2$ (the same expansion parameters as Danielsen
uses) and performing the integral gives
\begin{equation}\label{i4}
I_4(\sigma) = \sum_{l = 0}^\infty C_{4l}\cos \bigl((2l+1)\sigma\bigr),
\end{equation}
where
\begin{align}
C_{40} &= \bigl(\tfrac2/3 - \tfrac1/15 e'^2 + \tfrac4/105 e'^4
        - \tfrac8/315 e'^6 + \tfrac64/3465 e'^8
        - \tfrac128/9009 e'^{10}\bigr)\notag\\&\quad
        - \bigl(\tfrac1/20 - \tfrac1/35 e'^2 + \tfrac2/105 e'^4
        - \tfrac16/1155 e'^6 + \tfrac32/3003 e'^8 \bigr) k^2\notag\\&\quad
        + \bigl(\tfrac1/42 - \tfrac1/63 e'^2 + \tfrac8/693 e'^4
        - \tfrac80/9009 e'^6\bigr) k^4\notag\\&\quad
        - \bigl(\tfrac1/72 - \tfrac1/99 e'^2 + \tfrac10/1287 e'^4
       \bigr) k^6\notag\\&\quad
        + \bigl(\tfrac1/110 - \tfrac1/143 e'^2\bigr) k^8
        - \tfrac1/156 k^{10} + \cdots,\displaybreak[0]\notag\\
C_{41} &= \bigl(\tfrac1/180 - \tfrac1/315 e'^2 + \tfrac2/945 e'^4
        - \tfrac16/10395 e'^6 + \tfrac32/27027 e'^8\bigr) k^2\notag\\&\quad
        - \bigl(\tfrac1/252 - \tfrac1/378 e'^2 + \tfrac4/2079 e'^4
        - \tfrac40/27027 e'^6\bigr) k^4\notag\\&\quad
        + \bigl(\tfrac1/360 - \tfrac1/495 e'^2 + \tfrac2/1287 e'^4
       \bigr) k^6\notag\\&\quad
        - \bigl(\tfrac1/495 - \tfrac2/1287 e'^2\bigr) k^8
        + \tfrac5/3276 k^{10} + \cdots,\displaybreak[0]\notag\\
C_{42} &= \bigl(\tfrac1/2100 - \tfrac1/3150 e'^2 + \tfrac4/17325 e'^4
        - \tfrac8/45045 e'^6\bigr) k^4\notag\\&\quad
        - \bigl(\tfrac1/1800 - \tfrac1/2475 e'^2 + \tfrac2/6435 e'^4
       \bigr) k^6\notag\\&\quad
        + \bigl(\tfrac1/1925 - \tfrac2/5005 e'^2\bigr) k^8
        - \tfrac1/2184 k^{10} + \cdots,\displaybreak[0]\notag\\
C_{43} &= \bigl(\tfrac1/17640 - \tfrac1/24255 e'^2 + \tfrac2/63063 e'^4
       \bigr) k^6\notag\\&\quad
        - \bigl(\tfrac1/10780 - \tfrac1/14014 e'^2\bigr) k^8
        + \tfrac5/45864 k^{10} + \cdots,\displaybreak[0]\notag\\
C_{44} &= \bigl(\tfrac1/124740 - \tfrac1/162162 e'^2\bigr) k^8
        - \tfrac1/58968 k^{10} + \cdots,\displaybreak[0]\notag\\
C_{45} &= \tfrac1/792792 k^{10} + \cdots.
\end{align}
I have included terms up to $O(f^5)$ so that the expression for
$S(\sigma)$ is accurate to $O(f^6)$.  This is consistent with the order
to which the distance and longitude integrals need to be evaluated to
give accuracy to the round-off limit for $f = 1/150$.  In contrast the
series given by \citet{danielsen89} gives $S$ accurate to $O(f^4)$.
\citet{clenshaw55} summation can be used to perform the (truncated) sum in
Eq.~(\ref{i4}), with
\[
\sum_{l=1}^L a_l \cos \bigl(l-\tfrac1/2 \bigr)x
     = (b_1 - b_2) \cos \tfrac1/2 x,
\]
where $b_l$ is given by Eq.~(\ref{clenshaw}).

Summing Eq.~(\ref{S12}) for each side of a polygon gives the total area
of the polygon, provided it does not include a pole.  If it does, then
$2\pi c^2$ should be added to the result.  The combined round-off and
truncation error in the evaluation of $I_4(\sigma)$ is about $5\times
10^{-3}\,\mathrm m^2$ for the WGS84 ellipsoid.  However, the bigger
source of errors is in the computation of the geodesic excess, i.e., the
first term in Eq.~(\ref{Sdef}); the sum of these terms gives the area of
the spherical polygon and, in Appendix \ref{spherearea}, I show how to
calculate this accurately.  The resulting errors in the area of polygon
can be estimated using the data for the errors in the azimuths given in
Sect.~\ref{errs}.  Typically, the error is approximately
$a\times15\,\mathrm{nm} = 0.1\,\mathrm m^2$ per vertex; however, it may
be greater if polygon vertices are very close to a pole or if a side is
nearly hemispherical.  Sometimes the polygon may include many thousands
of vertices, e.g., determining the area of the Japan.  In this case an
additional source of round-off error occurs when summing the separate
contributions $S_{12}$ from each edge; this error can be controlled by
using \citet{kahan65} summation.

This method of area computation requires that the sides of the polygon
be geodesics.  If this condition is fulfilled then the work of computing
the area is proportional to the number of sides of the polygon.  If the
sides are some other sort of lines, then additional vertices must be
inserted so that the individual sides are well approximated by
geodesics.  Alternatively the lines can be mapped to an equal-area
projection, such as the Albers conic projection \citep[\S14]{snyder87},
as suggested by \citet{gillissen93}, and the area computed in the
projected space.  In either of these cases, the work of computing the
area will be proportional to the perimeter of the polygon.

\section{Conclusions}\label{conclusions}

This paper presents solutions for the direct and inverse geodesic
problems which are accurate to close to machine precision; the errors
are less than $15\,\mathrm{nm}$ using double-precision arithmetic.  The
algorithm for the inverse problem always converges rapidly.  The
algorithms also give the reduced length and geodesic scale; these
provide scale factors for geodesic projections, allow Newton's method to
be used to solve various problems in ellipsoidal trigonometry, and enable
instrumental errors to be propagated through geodesic calculations.  I
introduce an ellipsoidal generalization of the gnomonic projection in
which geodesics project to approximately straight lines.  I discuss the
solution of several geodesic problems involving triangulation and the
determination of median lines.  I simplify and extend the formulas given
by \citet{danielsen89} for the area of geodesic polygons to arbitrary
precision.  The solution of the inverse geodesic problem uses a general
solution for converting from geocentric to geodetic coordinates
(Appendix \ref{geocent}).

Reviewing the list of references, it is remarkable the extent to which
this paper relies on 19th century authors.  Indeed my solution of the
direct geodesic problems is a straightforward extension of that given
by \citet{helmert80} to higher order.  The solution for the inverse
problem relies on two relatively modest innovations, the use of Newton's
method to accelerate convergence and a careful choice of the starting
guess to ensure convergence; however, the necessary machinery is all
available in \citet{helmert80}.  My advances, such as they are, rely on
a few 20th century innovations.  The most obvious one is the
availability of cheap hardware and software for flexibly carrying out
numerical calculations.  However, equally important for algorithm
development are software packages for algebraic manipulation and
arbitrary precision arithmetic, both of which are provided
by \citet{maxima}---these facilities have been available in Maxima since
the mid 1970s.

Computer code implementing much of this work is incorporated into
GeographicLib \citep{geographiclib17}.  This includes (a)~C++
implementations of the solutions for the direct and inverse geodesic
problem, (b)~methods for computing points along a geodesic in terms of
distance or spherical arc length, (c)~computation of the reduced length
$m_{12}$, the geodesic scales $M_{12}$ and $M_{21}$, and the area
$S_{12}$, (d)~implementations of the azimuthal equidistant,
Cassini--Soldner, and ellipsoidal gnomonic projections (all of which
return projection scales), (e)~command-line utilities for solving the
main geodesic problems, computing geodesic projections, and finding the
area of a geodesic polygon, (f)~Maxima code to generate the series
$I_j(\sigma)$ extract the coefficients $A_j$ and $C_{jl}$, and ``write''
the C++ code to evaluate the coefficients, (g)~the series expansions
carried out to 30th order, and (h)~the geodesic test data used in
Sect.~\ref{errs}.  The web
page \url{http://geographiclib.sf.net/geod.html} provides quick links to
all these resources.  In this paper, I have tried to document the
geodesic algorithms in GeographicLib accurately; the source code should
be consulted in case of any ambiguity.

I have been questioned on the need for nanometer accuracy when geodetic
measurements are frequently only accurate to about a centimeter.  I can
give four possible answers.  (1)~Geodesic routines which are accurate to
$1\,\mathrm{mm}$, say, can yield satisfactory results for simple
problems.  However, more complicated problems typically require much
greater accuracy; for example, the two-point equidistant projection may
entail the solution of ill-conditioned triangles for which millimeter
errors in the geodesic calculation would lead to much larger errors in
the results.  With accurate geodesic routines packaged as
``subroutines'', the azimuthal equidistant and Cassini--Soldner
projections (which are usually expressed as series with limited
applicability) can be easily and accurately computed for nearly the
whole earth.  The need for accuracy has has become more pressing with
the proliferation of ``geographic information systems'' which allow
users (who may be unaware of the pitfalls of error propagation) access
to geographic data.  (2)~Even if millimeter errors are tolerable, it is
frequently important that other properties of geodesics are well
satisfied, and this is best achieved by ensuring the geodesic
calculations are themselves very accurate.  An example of such a
property is the triangle inequality; this implies that the shortest path
between a point and a geodesic intersects the geodesic at right angles
and it also ensures the orthogonality of the polar graticule of the
azimuthal equidistant projection.  (3)~Accurate routines may be just as
fast as inaccurate ones.  In particular, the use of Clenshaw summation
means that there is little penalty to going to 6th order in the
expansions in Sect.~\ref{intser}.  On a $2.66\,\mathrm{GHz}$ Intel
processor with the g++ compiler, version 4.4.4, solving the direct
geodesic problem takes $0.88\,\mathrm{\mu s}$, while the inverse problem
takes $2.62\,\mathrm{\mu s}$ (on average).  Points along a geodesic can
be computed at the rate of $0.37\,\mathrm{\mu s}$
(resp.~$0.31\,\mathrm{\mu s}$) per point when the geodesic is
parametrized by distance (resp.~spherical arc length).  Thus the time to
perform a forward and reverse azimuthal equidistant projection
(equivalent to solving the inverse and direct geodesic problems) is
$3.4\,\mathrm{\mu s}$, which is only about $2$ times slower than
computing the transverse Mercator projection using Kr\"uger's
series \citep{karney11}.  The object code in geodesic code of
GeographicLib is substantially longer than an implementation of the
method of \citet{vincenty75a}, but, at about $30\,\mathrm{kbytes}$, it
is negligible compared to the available memory on most computers.
(4)~Finally, it is desirable that a well defined mathematical problem
have an accurate and complete computational solution;
paraphrasing \citet[p.~378]{gauss03} in a letter to Olbers (in 1827, on
the ellipsoidal corrections to the distribution of the geodesic excess
between the angles of a spheroidal triangle), ``the dignity of science
({\it die W\"urde der Wissenschaft\/})'' requires it.

\section*{Acknowledgments}

I would like to thank Rod Deakin and Dick Rapp for comments on a
preliminary version of this paper.  John Nolton kindly furnished me with
a copy of \citet{vincenty75b}.

\appendix

\section{Equations for a geodesic}\label{spheroid}

\begin{figure}[tb]
\begin{center}
\includegraphics[scale=0.75,angle=0]{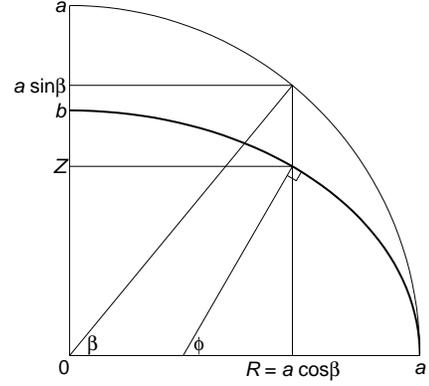}
\end{center}
\caption{\label{figredlat}
The construction for the reduced latitude.  The heavy curve shows a
quarter meridian of the spheroid for which the latitude $\phi$ is
defined as the angle between the normal and the equator.  Points are
transferred from the ellipsoid to the auxiliary sphere, shown as a light
curve, by preserving the radius of the circle of latitude $R$.  The
latitude $\beta$ on the auxiliary sphere is the reduced latitude defined
by $R = a \cos\beta$.}
\end{figure}%
Here, I give a derivation of Eqs.~(\ref{geodeq}) following, for the most
part, the presentation
of \citet{bessel25}.  \citet[Book~1, \S8]{laplace99} shows that the path
of a geodesic on a surface is the same as the motion of a particle
constrained to the surface but subject to no external forces.  For a
spheroid of revolution, conservation of angular momentum gives the
relation found by \citet{clairaut35},
\begin{equation}\label{clairaut0}
R\sin\alpha = a\sin\alpha_0,
\end{equation}
where $R$ is the distance from the axis of revolution (i.e., the radius
of the circle of latitude), $a$ is the maximum radius of the body,
$\alpha$ is the azimuth of the geodesic with respect to a meridian, and
$\alpha_0$ is the azimuth at the latitude of maximum radius.  Because
$R\le a$, $R$ can be written as $a\cos\beta$, where $\beta$ is the
latitude on the auxiliary sphere (see Fig.~\ref{figredlat}), and
Eq.~(\ref{clairaut0}) becomes
\begin{equation}\label{clairaut}
\cos\beta\sin\alpha = \sin\alpha_0.
\end{equation}
This is the sine rule applied to the angles $\alpha_0$ and $\pi-\alpha$
in the triangle $N\!EP$ on the auxiliary sphere in Fig.~\ref{figtri} and
establishes the correspondence with a geodesic on a spheroid with a
great circle on the auxiliary sphere.  It remains to establish the
relations between $\lambda$ and $s$ and their counterparts on the sphere
$\omega$ and $\sigma$.  For a given geodesic on the spheroid, an
elementary distance $\d s$ is related to changes in latitude and longitude
by \citep[Eqs.~(1)]{bessel25}
\begin{equation}\label{dseq}
\cos \alpha\,\d s = \rho\,\d\phi = - \d R/\sin\phi,\quad
\sin \alpha\,\d s = R\,\d\lambda,
\end{equation}
where $\rho$ is the meridional radius of curvature.  The corresponding
equations on the auxiliary sphere are
\begin{equation}\label{dsigeq}
a\cos \alpha\,\d\sigma = - \d R/\sin\beta,\quad
a\sin \alpha\,\d\sigma = R\,\d\omega.
\end{equation}
Dividing Eqs.~(\ref{dseq}) by Eqs.~(\ref{dsigeq})
gives \citep[Eqs.~(4)]{bessel25}
\begin{equation}\label{geodeqa}
\frac1a \frac{\d s}{\d\sigma} = \frac{\d\lambda}{\d\omega}
= \frac{\sin\beta}{\sin\phi}.
\end{equation}
These relations hold for geodesics on any spheroid of revolution.
Specializing now to an ellipsoid of revolution, parametrically given by
$R = a\cos\beta$ and $Z = b\sin\beta$.  The slope of the meridian
ellipse is given by
\[
\frac{\d Z}{\d R} = - \frac{b\cos\beta}{a\sin\beta} = - \frac{\cos\phi}{\sin\phi}.
\]
This gives the formula for the reduced latitude, Eq.~(\ref{redlat}), and
leads to
\[
\frac{\sin\beta}{\sin\phi} = \sqrt{1 - e^2\cos^2\beta} = w.
\]
Substituting this into Eqs.~(\ref{geodeqa}) gives Eqs.~(\ref{geodeq}).

\section{Transforming geocentric coordinates}\label{geocent}

\citet{vermeille02} presented a closed-form transformation
from geocentric to geodetic coordinates.  However, his solution does not
apply near the center of the earth.  Here, I remove this restriction and
improve the numerical stability of the method so that the method is
valid everywhere.  A key equation in Vermeille's method is the same as
Eq.~(\ref{kapeq}) and this method given here can therefore by used to
solve this equation.  While this paper was being
prepared, \citet{vermeille11} published an update on his earlier paper
which addresses some of the same problems.

As in the main body of this paper, the earth is treated as an ellipsoid
with equatorial radius $a$ and eccentricity $e$.  The geocentric
coordinates are represented by $(X,Y,Z)$ and the method transforms this
to geodetic coordinates $(\lambda,\phi,h)$ where $h$ is the height
measured normally from the surface of the ellipsoid.  Geocentric
coordinates are given in terms of geographic coordinates by
\begin{align*}
X &= (a\cos\beta + h\cos\phi) \cos\lambda,\\
Y &= (a\cos\beta + h\cos\phi) \sin\lambda,\\
Z &= b\sin\beta + h\sin\phi,
\end{align*}
where $\sin\beta = w\sin\phi$, $\cos\beta=w\cos\phi/\sqrt{1 - e^2}$ and
$w$ is given by Eq.~(\ref{weq}).  In inverting these equations, the
determination of $\lambda$ is trivial,
\[
\lambda = \ph(X + iY).
\]
This reduces the problem to a two-dimensional one, converting $(R,Z)$ to
$(\phi,h)$ where $R=\sqrt{X^2+Y^2}$.  Vermeille reduces the problem to
the solution of an algebraic equation
\begin{equation}\label{vkeq}
\kappa^4 + 2 e^2 \kappa^3 - (x^2 + y^2 - e^4) \kappa^2 -
2 e^2 y^2 \kappa - e^4 y^2 = 0,
\end{equation}
where
\[
x = R/a,\quad y = \sqrt{1-e^2} Z/a.
\]
Descartes' rule of signs shows that for $y \ne 0$, Eq.~(\ref{vkeq}) has
one positive root \citep[\S\dlmf{1.11(ii)}{1.11.ii}]{dlmf10}.  Similarly
for $x \ne 0$, it has one root satisfying $\kappa < -e^2$.  Inside the
astroid, $x^{2/3} + y^{2/3} < e^{4/3}$, there are two additional roots
satisfying $-e^2 < \kappa < 0$.

The geodetic coordinates are given by substituting the
real solutions for $\kappa$ into
\begin{align}
\phi &= \ph\bigl(R/(\kappa + e^2) + iZ/\kappa\bigr),\label{lateq}\\
h &= \biggl(1-\frac{1 - e^2}\kappa\biggr) \sqrt{D^2 + Z^2},\label{heq}
\end{align}
where $D = \kappa R/(\kappa + e^2)$.  The positive real root gives the
largest value of $h$ and I call this the ``standard solution''.

Equation~(\ref{vkeq}) may be solved by standard
methods \citep[\S\dlmf{1.11(iii)}{1.1.iii}]{dlmf10}.  Here, I summarize
Vermeille's solution modifying it to extend its range of validity and to
improve the accuracy.  The solution proceeds as follows
\begin{align*}
r &= \tfrac1/6 (x^2 + y^2 - e^4),\\
S &= \tfrac1/4 e^4 x^2 y^2,\\
d &= S (S + 2 r^3),\\
T &= \bigl(S + r^3 \pm \sqrt d\bigr)^{1/3},\\
u &= r + T + r^2/T.
\end{align*}
For $d \ge 0$, the sign of the square root in the expression for $T$
should match the sign of $S + r^3$ in order to minimize round-off
errors; also, the real cube root should be taken.  If $T = 0$, then take
$u = 0$.  For $d < 0$, $T$ is complex, and $u$ is given by
\begin{align*}
\psi &= \ph\bigl(-S - r^3 + i\sqrt{-d}\bigr),\\
T &= r \exp \tfrac1/3 i\psi,\\
u &= r \bigl(1 + 2 \cos\tfrac1/3 \psi\bigr).
\end{align*}

The right-hand side of Eq.~(\ref{vkeq}) may now be factored
into 2 quadratic terms in terms of $u$
\begin{equation}\label{kfac}
\kappa^2 + \frac{(v \pm u) \mp y^2}v e^2 \kappa \mp (v \pm u),
\end{equation}
where
\begin{align*}
v &= \sqrt{u^2 + e^4 y^2},\\
v \pm u &= \frac{e^4 y^2}{v \mp u}, \quad\text{for $u \lessgtr 0$},
\end{align*}
and where the latter equation merely gives a way to avoid round-off
error in the computation of $v \pm u$.  Only the factor with the upper
signs in Eq.~(\ref{kfac}) has a positive root given by
\begin{equation}
\kappa = \frac{v + u}{\sqrt{(v + u) + w^2} + w},
\end{equation}
where
\[
w = \frac{(v + u) - y^2}{2v}e^2.
\]
The number of real roots of Eq.~(\ref{vkeq}) is determined as follows.
The condition $d \gtrless 0$ is equivalent to $x^{2/3} +
y^{2/3} \gtrless e^{4/3}$.  For $d > 0$, only the quadratic factor with
the upper signs in Eq.~(\ref{kfac}) has real roots (satisfying $\kappa >
0$ and $\kappa < -e^2$, respectively).  For $d < 0$, both factors have
real roots yielding the four real roots of Eq.~(\ref{vkeq}).

Equations (\ref{lateq}) and (\ref{heq}) may become ill-defined if $x$
or $y$ vanishes.  Solving Eq.~(\ref{vkeq}) in the limit $x \rightarrow
0$ gives
\begin{equation}\label{xsmall}
\kappa = \pm y, \quad \kappa = -e^2 \pm e^2 x / \sqrt{e^4 - y^2}.
\end{equation}
Similarly in the limit $y \rightarrow 0$, Eq.~(\ref{vkeq}) yields
\begin{equation}\label{ysmall}
\kappa = -e^2 \pm x, \quad \kappa = \pm e^2 y / \sqrt{e^4 - x^2}.
\end{equation}
The only case where this these limiting forms are needed in
determining the standard solution are for $y = 0$ and $x \le e^2$.
Substituting the roots given by the second equation (\ref{ysmall})
into Eqs.~(\ref{lateq}) and (\ref{heq}) gives
\begin{align*}
\phi &= \ph\bigl(\sqrt{1 - e^2} x \pm i \sqrt{e^4 - x^2}\bigr),\\
h &= - b \sqrt{1 - x^2/e^2}.
\end{align*}

In the solution given here, I assumed that the ellipsoid is oblate.
This solution encompasses also the spherical limit, $e \rightarrow 0$;
the solution becomes $\kappa^2 \rightarrow x^2 + y^2$.  The method may
also be applied to a prolate ellipsoid, $e^2 < 0$.  Substituting $x =
y'$, $y = x'$, $\kappa = \kappa' - e^2$ in Eq.~(\ref{vkeq}) gives
\[
\kappa'^4 - 2 e^2 \kappa'^3 - (x'^2 + y'^2 - e^4) \kappa'^2 +
2 e^2 y'^2 \kappa' - e^4 y'^2 = 0,
\]
which transforms the problem for a prolate ellipsoid into an equivalent
problem for an oblate one.

In applying the results of this appendix to the inverse geodesic
problem, set $e = 1$ in order to convert Eq.~(\ref{vkeq}) into
Eq.~(\ref{kapeq}).

\section{Area of a spherical polygon}\label{spherearea}

The area $S_{12}$, Eq.~(\ref{area3}), includes the term $c^2 (\alpha_2
- \alpha_1)$.  Round-off errors in the evaluation of this term is a
potential source of error in determining $S_{12}$.  In this appendix, I
investigate ways to compute this term accurately.  This term
\begin{equation}\label{excess0}
E_{12} = \alpha_2 - \alpha_1
\end{equation}
is of course merely the spherical excess for the quadrilateral $AFGB$ in
Fig.~\ref{figtrig} transferred to the auxiliary sphere.  Thus $E_{12}$ is
the spherical excess for the quadrilateral with vertices
$(\beta_1,\omega_1)$, $(0,\omega_1)$, $(0,\omega_2)$, and
$(\beta_2,\omega_2)$.

If the geodesic $AB$ is determined by its arc
length $\sigma_{12}$ and its azimuth $\alpha_1$ at $A$ then use
Eq.~(\ref{napsb}) to determine $\alpha_2$ and so write
\begin{equation}\label{excess1}
\tan E_{12} = \frac
{\sin\alpha_0 \cos\alpha_0 (\cos\sigma_1 - \cos\sigma_2)}
{\sin^2\alpha_0 + \cos^2\alpha_0 \cos\sigma_1 \cos\sigma_2},
\end{equation}
with
\begin{multline*}
\cos\sigma_1 - \cos\sigma_2 =\\
\;\begin{cases}
\displaystyle
\sin\sigma_{12}\biggl(\frac{\cos\sigma_1 \sin\sigma_{12}}{1+\cos\sigma_{12}}
 + \sin\sigma_1\biggr), &\text{if $\cos\sigma_{12} > 0$},\\
\cos\sigma_1 (1 - \cos\sigma_{12}) + \sin\sigma_{12} \sin\sigma_1,
&\text{otherwise}.
\end{cases}
\end{multline*}
Here, $\alpha_0$ and $\sigma_1$ can be determined as described in
Sect.~\ref{direct}.

If the geodesic $AB$ is determined by the latitude and longitude of its
end points, then, for long arcs, determine $\alpha_1$ and $\alpha_2$
from Eqs.~(\ref{alpha1-spher}) and (\ref{alpha2-spher}), and substitute
these values into Eq.~(\ref{excess0}).  If, on the other hand, the arc
is short, use
\begin{equation}\label{excess2}
\tan\frac{E_{12}}2 =
\frac{\tan\tfrac1/2 \beta_1 + \tan\tfrac1/2 \beta_2}
{1 + \tan\tfrac1/2 \beta_1 \tan\tfrac1/2 \beta_2}
\tan\frac{\omega_{12}}2,
\end{equation}
where, if $\sin\theta$ and $\cos\theta$ are already known,
$\tan\frac12 \theta$ may be evaluated as $\sin\theta/(1+\cos\theta)$,.
This relation is the spherical generalization of the trapezoidal area;
in the limit $\beta_1 \rightarrow 0$, $\beta_2 \rightarrow 0$,
$\omega_{12} \rightarrow 0$, Eq.~(\ref{excess2}) becomes
\[
E_{12} \rightarrow \frac{\beta_1 + \beta_2}2 \, \omega_{12}.
\]
Equation (\ref{excess2}) takes on a simpler form if the latitude is
expressed in terms of the so-called isometric latitude, $\psi =
2\tanh^{-1}\tan\frac12 \beta = \sinh^{-1}\tan\beta$, namely
\[
\tan\frac{E_{12}}2 =
\tanh\frac{\psi_1 + \psi_2}2
\tan\frac{\omega_{12}}2.
\]

I obtained Eq.~(\ref{excess2}) from the formula for the area of a
spherical triangle, $E$, in terms of two of the sides, $a$ and $b$, and
their included angle $\gamma$ \citep[\S103]{todhunter71},
\[
\tan\frac E2 = \frac
{\tan\tfrac1/2 a \tan\tfrac1/2 b \sin\gamma}
{1+\tan\tfrac1/2 a \tan\tfrac1/2 b \cos\gamma},
\]
by substituting $a = \frac12 \pi + \beta_1$, $b = \frac12 \pi + \beta_2$
$\gamma = \omega_{12}$, and forming $E_{12} = E - \omega_{12}$.
However, it can also be simply found by using a formula
of \citet[\S11]{bessel25},
\[
\tan\frac{\alpha_2 - \alpha_1}2 =
\frac{\sin\tfrac1/2 (\beta_2 + \beta_1)}
{\cos\tfrac1/2 (\beta_2 - \beta_1)}
 \tan\frac{\omega_{12}}2,
\]
which, in turn, is derived from one of Napier's
analogies \citep[\S52]{todhunter71}.

The area for an $N$-sided spherical polygon is obtained by
\[
2\pi n - \sum_{i=1}^N E_{i-1,i},
\]
where $n$ is the number of times the polygon encircles the sphere in the
easterly direction.

\citet{miller94} proposed a formula for the area
of a spherical polygon which made use of L'Huilier's theorem for the
area of a spherical triangle in terms of its three
sides \citep[\S102]{todhunter71}.  However, any edge of the polygon
which is nearly aligned with a meridian leads to an ill-conditioned
triangle which results in about half of the precision of the
floating-point numbers being lost.

\section{Geodesics on a prolate ellipsoid}\label{prolate}

The focus in the paper has been on oblate ellipsoids.  However, most of
the analysis applies also to prolate ellipsoids ($f < 0$).  In the
appendix, I detail those aspects of the problem which need to be treated
differently in the two cases.

\begin{figure}[tb]
\begin{center}
\includegraphics[scale=0.75,angle=0]{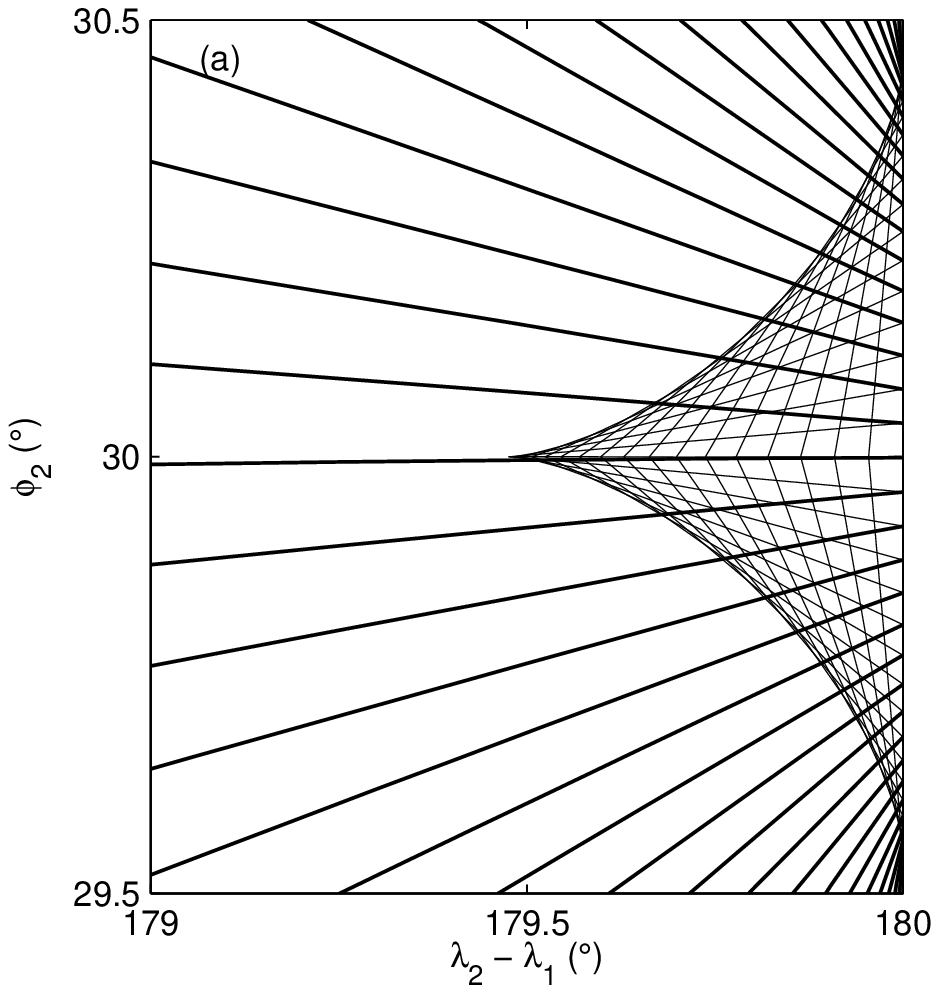}\\[10pt]
\includegraphics[scale=0.75,angle=0]{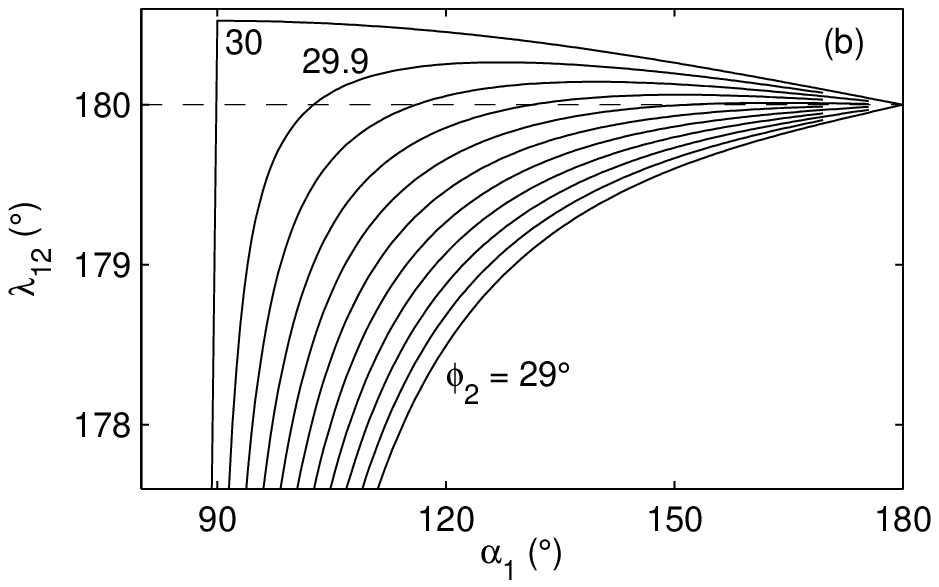}
\end{center}
\caption{\label{antipodalpro}
(a)~ Geodesics in antipodal region for a prolate ellipsoid.  This figure
is similar to Fig.~\ref{antipodalfig}a, except that $f = -1/297$.  In
addition the light lines are the continuation of the symmetric set of
west-going geodesics beyond the meridian $\lambda_{12} = 180^\circ$.
(b)~The dependence of $\lambda_{12}$ on $\alpha_1$ for the near
antipodal case with the same value of the flattening; compare with
Fig.~\ref{figalp2}b.}
\end{figure}%
All the series expansions given in Sect.~\ref{intser} and the expansions
for $S_{12}$ given in Sect.~\ref{areasec} are in terms of $f$, $n$,
$e^2$ or $e'^2$; prolate ellipsoids may be treated easily by allowing
these quantities to become negative.  The method of solving the direct
geodesic problem requires no alteration.  The solution of inverse
problem, on the other hand, is slightly different.  From
Eq.~(\ref{lameq}), it can be seen that the longitude difference for a
geodesic encircling the auxiliary sphere exceeds $2\pi$.  As a
consequence, the shortest geodesic between any two points on the equator
is equatorial; however, the shortest geodesics between two points on the
same meridian may not run along the meridian if the points are nearly
antipodal.  The test for meridional geodesics needs therefore to include
the requirement $m_{12} \ge 0$.  The solution for the inverse geodesic
problem is also unchanged except that the method of choosing starting
points for Newton's method needs to be altered in case 3 in
Fig.~\ref{inversefig}.  The envelope of the geodesics forms an astroid,
Fig.~\ref{antipodalpro}a, however, the $x$ and $y$ axes need to be
interchanged to match Fig.~\ref{antipodalfig}a, and with this
substitution, the derivation of the starting point depicted in
Fig.~\ref{astroidfig} applies.  Similarly region 3b in
Fig.~\ref{inversefig}, now lies along the meridian $\lambda_{12} = \pi$.
The techniques for solving the antipodal problem for a prolate ellipsoid
mirror closely those needed to transform geocentric coordinates as
described at the end of Appendix~\ref{geocent}.  The longitude
difference $\lambda_{12}$ as a function of $\alpha_1$ is no longer
monotonic when the ellipsoid is prolate; see Fig.~\ref{antipodalpro}b.
This potentially complicates the determination of $\alpha_1$;
nevertheless, the starting points used in region 3 are sufficiently
accurate that Newton's method converges.  The geodesic classes in
GeographicLib handle prolate ellipsoids; however, the algorithms have
not been thoroughly tested with geodesics on prolate ellipsoids.

Some of the closed form expressions can be recast into real terms for
prolate ellipsoids.  Thus Eqs.~(\ref{ellfun1})--(\ref{ellipt3}) should
be replaced by
\begin{align}
I_1(\sigma) &=\int_0^{u_2} \dn^2(u', k_2) \,\d u' =
E(\sigma, k_2),
\label{ellipt1p}
\displaybreak[0]\\
I_2(\sigma) &= u_2 = F(\sigma, k_2),
\label{ellipt2p}
\displaybreak[0]\\
I_3(\sigma) &= - \frac{1-f}f
\int_0^{u_2} \frac{\dn^2(u', k_2)}{1 - \cos^2\alpha_0\sn^2(u', k_2)}\,\d u'
\notag\\&\qquad
+\frac{\tan^{-1}(\sin\alpha_0\tan\sigma)}{f\sin\alpha_0}\notag\\
&=- \frac{1-f}f G(\sigma, \cos^2\alpha_0, k_2)
\notag\\&\qquad
+\frac{\tan^{-1}(\sin\alpha_0\tan\sigma)}{f\sin\alpha_0},
\label{ellipt3p}
\end{align}
where $k_2 = \sqrt{-k^2}$,
\[
\am(u_2, k_2) = \sigma,
\]
$\sn(x,k)$ and $\dn(x,k)$ are Jacobian elliptic functions
\citep[\S\dlmf{22.2}{22.2}]{dlmf10}, and $G(\phi,\alpha^2,k)$ is defined
by Eq.~(\ref{G-eq}), as before.  In this form, the integration constants
vanish.  Finally, Eq.~(\ref{authalic}) becomes
\begin{equation}
c^2 = \frac{a^2}2 + \frac{b^2}2 \frac{\tan^{-1}\sqrt{-e^2}}{\sqrt{-e^2}},
\end{equation}
and Eq.~(\ref{t-eq}), which appears in the integral for the geodesic
area, should be replaced by
\begin{equation}
t(-y) = -y + \sqrt{y^{-1} - 1}\sin^{-1}\sqrt y.
\end{equation}

\bibliography{geod}

\begin{thebibliography}{76}
\providecommand{\natexlab}[1]{#1}
\providecommand{\url}[1]{\texttt{#1}}
\providecommand{\urlprefix}{URL }
\expandafter\ifx\csname urlstyle\endcsname\relax
  \providecommand{\doi}[1]{doi:\discretionary{}{}{}#1}\else
  \providecommand{\doi}{doi:\discretionary{}{}{}\begingroup
  \urlstyle{rm}\Url}\fi
\providecommand{\eprint}[2][]{\url{#2}}

\bibitem[{Bagratuni(1967)}]{bagratuni67}
G.~V. Bagratuni, 1967, \emph{Course in spheroidal geodesy}, Technical Report
  FTD-MT-64-390, US Air Force, translation of {\it Kurs sferoidicheskoi
  geodezii} (Geodezizdat, Moscow, 1962), by Foreign Technology Division,
  Wright-Patterson AFB, \urlprefix\url{http://handle.dtic.mil/100.2/AD650520}.

\bibitem[{Beltrami(1865)}]{beltrami65}
E.~Beltrami, 1865, \emph{Risoluzione del problema: Riportare i punti di una
  superficie sopra un piano in modo che le linee geodetiche vengano
  rappresentate da linee rette}, Annali Mat. Pura App., \textbf{7}, 185--204,
  \urlprefix\url{http://books.google.com/books?id=dfgEAAAAYAAJ&pg=PA185}.

\bibitem[{Bessel(1825)}]{bessel25}
F.~W. Bessel, 1825, \emph{{\"U}ber die {B}erechnung der geographischen
  {L}\"angen und {B}reiten aus geod\"atischen {V}ermessungen}, Astron. Nachr.,
  \textbf{4}(86), 241--254, \doi{10.1002/asna.201011352}, translated into
  English by C. F. F. Karney and R. E. Deakin as {\it The calculation of
  longitude and latitude from geodesic measurements}, Astron. Nachr. {\bf
  331}(8), 852--861 (2010), \eprint{0908.1824},
  \urlprefix\url{http://adsabs.harvard.edu/abs/1825AN......4..241B}.

\bibitem[{Bonnet(1848)}]{bonnet48}
P.~O. Bonnet, 1848, \emph{M\'emoire sur la th\'eorie g\'en\'erale des
  surfaces}, J. l'\'Ecole Polytechnique, \textbf{19}(32), 1--146,
  \urlprefix\url{http://books.google.com/books?id=VGo_AAAAcAAJ&pg=PA1}.

\bibitem[{Bowring(1996)}]{bowring96}
B.~R. Bowring, 1996, \emph{Total inverse solutions of the geodesic and great
  elliptic}, Survey Review, \textbf{33}(261), 461--476.

\bibitem[{Bowring(1997)}]{bowring97}
---, 1997, \emph{The central projection of the spheroid and surface lines},
  Survey Review, \textbf{34}(265), 163--173.

\bibitem[{Bugayevskiy and Snyder(1995)}]{bugayevskiy95}
L.~M. Bugayevskiy and J.~P. Snyder, 1995, \emph{Map Projections: A Reference
  Manual} (Taylor \& Francis, London),
  \urlprefix\url{http://www.worldcat.org/oclc/31737484}.

\bibitem[{Bulirsch(1965)}]{bulirsch65}
R.~Bulirsch, 1965, \emph{Numerical calculation of elliptic integrals and
  elliptic functions}, Num. Math., \textbf{7}(1), 78--90,
  \doi{10.1007/BF01397975}.

\bibitem[{Carlson(1995)}]{carlson95}
B.~C. Carlson, 1995, \emph{Numerical computation of real or complex elliptic
  integrals}, Numerical Algorithms, \textbf{10}(1), 13--26,
  \doi{10.1007/BF02198293}, \eprint{math/9409227}.

\bibitem[{Cayley(1870)}]{cayley70}
A.~Cayley, 1870, \emph{On the geodesic lines on an oblate spheroid}, Phil. Mag.
  (4th ser.), \textbf{40}(268), 329--340,
  \urlprefix\url{http://books.google.com/books?id=Zk0wAAAAIAAJ&pg=PA329}.

\bibitem[{Christoffel(1910)}]{christoffel68}
E.~B. Christoffel, 1910, \emph{Allgemeine {T}heorie der geod\"atischen
  {D}reiecke (1868)}, in L.~Maurer, editor, \emph{Gesammelte Mathematische
  {A}bhandlungen}, volume~1, chapter~16, pp. 297--346 (Teubner, Leipzig),
  reprint of Math. Abhand. K\"onig. Akad. der Wiss. zu Berlin {\bf 8}, 119--176
  (1868),
  \urlprefix\url{http://books.google.com/books?id=9W9tAAAAMAAJ&pg=PA297}.

\bibitem[{Clairaut(1735)}]{clairaut35}
A.~C. Clairaut, 1735, \emph{D\'etermination g\'eometrique de la perpendiculaire
  \`a la m\'eridienne trac\'ee par {M}. {C}assini}, M\'em. de l'Acad. Roy. des
  Sciences de Paris 1733, pp. 406--416,
  \urlprefix\url{http://books.google.com/books?id=GOAEAAAAQAAJ&pg=PA406}.

\bibitem[{Clenshaw(1955)}]{clenshaw55}
C.~W. Clenshaw, 1955, \emph{A note on the summation of {C}hebyshev series},
  Math. Tables Aids Comput., \textbf{9}(51), 118--120,
  \urlprefix\url{http://www.jstor.org/stable/2002068}.

\bibitem[{Cox(1946)}]{cox46}
J.~F. Cox, 1946, \emph{The double equidistant projection}, Bull. G\'eod.,
  \textbf{2}(1), 74--76, \doi{10.1007/BF02521618}.

\bibitem[{Cox(1951)}]{cox51}
---, 1951, \emph{La projection \'equidistante pour deux points}, Bull. G\'eod.,
  \textbf{19}(1), 57--60, \doi{10.1007/BF02527404}.

\bibitem[{Danielsen(1989)}]{danielsen89}
J.~Danielsen, 1989, \emph{The area under the geodesic}, Survey Review,
  \textbf{30}(232), 61--66.

\bibitem[{Darboux(1894)}]{darboux94}
J.~G. Darboux, 1894, \emph{Le{\c c}ons sur la Th\'eorie G\'en\'erale des
  Surfaces}, volume~3 (Gauthier-Villars, Paris),
  \urlprefix\url{http://books.google.com/books?id=hGMSAAAAIAAJ}.

\bibitem[{Dionis~du S\'ejour(1789)}]{sejour89}
A.~P. Dionis~du S\'ejour, 1789, \emph{Trait\'e Analytique des Mouvemens
  apparens des Corps C\'elestes}, volume~2 (Valade, Paris),
  \urlprefix\url{http://books.google.com/books?id=tnHOAAAAMAAJ}.

\bibitem[{Eisenhart(1909)}]{eisenhart09}
L.~P. Eisenhart, 1909, \emph{A Treatise on the Differential Geometry of Curves
  and Surfaces} (Ginn \& Co., Boston),
  \urlprefix\url{http://books.google.com/books?id=hkENAAAAYAAJ}.

\bibitem[{Eisenhart(1940)}]{eisenhart40}
---, 1940, \emph{An Introduction to Differential Geometry} (Princeton Univ.
  Press), \urlprefix\url{http://www.worldcat.org/oclc/3165048}.

\bibitem[{Euler(1755)}]{euler55b}
L.~Euler, 1755, \emph{{\'E}l\'emens de la trigonom\'etrie sph\'ero\"{\i}dique
  tir\'es de la m\'ethode des plus grands et plus petits}, M\'em. de l'Acad.
  Roy. des Sciences de Berlin, \textbf{9}, 258--293,
  \urlprefix\url{http://books.google.com/books?id=QIIfAAAAYAAJ&pg=PA258}.

\bibitem[{Finkel and Bentley(1974)}]{finkel74}
R.~A. Finkel and J.~L. Bentley, 1974, \emph{Quad {T}rees, a data structure for
  retrieval on composite keys}, Acta Informatica, \textbf{4}(1), 1--9,
  \doi{10.1007/BF00288933}.

\bibitem[{Forsyth(1896)}]{forsyth96a}
A.~R. Forsyth, 1896, \emph{Geodesics on an oblate spheroid}, Mess. Math.,
  \textbf{25}, 81--124,
  \urlprefix\url{http://books.google.com/books?id=YsAKAAAAIAAJ&pg=PA81}.

\bibitem[{Gauss(1902)}]{gauss27}
C.~F. Gauss, 1902, \emph{General Investigations of Curved Surfaces of 1827 and
  1825} (Princeton Univ. Lib.), translation of {\it Disquisitiones generales
  circa superficies curvas}, by J. C. Morehead and A. M. Hiltebeitel,
  \urlprefix\url{http://books.google.com/books?id=a1wTJR3kHwUC}.

\bibitem[{Gauss(1903)}]{gauss03}
---, 1903, \emph{Werke}, volume~9 (Teubner, Leipzig),
  \urlprefix\url{http://books.google.com/books?id=ICwPAAAAIAAJ}.

\bibitem[{Gillissen(1993)}]{gillissen93}
I.~Gillissen, 1993, \emph{Area computation of a polygon on an ellipsoid},
  Survey Review, \textbf{32}(248), 92--98.

\bibitem[{Helmert(1880)}]{helmert80}
F.~R. Helmert, 1880, \emph{Die Mathematischen und Physikalischen {T}heorieen
  der H\"oheren {G}eod\"asie}, volume~1 (Teubner, Leipzig), translated into
  English by Aeronautical Chart and Information Center (St. Louis, 1964) as
  {\it Mathematical and Physical Theories of Higher Geodesy, Part 1}
  (\url{http://geographiclib.sf.net/geodesic-papers/helmert80-en.html}),
  \urlprefix\url{http://books.google.com/books?id=qt2CAAAAIAAJ}.

\bibitem[{Hilbert and Cohn-Vossen(1952)}]{hilbert52}
D.~Hilbert and S.~Cohn-Vossen, 1952, \emph{Geometry and the Imagination}
  (Chelsea, New York), translation of {\it Anschauliche Geometrie} (1932), by
  P. Nemenyi, \urlprefix\url{http://www.worldcat.org/oclc/301610346}.

\bibitem[{Hinks(1929)}]{hinks29}
A.~R. Hinks, 1929, \emph{A retro-azimuthal equidistant projection of the whole
  sphere}, Geog. J., \textbf{73}(3), 245--247,
  \urlprefix\url{http://www.jstor.org/stable/1784715}.

\bibitem[{Jacobi(1855)}]{jacobi55}
C.~G.~J. Jacobi, 1855, \emph{Solution nouvelle d'un probl\`eme de g\'eod\'esie
  fondamental}, Astron. Nachr., \textbf{41}(974), 209--216,
  \doi{10.1002/asna.18550411401}, op. post., communicated by E. Luther,
  \urlprefix\url{http://adsabs.harvard.edu/abs/1855AN.....41..209J}.

\bibitem[{Jacobi(1891)}]{jacobi91}
---, 1891, \emph{{\"U}ber die {C}urve, welche alle von einem {P}unkte
  ausgehenden geod\"atischen {L}inien eines {R}otationsellipsoides ber\"uhrt},
  in K.~T.~W. Weierstrass, editor, \emph{Gesammelte Werke}, volume~7, pp.
  72--87 (Reimer, Berlin), op. post., completed by A. Wangerin,
  \urlprefix\url{http://books.google.com/books?id=_09tAAAAMAAJ&pg=PA72}.

\bibitem[{Kahan(1965)}]{kahan65}
W.~M. Kahan, 1965, \emph{Further remarks on reducing truncation errors}, Comm.
  ACM, \textbf{8}(1), 40, \doi{10.1145/363707.363723}.

\bibitem[{Karney(2009)}]{geodbib}
C.~F.~F. Karney, 2009, \emph{An online geodesic bibliography},
  \urlprefix\url{http://geographiclib.sf.net/geodesic-papers/biblio.html}.

\bibitem[{Karney(2010)}]{geographiclib17}
---, 2010, \emph{Geographic{L}ib, version 1.7},
  \urlprefix\url{http://geographiclib.sf.net}.

\bibitem[{Karney(2011)}]{karney11}
---, 2011, \emph{Transverse {M}ercator with an accuracy of a few nanometers},
  J. Geod., \doi{10.1007/s00190-011-0445-3}, in press, \eprint{1002.1417}.

\bibitem[{Lagrange(1869)}]{lagrange70}
J.~L. Lagrange, 1869, \emph{Nouvelle m\'ethode pour r\'esoudre les
  \'equa\-tions litt\'erales par le moyen des s\'eries (1770)}, in
  \emph{Oeuvres}, volume~3, pp. 5--73 (Gauthier-Villars, Paris), reprint of
  M\'em. de l'Acad. Roy. des Sciences de Berlin {\bf 24}, 251--326 (1770),
  \urlprefix\url{http://books.google.com/books?id=YywPAAAAIAAJ&pg=PA5}.

\bibitem[{Laplace(1829)}]{laplace99}
P.~S. Laplace, 1829, \emph{Celestial Mechanics (1799)}, volume~1 (Hillard,
  Gray, Little, \& Wilkins, Boston), translation with commentary of {\it
  M\'ecanique C\'eleste}, by N. I. Bowditch,
  \urlprefix\url{http://books.google.com/books?id=k-cRAAAAYAAJ}.

\bibitem[{Legendre(1789)}]{legendre89}
A.~M. Legendre, 1789, \emph{M\'emoire sur les op\'erations trigonom\'etriques,
  dont les r\'esultats d\'ependent de la figure de la {T}erre}, M\'em. de
  l'Acad. Roy. des Sciences de Paris, 1787, pp. 352--383,
  \urlprefix\url{http://books.google.com/books?id=0uIEAAAAQAAJ&pg=PA352}.

\bibitem[{Legendre(1806)}]{legendre06}
---, 1806, \emph{Analyse des triangles trac\'es sur la surface d'un
  sph\'ero\"{\i}de}, M\'em. de l'Inst. Nat. de France, 1st sem., pp. 130--161,
  \urlprefix\url{http://books.google.com/books?id=-d0EAAAAQAAJ&pg=PA130-IA4}.

\bibitem[{Legendre(1811)}]{legendre11}
---, 1811, \emph{Exercices de Calcul Int\'egral sur Divers Ordres de
  Transcendantes et sur les Quadratures}, volume~1 (Courcier, Paris),
  \urlprefix\url{http://books.google.com/books?id=riIOAAAAQAAJ}.

\bibitem[{Letoval'tsev(1963)}]{letovaltsev63}
I.~G. Letoval'tsev, 1963, \emph{Generalization of the gnomonic projection for a
  spheroid and the principal geodetic problems involved in the alignment of
  surface routes}, Geodesy and Aerophotography, \textbf{5}, 271--274,
  translation of Geodeziya i Aerofotos'emka {\bf 5}, 61--68 (1963).

\bibitem[{Levallois(1970)}]{levallois70}
J.-J. Levallois, 1970, \emph{G\'eod\'esie G\'en\'erale}, volume~2 (Eyrolles,
  Paris), \urlprefix\url{http://www.worldcat.org/oclc/3870605}.

\bibitem[{Levallois and Dupuy(1952)}]{levallois52}
J.-J. Levallois and M.~Dupuy, 1952, \emph{Note sur le calcul des grandes
  g\'eod\'esiques}, Technical report, IGN, Paris,
  \urlprefix\url{http://www.worldcat.org/oclc/31726404}.

\bibitem[{Luther(1856)}]{luther56}
E.~Luther, 1856, \emph{Jacobi's {A}bleitung in seinem {A}ufsatze: ``{S}olution
  nouvelle d'un probl\`eme de g\'eod\'esie fondamental'' enthaltenen
  {F}ormeln}, Astron. Nachr., \textbf{42}(1006), 337--358,
  \doi{10.1002/asna.18550422201},
  \urlprefix\url{http://adsabs.harvard.edu/abs/1856AN.....42..337J}.

\bibitem[{Maxima(2009)}]{maxima}
Maxima, 2009, \emph{A computer algebra system, version 5.20.1},
  \urlprefix\url{http://maxima.sf.net}.

\bibitem[{Miller(1994)}]{miller94}
R.~D. Miller, 1994, \emph{Computing the area of a spherical polygon}, in P.~S.
  Heckbert, editor, \emph{Graphics Gems IV}, pp. 132--137 (Academic Press),
  \urlprefix\url{http://www.worldcat.org/oclc/29565566}.

\bibitem[{Minding(1830)}]{minding30}
F.~Minding, 1830, \emph{{\"U}ber die {C}urven des k\"urzesten {P}erimeters auf
  krummen {F}l\"achen}, J. Reine Angew. Math., \textbf{5}, 297--304,
  \doi{10.1515/crll.1830.5.297},
  \urlprefix\url{http://books.google.com/books?id=Y6wGAAAAYAAJ&pg=PA297}.

\bibitem[{Olver \emph{et~al.}(2010)Olver, Lozier, Boisvert, and Clark}]{dlmf10}
F.~W.~J. Olver, D.~W. Lozier, R.~F. Boisvert, and C.~W. Clark, editors, 2010,
  \emph{{NIST} Handbook of Mathematical Functions} (Cambridge Univ. Press),
  \urlprefix\url{http://dlmf.nist.gov}.

\bibitem[{Oriani(1806)}]{oriani06}
B.~Oriani, 1806, \emph{Elementi di trigonemetria sferoidica, {P}t. 1}, Mem.
  dell'Ist. Naz. Ital., \textbf{1}(1), 118--198,
  \urlprefix\url{http://www.archive.org/stream/memoriedellistit11isti#page/118%
}.

\bibitem[{Oriani(1808)}]{oriani08}
---, 1808, \emph{Elementi di trigonemetria sferoidica, {P}t. 2}, Mem. dell'Ist.
  Naz. Ital., \textbf{2}(1), 1--58,
  \urlprefix\url{http://www.archive.org/stream/memoriedellistit21isti#page/1}.

\bibitem[{Oriani(1810)}]{oriani10}
---, 1810, \emph{Elementi di trigonemetria sferoidica, {P}t. 3}, Mem. dell'Ist.
  Naz. Ital., \textbf{2}(2), 1--58,
  \urlprefix\url{http://www.archive.org/stream/memoriedellistit22isti#page/1}.

\bibitem[{Oriani(1833)}]{oriani33}
---, 1833, \emph{Nota aggiunta agli elementi della trigonometria sferoidica},
  Mem. dell'Imp. Reg. Ist. del Regno Lombardo-Veneto, \textbf{4}, 325--331,
  \urlprefix\url{http://books.google.com/books?id=6bsAAAAAYAAJ&pg=PA325}.

\bibitem[{Pittman(1986)}]{pittman86}
M.~E. Pittman, 1986, \emph{Precision direct and inverse solutions of the
  geodesic}, Surveying and Mapping, \textbf{46}(1), 47--54.

\bibitem[{Puissant(1831)}]{puissant31}
L.~Puissant, 1831, \emph{Nouvel essai de trigonom\'etrie sph\'ero\"{\i}dique},
  M\'em. de l'Acad. Roy. des Sciences de l'Inst. de France, \textbf{10},
  457--529,
  \urlprefix\url{http://books.google.com/books?id=KcjOAAAAMAAJ&pg=RA2-PA457}.

\bibitem[{Rainsford(1955)}]{rainsford55}
H.~F. Rainsford, 1955, \emph{Long geodesics on the ellipsoid}, Bull. G\'eod.,
  \textbf{37}(1), 12--22, \doi{10.1007/BF02527187}.

\bibitem[{Rapp(1991)}]{rapp91}
R.~H. Rapp, 1991, \emph{Geometric geodesy, part {I}}, Technical report, Ohio
  State Univ., \urlprefix\url{http://hdl.handle.net/1811/24333}.

\bibitem[{Rapp(1993)}]{rapp93}
---, 1993, \emph{Geometric geodesy, part {II}}, Technical report, Ohio State
  Univ., \urlprefix\url{http://hdl.handle.net/1811/24409}.

\bibitem[{RNAV(2007)}]{rnav07}
RNAV, 2007, \emph{Order 8260.54A, The {U}nited {S}tates Standard for Area
  Navigation}, U.S. Federal Aviation Administration, Washington, DC,
  \urlprefix\url{http://www.faa.gov/documentLibrary/media/Order/8260_54A.pdf}.

\bibitem[{Saito(1970)}]{saito70}
T.~Saito, 1970, \emph{The computation of long geodesics on the ellipsoid by
  non-series expanding procedure}, Bull. G\'eod., \textbf{98}(1), 341--373,
  \doi{10.1007/BF02522166}.

\bibitem[{Saito(1979)}]{saito79}
---, 1979, \emph{The computation of long geodesics on the ellipsoid through
  {G}aussian quadrature}, J. Geod., \textbf{53}(2), 165--177,
  \doi{10.1007/BF02521087}.

\bibitem[{Schmidt(2000)}]{schmidt00}
H.~Schmidt, 2000, \emph{Berechnung geod\"atischer {L}inien auf dem
  {R}otationsellipsoid im {G}renzbereich diametraler {E}ndpunkte}, Z. f.
  Vermess., \textbf{125}(2), 61--64,
  \urlprefix\url{http://www.gia.rwth-aachen.de/Forschung/AngwGeodaesie/geodaet%
ische_linie/artikel2/}.

\bibitem[{Sj\"oberg(2002)}]{sjoeberg02}
L.~E. Sj\"oberg, 2002, \emph{Intersections on the sphere and ellipsoid}, J.
  Geod., \textbf{76}(2), 115--120, \doi{10.1007/s00190-001-0230-9}.

\bibitem[{Sj\"oberg(2006)}]{sjoeberg06}
---, 2006, \emph{Determination of areas on the plane, sphere, and ellipsoid},
  Survey Review, \textbf{38}(301), 583--593.

\bibitem[{Snyder(1987)}]{snyder87}
J.~P. Snyder, 1987, \emph{Map projection---a working manual}, Professional
  Paper 1395, U.S. Geological Survey,
  \urlprefix\url{http://pubs.er.usgs.gov/publication/pp1395}.

\bibitem[{Sodano(1958)}]{sodano58}
E.~M. Sodano, 1958, \emph{A rigorous non-iterative procedure for rapid inverse
  solution of very long geodesics}, Bull. G\'eod., \textbf{48}(1), 13--25,
  \doi{10.1007/BF02537675}.

\bibitem[{TALOS(2006)}]{talos06}
TALOS, 2006, \emph{A Manual on Technical Aspects of the United Nations
  Convention on the Law of the Sea, 1982}, International Hydrographic Bureau,
  Monaco, 4th edition,
  \urlprefix\url{http://www.iho-ohi.net/iho_pubs/CB/S-51_Ed4-EN.pdf}.

\bibitem[{Tobey(1928)}]{tobey28}
W.~M. Tobey, 1928, \emph{Geodesy}, Technical Report~11, Geodetic Survey of
  Canada, Ottawa, \urlprefix\url{http://www.worldcat.org/oclc/839082}.

\bibitem[{Todhunter(1871)}]{todhunter71}
I.~Todhunter, 1871, \emph{Spherical Trigonometry} (Macmillan, London), 3rd
  edition, \urlprefix\url{http://books.google.com/books?id=3uBHAAAAIAAJ}.

\bibitem[{Vermeille(2002)}]{vermeille02}
H.~Vermeille, 2002, \emph{Direct transformation from geocentric coordinates to
  geodetic coordinates}, J. Geod., \textbf{76}(9), 451--454,
  \doi{10.1007/s00190-002-0273-6}.

\bibitem[{Vermeille(2011)}]{vermeille11}
---, 2011, \emph{An analytical method to transform geocentric into geodetic
  coordinates}, J. Geod., \textbf{85}(2), 105--117,
  \doi{10.1007/s00190-010-0419-x}.

\bibitem[{Vincenty(1975{\natexlab{a}})}]{vincenty75a}
T.~Vincenty, 1975{\natexlab{a}}, \emph{Direct and inverse solutions of
  geodesics on the ellipsoid with application of nested equations}, Survey
  Review, \textbf{23}(176), 88--93, addendum: Survey Review {\bf 23}(180), 294
  (1976), \urlprefix\url{http://www.ngs.noaa.gov/PUBS_LIB/inverse.pdf}.

\bibitem[{Vincenty(1975{\natexlab{b}})}]{vincenty75b}
---, 1975{\natexlab{b}}, \emph{Geodetic inverse solution between antipodal
  points}, unpublished report dated Aug. 28,
  \urlprefix\url{http://geographiclib.sf.net/geodesic-papers/vincenty75b.pdf}.

\bibitem[{Weingarten(1863)}]{weingarten63}
J.~Weingarten, 1863, \emph{{\"U}ber die {O}berfl\"achen f\"ur welche einer der
  beiden {H}auptkr\"ummungshalbmesser eine function des anderen ist}, J. Reine
  Angew. Math., \textbf{62}, 160--173, \doi{10.1515/crll.1863.62.160},
  \urlprefix\url{http://books.google.com/books?id=ggRCAAAAcAAJ&pg=PA160}.

\bibitem[{Wessel and Smith(2010)}]{gmt455}
P.~Wessel and W.~H.~F. Smith, 2010, \emph{Generic mapping tools, 4.5.5},
  \urlprefix\url{http://gmt.soest.hawaii.edu/}.

\bibitem[{Williams(2002)}]{williams02}
E.~A. Williams, 2002, \emph{Navigation on the spheroidal earth},
  \urlprefix\url{http://williams.best.vwh.net/ellipsoid/ellipsoid.html}.

\bibitem[{Williams(1997)}]{williams97}
R.~Williams, 1997, \emph{Gnomonic projection of the surface of an ellipsoid},
  J. Nav., \textbf{50}(2), 314--320, \doi{10.1017/S0373463300023936}.

\end{thebibliography}
\end{document}